\documentclass[backref, breaklinks, colorlinks, onecolumn]{aastex63}   
\hypersetup{linkcolor=blue,citecolor=blue,filecolor=cyan,urlcolor=magenta}

\def\lambar{\lambda\llap {--}}
\def\wcyc{\omega_{\rm B}}

\newcommand{\beq}{\begin{equation}}
\newcommand{\eeq}{\end{equation}}
\newcommand{\beqa}{\begin{eqnarray}}
\newcommand{\eeqa}{\end{eqnarray}}

\usepackage{bm}

\usepackage{enumitem}
\usepackage{mathtools}
\usepackage{graphicx}
\usepackage{amssymb}
\usepackage{amsmath}
\usepackage{ulem}
\usepackage{bm}
\usepackage{epstopdf}
\usepackage{color}
\usepackage[caption=false]{subfig}

\shorttitle{X-RAY AND GAMMA-RAY SHOCK EMISSION FROM SPIDER BINARIES}

\begin{document}
%\tighten

\title{X-ray Through Very-High-Energy Intrabinary Shock Emission from Black Widows and Redbacks}

\author[0000-0002-6363-1139]{C.~J.~T. van der Merwe}
\affiliation{Centre for Space Research, North-West University, Potchefstroom Campus, Private Bag X6001, Potchefstroom 2520, South Africa}

\author[0000-0002-9249-0515]{Z. Wadiasingh}
\affiliation{Astrophysics Science Division, NASA Goddard Space Flight Center, Greenbelt, MD 20771, USA}
\affiliation{Universities Space Research Association (USRA), Columbia, MD 21046, USA}
\affiliation{Centre for Space Research, North-West University, Potchefstroom Campus, Private Bag X6001, Potchefstroom 2520, South Africa}

\author[0000-0002-2666-4812]{C. Venter}
\affiliation{Centre for Space Research, North-West University, Potchefstroom Campus, Private Bag X6001, Potchefstroom 2520, South Africa}

\author[0000-0001-6119-859X]{A.~K. Harding}
\affiliation{Astrophysics Science Division, NASA Goddard Space Flight Center, Greenbelt, MD 20771, USA}

\author[0000-0003-4433-1365]{M.~G. Baring}
\affiliation{Department of Physics and Astronomy - MS 108, Rice University, 6100 Main St., Houston, TX 77251-1892, USA}

%%%%%%%%%%%%%%%%%%%%%%%%%%%%%%%%%%%%%%%%%%%%%%%%%%%%%%%%%%%%%%%%%%%%%%%%%%%%%%%%%%%%%%%%%%%%%%%%
\begin{abstract}
%%%%%%%%%%%%%%%%%%%%%%%%%%%%%%%%%%%%%%%%%%%%%%%%%%%%%%%%%%%%%%%%%%%%%%%%%%%%%%%%%%%%%%%%%%%%%%%%

Black widow and redback systems are compact binaries in which a millisecond pulsar heats and may even ablate its low-mass companion by its intense wind of relativistic particles and radiation. 
In such systems, an intrabinary shock can form as a site of particle acceleration and associated non-thermal emission.
We model the X-ray and gamma-ray synchrotron and inverse-Compton spectral components for select spider binaries, including diffusion, convection and radiative energy losses in an axially-symmetric, steady-state approach. Our new multi-zone code simultaneously yields energy-dependent light curves and orbital phase-resolved spectra. Using parameter studies and matching the observed X-ray spectra and light curves, and \textit{Fermi} Large Area Telescope spectra where available, with a synchrotron component, we can constrain certain model parameters. For PSR~J1723--2837 these are notably the magnetic field and bulk flow speed of plasma moving along the shock tangent, the shock acceleration efficiency, and the multiplicity and spectrum of pairs accelerated by the pulsar. This affords a more robust prediction of the expected high-energy and very-high-energy gamma-ray flux. We find that nearby pulsars with hot or flaring companions may be promising targets for the future Cherenkov Telescope Array. Moreover, many spiders are likely to be of significant interest to future MeV-band missions such as \textit{AMEGO} and \textit{e-ASTROGAM}.

\end{abstract}

\keywords{binaries: close -- pulsars: individual (PSR~B1957+20, PSR J1723$-$2837, PSR~J1311$-$3430, PSR~J2339$-$0533) --  radiation mechanisms: non-thermal -- X-rays: binaries -- gamma rays: stars -- stars: winds, outflows}

%%%%%%%%%%%%%%%%%%%%%%%%%%%%%%%%%%%%%%%%%%%%%%%%%%%%%%%%%%%%%%%%%%%%%%%%%%%%%%%%%%%%%%%%%%%%%%%%
\section{Introduction} \label{sec:Intro}
\defcitealias{Wadiasingh2017}{W17}
%%%%%%%%%%%%%%%%%%%%%%%%%%%%%%%%%%%%%%%%%%%%%%%%%%%%%%%%%%%%%%%%%%%%%%%%%%%%%%%%%%%%%%%%%%%%%%%%

Rotation-powered millisecond pulsars (MSPs) that have retained a low-mass stellar binary companion can exhibit a rich set of phenomenology and physics in a relatively well-constrained system. 
The \textit{Fermi} Large Area Telescope (LAT) has detected about two dozen\footnote{See the \href{https://confluence.slac.stanford.edu/display/GLAMCOG/Public+List+of+LAT-Detected+Gamma-Ray+Pulsars}{public {\it Fermi} LAT pulsar list} {and candidates in \cite{2020arXiv200403128T}.}} of these systems, and the number of MSPs is also expected to grow significantly with continued operation of {\it Fermi} and with next-generation radio facilities. Two subsets of binary MSPs in nearly-circular, short (typically\footnote{More rarely, some redbacks exceed this timescale, e.g., 1FGL J1417.7$-$4402 at 5.4  days \citep{2018ApJ...866...83S} and PSR J1740$-$5340 at 1.4 days \citep{2010ApJ...709..241B}.} $\lesssim 1$ day) orbital periods are the ``spiders": black widows (BWs) and redbacks \citep[RBs;][]{2011AIPC.1357..127R}. These spiders\footnote{For a recent observationally-focused review, see \cite{2019Galax...7...93H}.} are chiefly differentiated by the mass $M_{\rm c}$ of their tidally-locked companion, with the mass of RB companions $M_{\rm c} \gtrsim 0.1 M_\odot$ while for BWs $M_{\rm c} \ll 0.1 M_\odot$. In RB systems, the companion is usually bloated to nearly the Roche limit, well beyond the radius of an isolated main-sequence star of similar mass. Since the MSP is often well-timed in the radio or by the LAT, the pulsar mass function is precisely known. Likewise, optical studies of the companion can constrain the companion mass function and orbital inclination \citep[e.g.,][]{2007MNRAS.379.1117R,2011ApJ...728...95V,2011ApJ...743L..26R,2013ApJ...769..108B,2015ApJ...804..115R,2016ApJ...816...74B,2018ApJ...862L...6D}.
Therefore all the orbital elements can be adequately constrained, enabling these systems to be excellent laboratories for the study of pulsar winds and particle acceleration in relativistic magnetized outflows.

When in a rotation-powered state, many spiders exhibit a non-thermal (synchrotron radiation; SR) orbitally-modulated emission component in the X-ray band, attributed to particle acceleration in an intrabinary pulsar wind termination shock (IBS) and Doppler-boosting of a bulk flow along the shock tangent \citep[e.g.,][]{2015hasa.confE..29W,2016arXiv160603518R,Wadiasingh2017}. {This type of geometry/radation picture was identified over two decades ago for the BW system involving the MSP B1957+20 \citep{Arons_Tavani_1993ApJ} and later for the gamma-ray emitting PSR B1259-63/Be star binary system \citep{Tavani_Arons_1997ApJ}.} The existence of a shock is also indicated by orbital-phase and frequency-dependent radio eclipses of the MSP (since plasma structures at the shock attenuates the radio), where in RBs the eclipse fraction can be $> 50\%$ \citep[e.g.,][]{2009Sci...324.1411A,2013arXiv1311.5161A,2016MNRAS.459.2681B,2018JPhCS.956a2004M} of the orbital phase (while the MSP remains largely uneclipsed at inferior conjunction of the pulsar). The hard power laws inferred in X-rays imply emission due to an energetic electron population. The X-ray spectra may furthermore extend to at least $\gtrsim50$~keV with no suggestion of spectral cutoffs \citep[e.g.,][]{2014ApJ...791...77T,2017ApJ...839..130K,2018ApJ...861...89A}, constraining the shock magnetic field to $B_{\rm sh}\gtrsim1$~G \citep{Wadiasingh2017}. For a given pulsar spin-down power $\dot{E}_{\rm SD} \sim 10^{34}-10^{35}$ erg s$^{-1}$, the magnetic field $B_{\rm sh}$ at the shock can be bounded using the Poynting flux $B_{\rm w}^2c/4\pi$ for a magnetic field $B_{\rm w} \propto R^{-1}$ in the striped wind outside the light cylinder.  The flux can be integrated over a spherical surface of area $4\pi R_{\rm sh}^2$ at the shock radius $R_{\rm sh}$ to yield an isotropic electromagnetic luminosity $B_{\rm w}^2R_{\rm sh}^2c \sim \dot{E}_{\rm SD}$.  {One therefore arrives at the constraint $B_{\rm sh} < B_{\rm w} \sim (\dot{E}_{\rm SD}/ c)^{1/2}\, /\, R_{\rm sh}$, which is detailed in Eq.~(\ref{eq:B_1}), being generally less than 100 Gauss.}

This may be recast as a condition that the ratio of electromagnetic energy density to particle pressure, $\sigma$, is less than unity, i.e., that $\dot{E}_{\rm SD}$ is dominated by the plasma wind contribution. As will be apparent in due course (cf.\ Section~\ref{sec:Results}), if the particle acceleration is as fast and efficient (attaining the synchrotron burn-off limit) as in pulsar wind nebulae, MeV synchrotron components may be observable by future medium-energy, gamma-ray Compton/pair telescopes, such as \textit{e-ASTROGAM} \citep{de2017astrogam} and \textit{AMEGO}\footnote{AMEGO: {\sl https://asd.gsfc.nasa.gov/amego/index.html} \citep{2019BAAS...51g.245M}}.

Owing to irradiation and heating powered by the pulsar, many companions are optically bright, implying a rich target photon field for inverse Compton (IC) scattering by the same population of relativistic electrons implied and constrained by the observed X-ray components. Yet, optical emission can also be highly variable \citep[e.g.,][]{2016ApJ...833L..12V}, providing opportunities for synergetic time-resolved coordinated studies among optical and imaging atmospheric Cherenkov telescope (IACT) TeV facilities. Detection of such an IC component would imply maximum particle energies in the TeV range; given high magnetic fields such as those in Eq.~(\ref{eq:B_1}), this may imply SR components extending into the MeV band prior to the synchrotron burn-off limit at a few hundred MeV \citep{de1996gamma}. Tentative LAT signals \citep{hui2014exploring,an2017high} for some systems can also be employed to constrain the maximum particle energies, as demonstrated in this work.

Recently, \cite{2020arXiv200507888A} reported on an orbitally-modulated gamma-ray component in RB J2339$-$0533 with modulation maximum at pulsar superior conjunction (i.e., $180^\circ$ phase offset from the X-ray double peaks) that is an enhancement of the MSP's pulsed GeV emission. This component must arise from a different energetic leptonic population than the shock emission we consider in this work. For instance, an upstream relativistic pulsar wind with low magnetization and bulk Lorentz factor of $\Gamma_{\rm w} \sim 10^4$ may produce such a component via anisotropic IC. 

Alternatively, the same ultrarelativistic (electron Lorentz factor $\gamma_{\rm e} \sim 10^7$) pairs which produce pulsed GeV emission in the current sheet via curvature radiation \citep{2019ApJ...883L...4K} may pass through the shock and enter the kilogauss \citep{wadiasingh2018pressure} magnetosphere of the companion and produce GeV emission via rapid synchrotron cooling. Phase coherence with GeV pulsations of the MSP would be maintained since the ultrarelativistic leptons cool in ${\cal O}(10^{-4})$ s in such kilogauss fields, a cooling time much shorter than the MSP spin period. 
We defer treatment such additional non-shock emission components to a different work.

A significant, now well-established and perhaps unusual feature of many\footnote{This is perhaps a selection effect for the brightest sources.} spiders is that the intrabinary shock appears to enshroud the pulsar, not the companion, as first noted in \cite{2015hasa.confE..29W}. Such a shock orientation is supported both by the centering of double-peaked X-ray orbital modulation at the pulsar inferior conjunction as well as large radio eclipse fractions of the MSP at pulsar superior conjunction. In this case, a large fraction of the pulsar's total wind output is captured, enabling high radiative efficiency for a given spin-down power. The low-mass companion's wind is far too feeble to overpower the MSP wind near the pulsar in such an orientation for timescales of months or years, as observed -- therefore other means such as companion magnetospheric pressure or irradiation feedback must be invoked for the global shock geometry and orientation \citep{wadiasingh2018pressure}. 

Geometrical models with Doppler-boosting for X-ray orbital modulation can generally explain the broad morphology of the light curves well. Yet, as noted in \citet{Wadiasingh2017}, any spatial dependence of the particle power-law index along the shock implies energy-dependent light curves, which simple geometric models cannot capture. In general, the particle acceleration spectral index is expected to vary along the shock tangent owing to differing magnetic obliquity of the shock normal and how the anisotropic striped wind is processed. Crucially, as geometric models do not capture the details of the electron population, they also cannot predict the energetics of various emission components and their relative normalization -- that is, the orbital-phase-dependent spectral energy distribution (SED). {{ We note that an earlier study of such IC components and SEDs \citep[][]{2014A&A...561A.116B} assumed a shock orientation appropriate for BWs, which has since been ruled out by later observations of X-ray orbital modulation \citep[see][]{Wadiasingh2017}. \cite{2014A&A...561A.116B} also assumed efficient mixing between the pulsar and companion winds, so that the downstream flow was nonrelativistic and therefore did not involve any Doppler boosting or anisotropic particle distributions; their model is thus superseded by that developed here. }}

In this paper, we take a first step beyond purely geometric models. We report on a new multi-zone code {\tt{UMBRELA}} (Unraveling the Multi-wavelength Beamed Radiation for Energetic Leptons in Arachnids) which solves for the steady-state particle distributions in spatial zones on the shock surface with variable bulk motion, and calculates SR and IC emission beamed toward an observer. This scheme may be regarded as similar to multi-zone blazar jet models \citep[e.g.,][]{2012MNRAS.423..756P} but with time and spatially-variable Doppler factors from a curved ``jet" comprising the locale of the intrabinary bow shock along with an external Compton component arising from the companion's dominant photon field. 
In \S\ref{sec:Model} we describe our model, we present a parameter study and predicted spectra and light curves in \S\ref{sec:Results}, and give our conclusions in \S\ref{sec:Conclusion}.

%%%%%%%%%%%%%%%%%%%%%%%%%%%%%%%%%%%%%%%%%%%%%%%%%%%%%%%%%%%%%%%%%%%%%%%%%%%%%%%%%%%%%%%%%%%%%%%%
\section{Formalism and Model Assumptions} \label{sec:Model}
%%%%%%%%%%%%%%%%%%%%%%%%%%%%%%%%%%%%%%%%%%%%%%%%%%%%%%%%%%%%%%%%%%%%%%%%%%%%%%%%%%%%%%%%%%%%%%%%

\begin{figure}[t]
    \centering
    \includegraphics[scale=0.5]{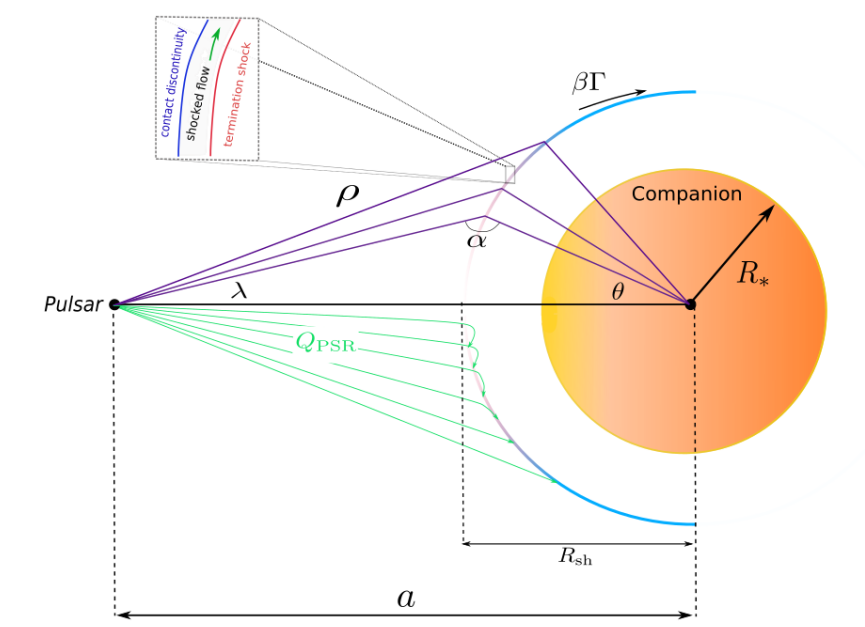}
    \caption{Schematic diagram indicating a cross-section of the shock wrapping around the companion, {putatively corresponding to the BW case,} with parameters defined as indicated.  The pulsar wind is emanating from the pulsar, indicated by green lines, and the particles are captured by and flow along the intrabinary shock. In this work, we approximate the shock as a 3D thin hemispherical shell, rather than one of finite thickness and multi-layer structure (as alluded to in the inset). The particles are accelerated at the shock locale to ultrarelativistic energies and acquire slight anisotropy in their steady-state distribution function (mildly relativistic ``bulk motion") along the shock tangent, as indicated by in increase in the bulk momentum $\beta\Gamma$ and the corresponding blue color.}
    \label{fig:schematic_BW}
\end{figure}

There is some debate about the source of pressure balance for the formation and orientation of IBSs in spider binaries, which impacts the three-dimensional shock shape. For instance, if wind ram pressures govern the shock, approximate thin-shell bow shock models such as those by \citet{wilkin1996exact,Canto1996,rodriguez1998compact} for the collision of two isotropic or isotropic-parallel winds might be decent approximations. Conversely, if a companion magnetosphere \citep[e.g.,][]{Harding1990,wadiasingh2018pressure} supports the pressure balance in the shock, the morphology may be dramatically different than the wind scenario. The anisotropy of the pulsar wind and its dependence on pulsar magnetic obliquity will also likely impact the final shock structures realized in the population of spider binaries. However, for expediency, we assume an infinitely thin, hemispherical shock morphology as a zeroth-order approximation for a shock surrounding either the pulsar or the companion (Fig.~\ref{fig:schematic_BW} and Fig.~\ref{fig:schematic_RB}). Such an ``umbrella''-like symmetry and hemispherical structure 

should be generic for any non-pathological shock structure close to the shock apex (``nose''), and therefore is a reasonable approximation at this juncture. We implement azimuthal symmetry about the line joining the pulsar and companion ($\partial/\partial\phi = 0$), as well as a steady-state regime ($\partial/\partial t = 0$). Relativistic electrons are injected by the pulsar and captured by the shock, where a substantial magnetic field $B_{\rm sh}$ leads to SR. Isotropic blackbody (BB) emission at a temperature $T$ from the facing hemisphere of the companion in the observer frame is presumed as a soft-photon target field for IC. Following our evaluation of the transport of particles,
we calculate SR and IC emissivities, but neglect synchrotron self-Compton (SSC) emission as it is expected to  be negligible for spider binaries (Appendix~\ref{sec:AppSSC}).

%%%%%%%%%%%%%%%%%%%%%%%%%%%%%%%%%%%%%%%%%%%%%%%%%%%%%%%%%%%%%%%%%%%%%%%%%%%%%%%%%%%%%%%%%%%%%%%%
\subsection{Treatment of Particle Acceleration and Injection in {\tt UMBRELA}}
%%%%%%%%%%%%%%%%%%%%%%%%%%%%%%%%%%%%%%%%%%%%%%%%%%%%%%%%%%%%%%%%%%%%%%%%%%%%%%%%%%%%%%%%%%%%%%%%

\label{sec:Injection}
{The upstream time-averaged Poynting flux in the pulsar wind is a moderate function of the magnetic inclination angle $\alpha$ \citep[e.g.,][]{Tchekhovskoy2016}, the variation being roughly a factor of two from an aligned to orthogonal rotator, noting that the plasma component loads the magnetosphere and contributes significantly to the spin-down torque on the pulsar \citep[e.g.,][]{Spitkovsky06}. 
The pulsar wind is ``striped'',} as field lines of opposite polarity make up the  undulating current sheet. At large distances the stripes may disrupt as magnetic field energy is converted to particle energy (thus, lowering $\sigma$).
Magnetohydrodynamic (MHD) and kinetic plasma instabilities generate hydromagnetic and plasma turbulence, and precipitate localized and dynamic magnetic reconnection and associated heating of charges. Thus the entropy increases, and these decoherence effects tend to isotropize the MHD flow in the wind frame. To simplify the treatment here, in particular for considerations of the energy budget in spatial zones along the IBS  (Fig.~\ref{fig:schematic_BW}), we assume that the pulsar wind is isotropic, {and omit detailed consideration of modest anisotropies identified in  \cite{Tchekhovskoy2016}. }

To reduce the number of free model parameters, we assume that a fraction $\eta_{\rm p} <1$ of the pulsar's spin-down luminosity $\dot{E}_{\rm SD}$ is converted into particle/pair luminosity $L_{\rm pair}$. We do not model the microphysics of particle acceleration, but  assume that this process takes place during the motion of particles from the pulsar to the downstream shock locale. This can include energization in pulsar ``gap'' potentials inside the light cylinder, and subsequent additional acceleration in the striped wind and at the pulsar termination shock.
Using the ratio $\sigma$ of magnetic to plasma energy densities, and the pulsar pair wind plus electromagnetic field energy density (luminosity)
$L_{\rm PSR} = L_{\rm pair} + L_{\rm field} = (1+\sigma )L_{\rm pair} = 4\pi R_{\rm sh}^2 n_{\rm pair}\gamma_{\rm pair} m_ec^3 (1+\sigma) \lesssim {\dot E}_{\rm SD}$,
one can estimate \citep[e.g.,][]{1984ApJ...283..710K} the magnetic field strength $B_1$ just upstream of the termination or {intrabinary shock:
\begin{equation}
    B_{1} \approx \left( \frac{\dot{E}_{\rm SD}}{c R^2_{\rm sh}} \frac{\sigma}{1 + \sigma} \right)^{1/2}  \sim 160 \left( \frac{\sigma}{1 + \sigma}
    \frac{\dot{E}_{\rm SD}}{10^{35} \,\,\rm erg \, s^{-1}}
     \right)^{1/2} \left(\frac{R_{\rm sh}}{10^{10} \, \, \rm cm} \right)^{-1}  \quad {\rm G}\;\; .
 \label{eq:B_1}
\end{equation}
The downstream} $B$-field in the shock rest frame can be obtained from the Rankine-Hugoniot jump conditions for energy and momentum flux conservation  \citep{1984ApJ...283..694K,2004ApJ...600..485D}.  For $\sigma < 0.1$, corresponding to a high Alfv\'enic Mach number shock, the result for an ultra-relativistic flow upstream is 
\begin{equation}
    B_{\rm sh} \approx 3(1 - 4\sigma)B_{1}\;\; .
\end{equation}
The field is thus compressed at the shock, by the same factor as the plasma density. This common compression applies to a so-called perpendicular shock, wherein the fields in the observer/shock frame lie approximately in the plane of the termination shock; it is the natural result of a relativistic transformation of fields from the wind frame. For the compression ratio of an ultra-relativistic shock of arbitrary field obliquity, see Eq.~(47) of \cite{2004ApJ...600..485D}. Combining the above expressions,
\begin{equation}
    B_{\rm sh} \approx 3(1 - 4\sigma)\left( \frac{\dot{E}_{\rm SD}}{c R^{2}_{\rm sh}} \frac{\sigma}{1 + \sigma} \right)^{1/2}
    \; =\; 3(1 - 4\sigma)\left( \frac{\sigma L_{\rm pair}}{c R^{2}_{\rm sh}} \right)^{1/2} \;\; ,
 \label{eq:B_shock}
\end{equation}
noting that $L_{\rm pair}$ is expressed in Eq.~(\ref{eq:L_pm_QPSR}) below, {wherein the ratio $\eta_{\rm p} = L_{\rm pair}/{\dot E}_{\rm SD} \equiv 1/(1+\sigma)$ is defined.  In the models presented in Section~3, both $B_{\rm sh}$ and $R_{\rm sh}$ will be free parameters. At face value, using Eq.~(\ref{eq:B_shock}), this leads to the specification of $\sigma$ since $\dot{E}_{\rm SD}$ is fixed by the pulsar spin down. Inserting optimal values for $B_{\rm sh}\sim1-10$~G required to match the data in the X-ray band (see Section~\ref{sec:Results} and Table~1)}, we find the following low values for $\sigma$ {at the intrabinary shock} for different systems modeled in Section~\ref{subsec:source_specific}: PSR~J1723$-$2837 - $\sigma \sim 0.001$; PSR~J2339$-$0533 - $\sigma \sim 0.0001$; PSR~J1311$-$3430 - $\sigma \sim 0.00007$; PSR~B1957+20 - $\sigma \sim 0.0003$. {Accordingly, the pulsar wind is plasma-dominated when it impinges upon the IBS, and $\sigma$ will not be explicitly specified in the models below that generate $\eta_{\rm p}\approx 1$.  We remark that these estimates captured from Eq.~(\ref{eq:B_shock}) apply to a constant, spherical wind, and in general $\sigma$ is a function of locale along the IBS, the details of which will be deferred to future work.}

We implement particle injection as follows. At the shock, a parametric form for the injection spectrum is employed:
\citep[e.g.,][]{Venter2015}:
\begin{equation}
 Q_{\rm PSR}(\gamma_{\rm e}) = Q_0 \gamma_{\rm e}^{-p}\exp\left(-\frac{\gamma_{\rm e}}{\gamma_{ \rm e, max}}\right),\label{eq:Q}
\end{equation}
with $p$ the spectral index, $\gamma_{\rm e}$ the electron Lorentz factor, and $Q_0$ the normalisation coefficient. This injection represents the cumulative input from electrodynamic energization in the pulsar magnetosphere and the striped wind, combined with acceleration at the bow shock. The maximum Lorentz factor $\gamma_{ \rm e, max}$ can be estimated or limited using three different methods.
The first two set restrictions on the particle acceleration in the shock, whereas the third estimate is the voltage across the open field lines, the maximum energy that is available in the system for particle acceleration. The estimates are:\\
(1) The Hillas criterion for when the shock radius $R_{\rm sh}$ and the lepton's gyroradius $r_g = pc/eB_{\rm sh}$ are equal, yielding \citep{hillas1984origin}
\begin{equation}
   \gamma_{\rm e,\,max}^{\rm H} \approx \frac{e R_{\rm sh}B _{\rm sh}}{m_{\rm e} c^2}\sim 2\times10^8\left(\frac{R_{\rm sh}}{10^{10}~{\rm cm}}\right)\left(\frac{B_{\rm sh}}{10~{\rm G}}\right).
\label{eq: Emax}
\end{equation}
(2) Balancing the diffusive acceleration rate at the shock with the SR loss-rate timescale \citep[e.g.,][]{de1996gamma}
\begin{equation}
   \gamma_{\rm e,\,max}^{\rm acc} = \frac{3}{2}\sqrt{ \frac{\epsilon_{\rm acc} B_{\rm cr}}{\alpha_{\rm f} B_{\rm sh}}} \sim4\times10^7 \epsilon_{\rm acc}^{1/2} \left(\frac{B_{\rm sh}}{10~{\rm G}}\right)^{-1/2}\!\!\!\!\!\!\!\!.
\end{equation}
(3) The pulsar polar cap (PC) voltage drop $\Phi_{\rm open}$, limiting the maximum primary electron energy to
\begin{equation}
   \gamma_{\rm e,\,max}^{\rm P} = \frac{e\Phi_{\rm open}}{m_{\rm e}c^2}\sim 5\times10^{8}\left(\frac{P}{5\times10^{-3}~\rm s} \right)^{-2} \left(\frac{R_{\rm PSR}}{10^{6}~\rm cm} \right)^3 \left( \frac{B_{\rm PSR}}{10^{9}~\rm G} \right).
   \label{eq:gammax_GJ}
\end{equation}
Here $e$ is the elementary charge, $m_{\rm e}$ is the electron mass, $c$ the speed of light in vacuum, { $\alpha_{\rm f}=e^2/\hbar c$ is the fine structure constant, and $B_{\rm cr}=m_{\rm e}^2c^3/e\hbar \approx 4.41\times10^{13}$~G is the quantum critical magnetic field, at which the cyclotron energy $\hbar \wcyc $ of the electron equals its rest mass energy $m_{\rm e}c^2$.   Here $\wcyc = eB/m_{\rm e}c$ is the electron  cyclotron frequency, which defines the scale for the rate of gyroresonant diffusive acceleration. The dimensionless parameter $\epsilon_{\rm acc} \equiv r_{\rm g}/ \lambda (\gamma_{\rm e}) \leq 1 $ expresses the diffusive mean free path $\lambda (\gamma_{\rm e})$ in units of the electron's gyroradius $r_{\rm g}$.  It describes the complexities of diffusion near the shock and how they impact the acceleration rate; $\epsilon_{\rm acc}$} is around unity in the so-called Bohm limit when lepton diffusion is quasi-isotropic and on the gyroradius scale, and can be much {less} than unity if field turbulence that seeds diffusion is at a low level \cite[e.g., see][for blazar contexts]{BBS_2017MNRAS_blazar}.

The third estimate applies for a pulsar of radius $R_{\rm PSR}$ and surface polar field strength $B_{\rm PSR}$. For a rotational period $P$, corresponding to an angular frequency $\Omega=2\pi/P$, \cite{goldreich1969pulsar}.

Evaluating the resulting potential drop for the open-field-line region leads to $\Phi_{\rm open}=\Omega^2R_{\rm PSR}^3B_{\rm PSR}/2c^2$, yielding a maximum $\gamma\sim10^8$; cf.\
Eq.~(\ref{eq:gammax_GJ}), which defines the most optimistic acceleration case.

We enforce $\gamma_{\rm e,max}~\equiv~\min \left(\gamma_{\rm e,\,max}^{\rm H}, \gamma_{\rm e,\,max}^{\rm acc}, \gamma_{\rm e,\,max}^{\rm P} \right)$.
The third estimate of the Lorentz factor is typically an absolute upper limit; however, in our case we choose $B_{\rm sh}$ independently, so we have to take the minimum of these factors as the maximum particle energy.

We normalize the particle injection spectrum by requiring \citep{sefako2003constraints} the pulsar wind to be the sole supplier of pair flux incident upon the IBS: 
\begin{equation}
   \int_{\gamma_{\rm e, min}}^\infty Q_{\rm PSR}\,d\gamma_{\rm e} =  (M_{\rm pair}+1)\dot{N}_{\rm GJ},\quad
    \dot{N}_{\rm GJ} \sim \frac{4\pi^2 B_{\rm PSR}R_{\rm PSR}^3}{P^2c\, e} \; .
 \label{eq:Ndot_GJ_QPSR}
\end{equation}
This pair flux is thereby benchmarked using the Goldreich-Julian {\it primary} particle injection rate $\dot{N}_{\rm GJ}\equiv I_{\rm GJ}/e$ for pulsar magnetospheres, which is expressed in terms of the current $I_{\rm GJ} \sim \sqrt{\dot{E}_{\rm SD}\, c}\;$. The Goldreich-Julian current $I_{\rm GJ}$ is determined as follows. The charge density is $\rho \approx - \boldsymbol{\Omega} \cdot \mathbf{B} /(2\pi c)$, so that $\vert \rho \vert \approx B_{\rm PSR}/Pc$ at the surface. For small PC sizes, the area of both PCs that are proximate to this charge is $A_{\rm cap} \approx 2\pi R_{\rm PC}^2 = (4 \pi^{2} R^3_{\rm PSR}/Pc )$. The flux of charge through this area is then $I_{\rm GJ} = \vert \rho\vert A_{\rm cap} c$ and reduces to $e {\dot N}_{\rm GJ}$ with ${\dot N}_{\rm GJ}$ as given in Eq.~(\ref{eq:Ndot_GJ_QPSR}).
The primary flux is enhanced by the pair multiplicity $M_{\rm pair}$ in a magnetospheric pair cascade, i.e., the number of pairs spawned per primary accelerated in the pulsar's PC $E_{\parallel}$ field. Typically $M_{\rm pair} \sim 10^{2}$ -- $10^{5}$ results from pulsar models; see just below.

A parallel constraint is that the power in pairs impinging upon the shock not exceed the pulsar spin-down power $\dot{E}_{\rm SD}$. Thus, we introduce a conversion efficiency $\eta_{\rm p}<1$ that is defined by the pair luminosity relation
\begin{equation}
   L_{\rm pair} \equiv m_{\rm e} c^2\int_{\gamma_{\rm e, min}}^\infty \gamma_{\rm e} Q_{\rm PSR}\,d\gamma_{\rm e}  =   \eta_{\rm p}\dot{E}_{\rm SD},\quad
   \dot{E}_{\rm SD} = \frac{4\pi^2I\dot{P}}{P^3},
 \label{eq:L_pm_QPSR}
\end{equation}
with $I$ the moment of inertia and $\dot{P}$ the time derivative of the pulsar period.
The value of $\eta_{\rm p}<1$ includes the density compression across the relativistic intrabinary shock. It is evident, given the form for $Q_{\rm PSR}$ in Eq.~(\ref{eq:Q}), that these pair flux and power equations are coupled. Thus,
by fixing $p$, $M_{\rm pair}$, $\eta_{\rm p}$, and integrating up to $5\gamma_{ \rm e, max}$, one may self-consistently solve for $Q_0$ as well as the minimum particle energy $\gamma_{\rm e,min}$~\citep{Venter2015}. Attributing the 
spin-down to losses in rotational kinetic energy in the usual manner, one can estimate the pulsar surface polar magnetic field using the standard vacuum orthogonal rotator formula: $B_{\rm PSR} \approx 6.4\times10^{19} \sqrt{P \dot{P}}$ G.  Since plasma contributions to the electromagnetic torque are generally comparable to the vacuum ones \citep[e.g., see][for force-free MHD considerations]{Spitkovsky06}, these formulae as estimates are generally secure. For instance, the magnetic obliquity of the MSP (or additional multipolar components) generally has an order of unity influence on inferred parameters \citep[e.g.,][]{Tchekhovskoy2016}.
Thus, given this closure, we may specify $Q_{\rm PSR}$.

%%%%%%%%%%%%%%%%%%%%%%%%%%%%%%%%%%%%%%%%%%%%%%%%%%%%%%%%%%%%%%%%%%%%%%%%%%%%%%%%%%%%%%%%%%%%%%%%
\subsection{Pair Cascades and Pair Multiplicity}
%%%%%%%%%%%%%%%%%%%%%%%%%%%%%%%%%%%%%%%%%%%%%%%%%%%%%%%%%%%%%%%%%%%%%%%%%%%%%%%%%%%%%%%%%%%%%%%%

PC pair cascades are thought to occur in all pulsars, both for generating coherent radio emission \citep{Philippov2020} and for supplying plasma for the current closure in the magnetosphere \citep{Brambilla2018}. 
Strong electric fields that develop above the PCs  accelerate particles to Lorentz factors exceeding $10^7$ and their curvature radiation photons can create electron-positron pairs via magnetic photon pair production ($\gamma \to e^\pm$) in the strong magnetic field \citep{DH1983}. The pairs screen the electric field and short out the gap potential, which again builds up in cycles as the pairs clear the gap \citep{Timokhin2013}. The bursts of pairs from the gap then continue to produce further generations of pairs through SR and IC radiation above the gap to build up the full multiplicity $M_{\rm pair}$  \citep{Daugherty1982,Timokhin2015}. The maximum multiplicity can reach $\sim 10^5$ \citep{Timokhin2019} and most likely increases with pulsar spin-down power, because both higher magnetic fields and smaller field line radii of curvature enhance the magnetic pair creation rate. However, the detailed dependence is still an open question.  

Monte Carlo simulations have modeled the pair cascade spectra, which are hard power laws with low- and high-energy cutoffs that depend on the pulsar period and period derivative \citep{Harding2011}. The pair spectral index is roughly $p_{\rm pair} \sim 1.5$ near the low-energy cutoff and the spectrum is curved so that the index increases with energy. Interestingly, such a hard index for the underlying electron population is suggested by spectrum of the orbitally-modulated X-ray emission in RBs. This index may or may not be modified by energization at the IBS, depending on the intrinsic acceleration characteristics of that shock. If the shock is weak enough to naturally generate power laws with indices $p > 1.6 - 1.7$, then the injection distribution from the pulsar will maintain its index $p_{\rm pair}$, but be boosted up to higher Lorentz factors \citep[e.g.,][]{Jones_Ellison_1991SSRv}. If instead, the shock naturally generates very flat non-thermal distributions with $1 < p < 1.5$, for example through efficient shock drift energization in weak shock layer turbulence \citep{Summerlin_2012ApJ}, then this is the index that is appropriate for the injection function $Q_{\rm PSR}$. Note that in this study the index $p$ is a free parameter, so that we do not consider specific details of the shock acceleration characteristics.
For young pulsars, low-energy cutoffs are around $\gamma_{\rm pair,min} \sim 100$ and high-energy cutoffs vary between $\gamma_{\rm pair,max} \sim 10^5$ and $\gamma_{\rm pair,max} \sim 10^6$.  For MSPs, low-energy cutoffs are around $\gamma_{\rm pair,min} \sim 10^4$ and high-energy cutoffs vary between $\gamma_{\rm pair,max} \sim 10^6$ and $\gamma_{\rm pair,max} \sim 10^7$. The different behaviors result from the much lower surface magnetic fields of MSPs, where the photon energies must be higher in order to produce pairs that have energies roughly equal to half of the parent photon energy.

Pair production by MSPs has been puzzling, since most have surface magnetic fields that are too low to initiate pair cascades in purely dipolar fields. As a result, studies of pair cascades have long suggested the presence of non-dipolar magnetic fields near the stellar surface to produce the pairs needed for radio emission \citep{Ruderman1975,Arons1979}.  \citet{Harding2011} found that deviations from a dipole field could enable pair cascades in older pulsars, including MSPs, and significantly increase the pair multiplicities. The recent result from the Neutron Star Interior Composition Explorer (\textit{NICER}) inferring a heated spot geometry on the surface of MSP J0030+0451 \citep{Riley2019,Miller2019} that is far from antipodal strongly suggests the presence of non-dipolar fields as well as a severely-offset dipole { \citep[][Kalapotharakos et al. 2020, submitted]{Bilous2019}}. This gives direct support to the idea that MSP surface fields are highly non-dipolar, giving them the ability to produce high pair multiplicity, though probably lower than for young rotation-powered pulsars (RRPs). As we demonstrate in Section~\ref{sec:Results}, the data seem to be consistent with theoretical constraints on particle energies and high pair multiplicities for pair injection from MSPs. This may have implications for the locally-measured energetic positron excess \citep{Venter2015}

%%%%%%%%%%%%%%%%%%%%%%%%%%%%%%%%%%%%%%%%%%%%%%%%%%%%%%%%%%%%%%%%%%%%%%%%%%%%%%%%%%%%%%%%%%%%%%%%
\subsection{Shock Orientation and Division into Spatial Zones}
%%%%%%%%%%%%%%%%%%%%%%%%%%%%%%%%%%%%%%%%%%%%%%%%%%%%%%%%%%%%%%%%%%%%%%%%%%%%%%%%%%%%%%%%%%%%%%%%

\begin{figure}[t]
    \centering
    \includegraphics[scale=0.5]{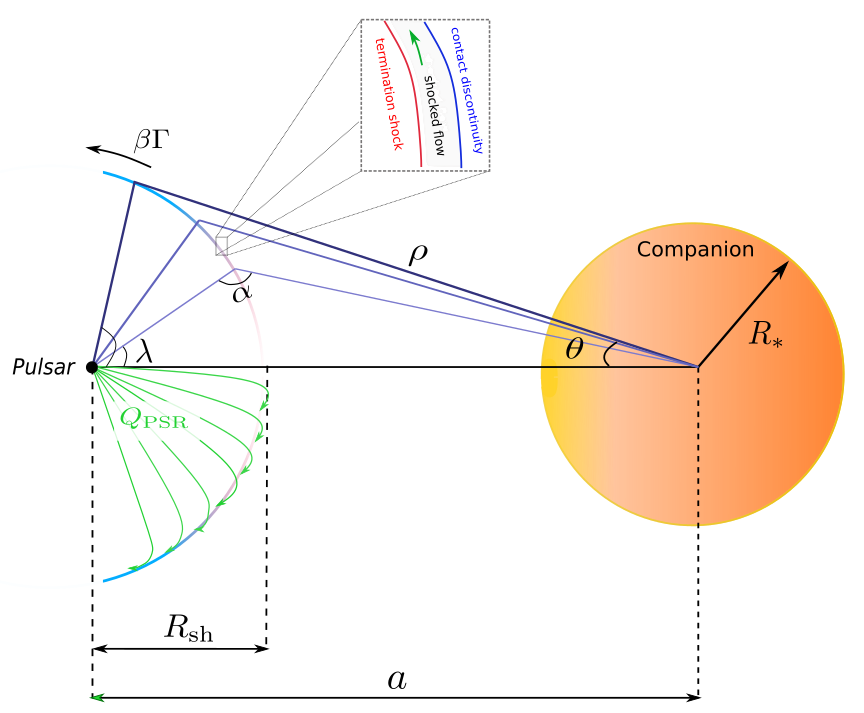}
    \caption{Schematic diagram indicating a cross-section of the shock wrapping around the pulsar, {putatively corresponding to the RB  case,} with parameters defined as indicated.  The pulsar wind is emanating from the pulsar, indicated by green lines, and the particles are captured by and flow along the intrabinary shock.}
    \label{fig:schematic_RB}
\end{figure}

Shock orientation is a key parameter choice in {\tt UMBRELA}. In BWs such as B1957+20, observations suggest that the shock wraps around the companion. In contrast, in RBs this orientation appears to be reversed. This influences how much of the pulsar wind is captured by and interacts with the shock, and hence affects the radiative power of the system in the X-ray band and beyond. 
Expectations for the RB case are then that the IC flux will be lower than in the BW case (given a lower photon energy density, but also depending on the companion temperature) and that the SR flux will be higher (for a higher magnetic field), and these are borne out by model output (Section~\ref{sec:Results}).
Furthermore, the velocity tangent to the spherical shock is changed from $u_x$ to $-u_x$ in the RB case (Section~\ref{sec:Beaming}). The effect of this is a 0.5 phase change in the position of the light curve peak compared to the BW case. We assume that the particle flow is initially radial upstream of the termination shock, but then along the shock tangent downstream, which is taken to be azimuthally symmetric (Fig.~\ref{fig:schematic_RB}). We denote the azimuthal angle about the $z$ axis that is oriented along the line joining the two stellar centers by $\phi_{\rm z}$.

\begin{figure}[t!]
    \centering
    \includegraphics[scale=0.45]{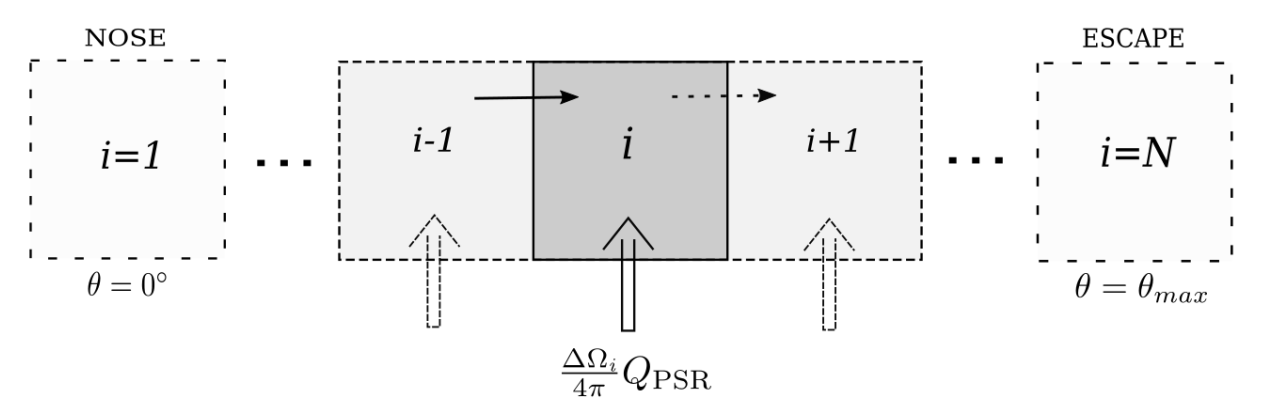}
    \caption{Schematic of particle injection and transport in zones along the shock colatitude sectors. A fraction of $Q_{\rm PSR}$ is injected into the $i^{\rm th}$ zone of the shock (as indicated by large arrows). The small arrows between cells signify the direction of bulk motion of particles flowing away from the nose of the shock.}
    \label{fig:Zones}
\end{figure}

We divide the shock into multiple zones of equal width in $\mu\equiv\cos\theta$, with $\theta$ the angle measured from the companion center with respect to the $z$ axis. When the shock is wrapped around the pulsar, let us label the latitude as measured from the pulsar by $\lambda$, the shock radius from the {pulsar} center by $R_{\rm sh}$, $a$ the orbital separation, and the distance from the companion to the shock is measured as $\rho$ (see Fig.~\ref{fig:schematic_BW}).
Using elementary trigonometry, we find 
\begin{eqnarray}
    \rho(\theta) & = & \sqrt{R_{\rm sh}^2 + a^2 -2R_{\rm sh}a\cos\theta}, \label{eq:rho}\\
    \lambda(\theta) & = & \sin^{-1}\left(\frac{\rho(\theta)\sin\theta}{R_{\rm sh}}\right).
    \label{eq:lambda}
\end{eqnarray}
In the BW case (shock around companion) the origin is centered on the companion with $\theta$ running clockwise and being used to define the spatial zones along the shock. The distance from the companion center to a particular zone is always $R_{\rm sh}$. However, when we consider the RB case (shock around pulsar; see Fig.~\ref{fig:schematic_RB}), the origin is centered on the pulsar 
using $\lambda(\theta)$ (Eq.~[\ref{eq:lambda}]) to indicate the spatial zones along the shock. To calculate the distance from the companion center to a particular zone we then use a cosine rule as seen in Eq.~(\ref{eq:rho}).
Most of the power of the pulsar wind (if spin and orbital axes are aligned) is confined to the equator at a pulsar co-latitude of $\theta_{\rm PSR}=\pi/2$ or $\lambda=0$, corresponding to the nose of the shock. For small values of $\lambda$, we approximate the injection spectrum assuming isotropy of the pulsar wind near the nose. Furthermore, the $i^{\rm th}$ zone in the shock intercepts a fraction of the total injected wind $Q_{\rm PSR}$ such that $Q_i=[d\Omega(\lambda_i)/4\pi ]Q_{\rm PSR}$, as illustrated by large arrows in Fig.~\ref{fig:Zones}, with $d\Omega(\lambda_{i})$ the solid angle subtended by the $i^{\rm th}$ zone.

For zones $i>1$, there is also a contribution of particles diffusing isotropically from the previous zone into the current one, as indicated by the small arrows in Fig.~\ref{fig:Zones}. It thus follows that the injection spectrum for the first zone is
\begin{eqnarray}
    Q_1  =  \left(\frac{1}{4\pi}\int_0^{2\pi} d\phi \int_{\lambda_1}^{\lambda_2}\!\!\sin\lambda\,d\lambda \right) Q_{\rm PSR}
     =   \frac{1}{2}\left(\cos\lambda_1 - \cos\lambda_2\right)Q_{\rm PSR},
     \label{eq:Q_PSR}
\end{eqnarray}
while the injection spectra for other zones ($i=2,\dots N$) are
\begin{eqnarray}
    Q_i & = & \frac{N_{{\rm e},i-1}}{t_{{\rm diff},i-1}} + \frac{1}{2}\left(\cos\lambda_i - \cos\lambda_{i+1}\right)Q_{\rm PSR},\label{eq:Q1}  
\end{eqnarray}
with $t_{\rm diff}$ the diffusion timescale defined in Section~\ref{sec:Transport}.

%%%%%%%%%%%%%%%%%%%%%%%%%%%%%%%%%%%%%%%%%%%%%%%%%%%%%%%%%%%%%%%%%%%%%%%%%%%%%%%%%%%%%%%%%%%
\subsection{The Convection-Diffusion and Particle Transport Equation} \label{sec:Transport}
%%%%%%%%%%%%%%%%%%%%%%%%%%%%%%%%%%%%%%%%%%%%%%%%%%%%%%%%%%%%%%%%%%%%%%%%%%%%%%%%%%%%%%%%%%%

We implement a bulk flow of relativistic particles along the shock tangent, and parameterize the bulk speed such that~\citep{Wadiasingh2017}
\begin{equation}
  (\beta\Gamma)_{i} = (\beta\Gamma)_{\rm max}\left(\frac{\theta_{i}}{\theta_{\rm max}}\right),
\label{eq:bulk_flow}
\end{equation}
with $\beta =(\beta\Gamma)/\sqrt{1+(\beta\Gamma)^2}$. The maximum bulk momentum required to describe the X-ray light curves is generally low, $(\beta\Gamma)_{\rm max} \lesssim 10$. Given that the particles in the comoving frame are themselves ultrarelativistic, $\gamma_e \gtrsim 10^4$, such a bulk flow may be regarded as a small anisotropy in momentum space of otherwise nearly-isotropic relativistic plasma.
We note that in thin-shell models \citep[e.g.,][]{wilkin1996exact,Canto1996}, a linear dependence on momentum is recovered by performing a Taylor series expansion near the shock nose, and therefore such a linear term is a zeroth-order approximation to a generalized curved shock tangent velocity.
Implementing a bulk flow along the shock tangent implies that the effects of convection and adiabatic losses in our transport equation are important.
Below, we approximately solve a Boltzmann-type convection-diffusion equation for particle transport, including the effect of radiative losses, isotropic in momentum-space in each zone in the comoving frame, applicable to relativistic particle flow and a spatially-independent diffusion coefficient $\kappa(\gamma_{\rm e})$~\citep[e.g.,][]{Carlo_M}:
\begin{equation}
\frac{\partial N_{\rm e}}{\partial t}  = -\vec{V}\cdot\left(\vec{\nabla}N_{\rm e}\right) + \kappa(\gamma_{\rm e})\nabla^2N_{\rm e} + \frac{\partial}{\partial \gamma_{\rm e}}\left(\dot{\gamma}_{\rm e,tot}N_{\rm e}\right)-\left(\vec{\nabla}\cdot\vec{V}\right)N_{\rm e}+Q,\label{eq:transport}
\end{equation}
with $N_{\rm e}$ the differential particle distribution function per energy interval in units of erg$^{-1}$ cm$^{-3}$, $\vec{V}=\vec{\beta} c$ the bulk velocity (assumed to be directed along the shock tangent), $\gamma_{\rm e}$ the particle energy, $\dot{\gamma}_{\rm e,tot}=\dot{\gamma}_{\rm e,ad}+\dot{\gamma}_{\rm e,rad}$, $\dot{\gamma}_{\rm e,ad}$ adiabatic losses, and $\dot{\gamma}_{\rm e,rad}$ the total radiation (SR and IC) losses.
%%%
In a steady-state approach, $\partial N_{\rm e}/\partial t \equiv 0$. Under various assumptions (see Appendix~\ref{app_Transport}), one may simplify the above transport equation to find 
\begin{equation*}
  0  =  -\frac{N_{{\rm e,}i}}{\tau_{{\rm ad},i}}-\frac{N_{{\rm e,}i}}{\tau_{{\rm diff,}i}}-\frac{N_{{\rm e,}i}}{\tau_{{\rm 1,}i}}-\frac{N_{{\rm e,}i}}{\tau_{{\rm 2,}i}}-\frac{N_{{\rm e,}i}}{\tau_{{\rm rad,}i}} + Q_{ i}
\end{equation*}
\begin{equation}
    \Rightarrow N_{{\rm e,}i}  =  Q_i\tau_{{\rm eff,}i},
    \label{eq:transporteq}
\end{equation}
where
\begin{eqnarray}
  \tau^{-1}_{{\rm eff,}i} & = & \tau^{-1}_{{\rm ad,}i}+\tau^{-1}_{{\rm diff,}i}+\tau^{-1}_{{\rm 1,}i}+ \tau^{-1}_{{\rm 2,}i} + \tau^{-1}_{{\rm rad,}i},
  \label{eq:t_eff}
\end{eqnarray}
and the respective timescales are defined in Appendix~\ref{app_Transport}.

We use a parametric form for the spatial diffusion coefficient
$\kappa(\gamma_{\rm e}) = \kappa_0 (\gamma_{\rm e})^{\alpha_{\rm D}}$. In this paper, we investigate Bohm diffusion ($\alpha_{\rm D}=1$ and $\kappa_0 = c\gamma_{\rm e}(3eB)^{-1}$), corresponding to very turbulent plasmas.

%%%%%%%%%%%%%%%%%%%%%%%%%%%%%%%%%%%%%%%%%%%%%%%%%%%%%%%%%%%%%%%%%%%%%%%%%%%%%%%%%%%%%%%%%%%%%%%%
\subsection{Radiation Losses} \label{sec:Radaition losses}
%%%%%%%%%%%%%%%%%%%%%%%%%%%%%%%%%%%%%%%%%%%%%%%%%%%%%%%%%%%%%%%%%%%%%%%%%%%%%%%%%%%%%%%%%%%%%%%%

The radiation loss term is comprised of SR and IC losses: $\dot{\gamma}_{\rm e,rad}= \dot{\gamma}_{\rm SR} + \dot{\gamma}_{\rm IC}.$ 
The SR loss rate for an isotropic distribution of pitch angles is 
  \begin{equation}
  \dot{\gamma}_{\rm SR} = \frac{4}{3}\frac{\sigma_{\rm T}c U_{\rm B}\gamma_{\rm e}^2}{m_{\rm e}c^2},\label{eq:SR}
  \end{equation}
with $\sigma_{\rm T}$ the Thomson cross section, 
$\gamma_{\rm e}$ the electron Lorentz factor, and $U_{\rm B}=B_{\rm sh}^2/8\pi,$ with $B_{\rm sh}$
assumed to be constant at the shock radius $R_{\rm sh}$. For this initial study, we assume that $B_{\rm sh}$ does not transform between comoving and lab frames (i.e., it is roughly parallel to the downstream comoving flow).
For the IC loss rate, we take the Klein-Nishina effects approximately into account using~\citep{Ruppel2010}:
\begin{equation}
\dot{\gamma}_{{\rm IC}} = \frac{4}{3}\frac{\sigma_{\rm T}c U\gamma_{\rm e}^2}{m_{\rm e}c^2}\frac{\gamma_{{\rm KN}}^2}{\gamma_{{\rm KN}}^2 + \gamma_{\rm e}^2},\label{eq:Ruppel}
  \end{equation}
with
\begin{equation}
	\gamma_{{\rm KN}} \equiv \frac{3\sqrt{5}}{8\pi}\frac{m_{\rm e}c^2}{k_{\rm B}T},
	\label{IC_T}
\end{equation}
where $k_{\rm B}$ is the Boltzmann constant and $T$ the soft-photon temperature. 
The total BB energy density in the observer frame is
\begin{equation}
  U = \frac{2 \sigma_{\rm SB} T^4}{c }\left(\frac{R_*}{\rho}\right)^2 \sim 5 \times 10^{10} \left( \frac{T}{5 \times 10^{3}~{\rm K}} \right)^4  \text{eV\,cm$^{-3}$} 
 \label{density_IC}
\end{equation}
for the day-side hemisphere and for $R_{*}/\rho \sim 0.2$, with $\sigma_{\rm SB} = 5.6704\times10^{-5}$~erg\,cm$^{-2}$ s$^{-1}$ K$^{-4}$ the Stefan-Boltzmann constant. {{For the BW case, $\rho = R_{\rm sh}$ since the shock is wrapped around the companion at a fixed radius. In the RB case, $\rho$ is determined using the cosine rule to calculate the distance the particular zone is from the source of soft photons.}}
We include the CMB as an additional source of soft photons for completeness (using an energy density $U_{\rm CMB}=0.265$~eV\,cm$^{-3}$), but this has little effect on the predicted IC spectrum. We use Eq.~(\ref{density_IC}) to normalize the photon number density in the comoving frame:
\begin{equation}
n_*(\epsilon_*) = \frac{30\sigma_{\rm SB} h}{\left(\pi k_{\rm B}\right)^4} \left(\frac{R_{*}}{R_{\rm sh}}\right)^2 \frac{\epsilon^{2}_{*}}{\delta^2} \left[\exp\left(\frac{\epsilon_{*}}{\delta \, k_{\rm B} T}\right) - 1 \right]^{-1}\!\!\!\!\!\!,
 \label{nBB}
\end{equation}
where $\epsilon_{*}$ is the photon energy. This is very similar to the expression by \citet{dubus2015modelling}, and we see that $n_*\sim U/\delta^2$ (see Section~\ref{sec:Beaming}). See Appendix~\ref{app:nBB} for the derivation of this expression. 

%%%%%%%%%%%%%%%%%%%%%%%%%%%%%%%%%%%%%%%%%%%%%%%%%%%%%%%%%%%%%%%%%%%%%%%%%%%%%%%%%%%%%%%%%%%%%%%%
\subsection{Beaming and Emission}\label{sec:Beaming}
%%%%%%%%%%%%%%%%%%%%%%%%%%%%%%%%%%%%%%%%%%%%%%%%%%%%%%%%%%%%%%%%%%%%%%%%%%%%%%%%%%%%%%%%%%%%%%%%

\begin{figure}[t]
    \centering
    \includegraphics[width=\textwidth]{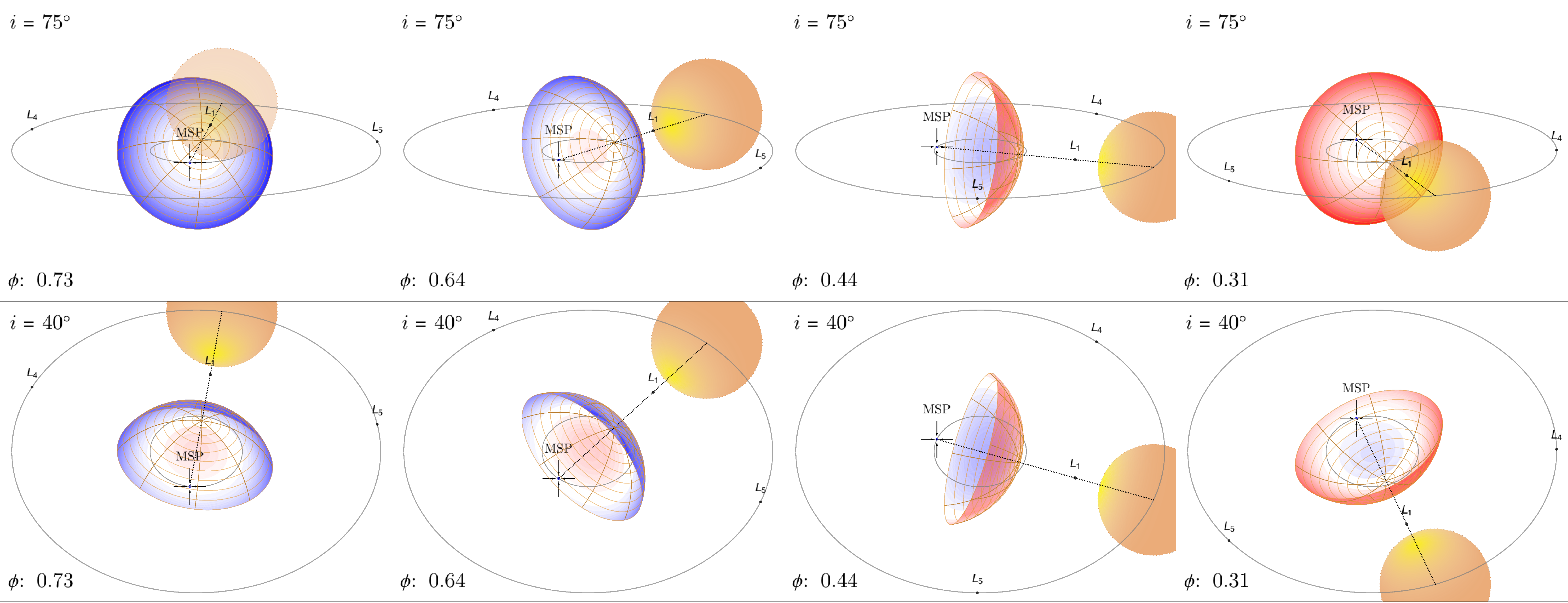}
    \caption{Schematic indicating the shock geometry and beamed emission from the shock. The shock is around the pulsar. In the top panel, the inclination $i=75^\circ$, while in the bottom panel $i=40^\circ$. Blue indicates flow along the shock surface directed toward the observer, and red indicates flow away from the observer. The normalized orbital phase is indicated in the left corner of each panel. For the BW case, refer to Fig.~3 in \citet{Wadiasingh2017}.}
    \label{fig:BeamingCartoon}
\end{figure}

In what follows, we denote observer-frame quantities without a prime, but quantities in the comoving frame will be primed.
We calculate the steady-state particle distributions in the observer frame by solving Eq.~(\ref{eq:transporteq}). The particle transport calculation is performed for the full shock, i.e., for $\theta\in(0,\theta_{\rm max})$ and for all azimuthal angles $\phi_{\rm z}$. The resulting particle spectrum is thus observer-independent. We neglect any change to the steady-state particle spectrum between the comoving and lab frames since the particle energies $\gamma_e$ are much larger than the imposed bulk motion:
\begin{equation}
	\frac{dN_{\rm e}}{d\gamma_{\rm e}} \approx \frac{dN^\prime_{\rm e}}{d\gamma_{\rm e}^\prime}.
\end{equation}
We also assume that the emitted flux is isotropic in the comoving frame, such that $\Omega^\prime_{\rm beam}=4\pi.$

Importantly, the eventual observed radiative flux will be observer-dependent, since the observer makes a particular slice through the shock and the emission is Doppler-boosted in the observer direction (Fig.~\ref{fig:BeamingCartoon}).

Let us define the unit velocity vector as the polar shock tangent
\begin{eqnarray}
	u^\prime_{\rm x} & = &  \sin\theta,\\
	u^\prime_{\rm y} & = &  \cos\theta\cos\phi_{\rm z},\\
	u^\prime_{\rm z} & = &  \cos\theta\sin\phi_{\rm z}.
\end{eqnarray}
We rotate the velocity vector by the binary phase $\Omega_{\rm b}t$ and inclination angle $i$:
  \begin{equation}
  \vec{u} = \Lambda_i\Lambda_{\Omega_bt}\vec{u}~^\prime = 
  \begin{pmatrix}
  \sin i & 0 & \cos i \\
  0 & 1 & 0 \\
  -\cos i & 0 & \sin i
  \end{pmatrix}
   \begin{pmatrix}
  \cos (\Omega_{\rm b}t) & -\sin (\Omega_{\rm b}t) & 0 \\
  \sin (\Omega_{\rm b}t) & \cos (\Omega_{\rm b}t) & 0 \\
  0 & 0 & 1
  \end{pmatrix}
  \begin{pmatrix}
  u^\prime_{\rm x} \\
  u^\prime_{\rm y}\\
  u^\prime_{\rm z}
  \end{pmatrix}.
  \end{equation}
Without loss of generality, we choose the vector pointing in the observer direction as $\vec{n}=(1,0,0)$.
Using the above, we calculate the Doppler beaming factor 
\begin{equation}
	\delta= \frac{1}{\Gamma\left(1 - \beta\vec{n}\cdot\vec{u}\right)}.
\label{eq:delta}
\end{equation}
We compute the emitted photon spectrum in each spatial zone in the comoving frame using the following expressions from \citet{Kopp2013}. For IC scattering
\begin{equation}
\frac{dN_{\rm \gamma, IC}}{dE_{\gamma}} (E_{\gamma},R_{\rm sh}) = \frac{g_{\rm IC}}{AE_0^2} \sum^{k-1}_{j=0} \iint n_{*, \rm j}(R_{\rm sh},\epsilon_{*}, T_{\rm j}) \times \frac{N_{\rm e}(\gamma_{\rm e}, R_{\rm sh})}{\epsilon_{*} \gamma^{2}_{\rm e}} \hat{\zeta}(\gamma_{\rm e}, E_{\gamma}, \epsilon_{*}) d\epsilon_{*} d\gamma_{\rm e},
\end{equation}
where $E_0 = m_{\rm e}c^2$, and $T_{j}$ is the BB temperature of the $j^{\rm th}$ component (we assume both the CMB and the heated companion surface as sources of target photons), $g_{\rm IC} = 2\pi e^{4} c$ (with $g_{\rm IC}/E_0^2\propto \sigma_{\rm T}$),  $n_*(\epsilon_*)$ is defined in Eq.~(\ref{nBB}), and $A = 4\pi d^2$ with $d$ being the distance from the source. The function $\hat{\zeta}(\gamma_{\rm e}, E_{\gamma}, \epsilon_{*}) $ is proportional to the collision rate and is split into four cases, depending on $E_{\gamma}$ \citep{1968PhRv..167.1159J}
\begin{eqnarray}
    \hat{\zeta}(\gamma_{\rm e}, E_{\gamma}, \epsilon_{*}) = 
    \begin{cases}
        0  &\text{if  }  E_{\gamma} \leq \frac{\epsilon_{*}}{4\gamma^{2}_{\rm e}}, \\
        \frac{E_{\gamma}}{\epsilon_{*}} - \frac{1}{4\gamma^{2}_{\rm e}}  &\text{if  } \frac{\epsilon_{*}}{4\gamma^{2}_{\rm e}} \leq E_{\gamma} \leq \epsilon_{*}, \\
        f(q,g_{0})  &\text{if  }  \epsilon \leq E_{\gamma} \leq \frac{4 \epsilon_{*} \gamma^{2}_{\rm e}}{1 + 4\epsilon_{*} \gamma_{\rm e}/E_{0}}, \\
        0  &\text{if  }  E_{\gamma} \geq \frac{4 \epsilon_{*} \gamma^{2}_{\rm e}}{1 + 4\epsilon_{*} \gamma_{\rm e}/E_{0}}.
    \end{cases}
\end{eqnarray}
The function $f(q,g_{0})$ is given by
\begin{equation}
    f(q,g_{0}) = 2q \ln q + (1-q)(1+ q(2+g_{0})),
\end{equation}
with $q = E^{2}_{0}E_{\gamma}/(4\epsilon_{*}E_{\rm e}(E_{\rm e} -E_{\gamma}))$ and $g_{0} = 2\epsilon_{*}E_{\gamma}/E^{2}_{0}$. The emitted SR photon spectrum for in each spatial zone is calculated using
\begin{equation}
     \frac{dN_{\rm \gamma, SR}}{dE_{\rm \gamma}}(E_{\gamma},R_{\rm sh}) = \frac{1}{A} \frac{1}{hE_{\gamma}} \frac{\sqrt{3} e^{3} B(R_{\rm sh})}{E_0} \iint^{\pi/2}_{0} N_{\rm e}(\gamma_{\rm e}, R_{\rm sh}) \kappa\left( \frac{\nu}{\nu_{\rm cr}(\gamma_{e}, \Theta)} \right) \sin^{2}\Theta d\Theta d\gamma_{\rm e},    
\end{equation}
where $\nu_{cr}$ stands for the critical frequency (with pitch angle $\Theta$)
\begin{equation}
    \nu_{\rm cr}(\gamma_{\rm e}, \Theta) = \frac{3ec}{4\pi E_{0}}\gamma_{\rm e}^{2} B \sin \Theta.
\end{equation}
The function $\kappa$ (where $K_{5/3}(y)$ is modified Bessel function of order 5/3) is
\begin{equation}
    \kappa(x) = x \int_{x}^{\infty} K_{5/3}(y)dy
\end{equation}
and is computed using the algorithm given by \citet{macleod2000accurate}. We note that as a matter of simplification, our IC calculation assumes isotropic radiation, rather than anisotropic, an improvement that we defer to future work. We lastly transform the photon energy flux (luminosity) from the comoving to the lab frame via the standard form \citep[e.g.,][]{bottcher2012relativistic}
\begin{equation}
\nu F_\nu = \delta^3\nu^\prime F_\nu^\prime.
\end{equation}
Similarly, the photon energy becomes $E_\gamma = \delta E^\prime_\gamma$. 

We ensure that the resulting observer-dependent flux (the flux in the observer frame that depends on the observer's line of sight) is grid-independent by scaling the (azimuthally-independent) flux by $d\phi_{\rm z}/2\pi$, with $d\phi_{\rm z}$ the bin size of the azimuthal coordinate measured about the line joining the two stars. The photon flux is already weighted by $\theta$ for each zone, since the injection spectrum reflects the zone size (see Eq.~[\ref{eq:Q}]). It is not necessary to weight this flux by the phase bin size $d\Omega_{\rm b}$, since we are calculating the $\nu F_\nu$ flux at a specific orbital phase. Our calculation utilizes $N_{\rm zones}=50$ zones along the shock surface to compute the steady-state spectrum $dN_{\rm e}/d\gamma_{\rm e}$, 300 bins for the SR and IC energies, 300 azimuthal bins, and 300 orbital phase bins. Our flux predictions do not change by more than $2\%$ when choosing a finer grid. Furthermore, the number of photon energy bins we use ensures a smooth IC high-energy tail that exhibits some numerical instabilities when this number is too low (see Fig.~\ref{fig:Tcomp&Bulk}). We have also tested energy conservation in this code by considering the energy input/output per zone due to particle injection, escape and radiation losses. We confirmed energy conservation to a good level, the discrepancy being lower than 20\% for all zones, which gives more confidence in the results the model produces. This discrepancy stems from the numerical approximations we make to solve the transport equation, as well as the fact that we imposed a parametric macroscopic bulk flow profile but did not take this into account in our energy conservation calculation, since we do not know all the details of the microphysics of the acceleration and bulk motion processes. We thus obtain the radiated energy flux as a function of orbital phase and photon energy. Making cuts along constant orbital phase yields spectra, while making cuts at certain constant photon energies yields light curves.

%%%%%%%%%%%%%%%%%%%%%%%%%%%%%%%%%%%%%%%%%%%%%%%%%%%%%%%%%%%%%%%%%%%%%%%%%%%%%%%%%%%%%%%%%%%%%%%%
\subsection{Shadowing}\label{sec:Shadow}
%%%%%%%%%%%%%%%%%%%%%%%%%%%%%%%%%%%%%%%%%%%%%%%%%%%%%%%%%%%%%%%%%%%%%%%%%%%%%%%%%%%%%%%%%%%%%%%%

Obscuration by the companion of the shock emission zones (``shadowing'') can be an important effect for cases where the shock surrounds the companion and the inclination is close to edge-on, especially around pulsar superior conjunction. The implementation of such shadowing by the companion follows a protocol similar to that promulgated in Appendix~A of \cite{Wadiasingh2017}. In this work, we adopt an expedient approximation that the companion is spherical, with a sharp terminator via a Heaviside step function. The shadowing function, which multiplies the observer-sampled emissivity from each zone of the shock, is
\begin{equation}
\Theta \;= \;
\begin{cases}
0 \qquad r_{x} - r_{*,x} < 0 \quad \land \quad (r_{y}-r_{*,y})^2 + (r_{z}-r_{*,z})^2 < R_*^2 \\
1 \qquad \rm otherwise,
\end{cases}
\end{equation}

where $r_j$ and $r_{*,j}$ are the transformed (via the matrices above) position vectors of each zone of the shock and the companion, respectively. This expression terminates emission when (i) the companion is in front of the shock, and (ii) emission is directed toward the observer, but is being blocked by the projection of the companion surface (at a particular orbital phase).

Although we have implemented this condition in what follows, we find this to be a negligible effect for most parameter choices. In particular, the shadowing effect is small for small $i$, when $\theta_{\rm max}$ is large, or when the ratio $R_{\rm comp}/R_{\rm sh}$ is small. This effect is also small in the RB case, when the shock is wrapped around pulsar. These conditions are typical for the fits below, so we include this small effect only for completeness.

%%%%%%%%%%%%%%%%%%%%%%%%%%%%%%%%%%%%%%%%%%%%%%%%%%%%%%%%%%%%%%%%%%%%%%%%%%%%%%%%%%%%%%%%%%%%%%%%
\section{RESULTS}\label{sec:Results}
%%%%%%%%%%%%%%%%%%%%%%%%%%%%%%%%%%%%%%%%%%%%%%%%%%%%%%%%%%%%%%%%%%%%%%%%%%%%%%%%%%%%%%%%%%%%%%%%

In this Section, we present several predictions and survey illustrative output from the model. First, we study the different timescales as a function of particle energy to aid us in interpreting the underlying physics. Second, we probe the parameter space using PSR~J1723$-$2837 as a case study, to uncover trends and diagnose the code's sensitivity to parameter choices. Finally, we present and discuss illustrative fits to the broadband spectra and light curves of four binaries. 

%%%%%%%%%%%%%%%%%%%%%%%%%%%%%%%%%%%%%%%%%%%%%%%%%%%%%%%%%%%%%%%%%%%%%%%%%%%%%%%%%%%%%%%%%%%%%%%%
\subsection{Timescales versus Particle Energy}
%%%%%%%%%%%%%%%%%%%%%%%%%%%%%%%%%%%%%%%%%%%%%%%%%%%%%%%%%%%%%%%%%%%%%%%%%%%%%%%%%%%%%%%%%%%%%%%%

In Fig.~4 we plot relevant timescales for the first and final spatial zones of the IBS for RB PSR~J1723--2837, where the shock wraps around the pulsar. We note that the IC timescale (blue line) is much shorter for the first zone than for the last, given the reduction of soft-photon energy density with distance from the companion star. The SR timescale (red dashed line) is the same for both zones, given the spatially-constant shock magnetic field we assumed. The green line is the effective radiation loss timescale. The yellow line indicates the diffusion timescale $t_{\rm diff,i}= R_{\rm sh}^2/2\kappa$. The adiabatic timescale (purple line) is also similar for similar zone sizes (our zones were chosen for constant $d\mu$ intervals, so zones at higher $\theta$ are slightly larger), but also depends on the local bulk flow velocity, which grows linearly with $\theta$ (cf.\ Eq.~[\ref{eq:bulk_flow}] and Eq.~[\ref{eq:t_ad}]). Thus, we see that the spatial timescales (i.e., the adiabatic one, which is very similar in magnitude to the convective timescales, as well as diffusion)  dominate the radiation (IC and SR) ones in this case. This leads to relatively short residence times of particles in the spatial zones, and thus reduced emission (especially for long IC timescales); conversely a relatively large bulk flow speed increases the Doppler factor in each zone, compensating for this effect, and leading to flux levels that should be accessible for some instruments. The total transport timescale (black line) is the shortest timescale between all those associated with the different processes (Eq.~[\ref{eq:t_eff}]). We have conducted a more exhaustive study of the behavior of the timescales for the different sources we have modeled below, detailed in Appendix~\ref{app:Timescales}, where the shock orientation is also seen to have a significant effect on the relative magnitudes of these timescales.

\begin{figure}[h!]    \label{fig:Timescale_zones}
    \gridline{\leftfig{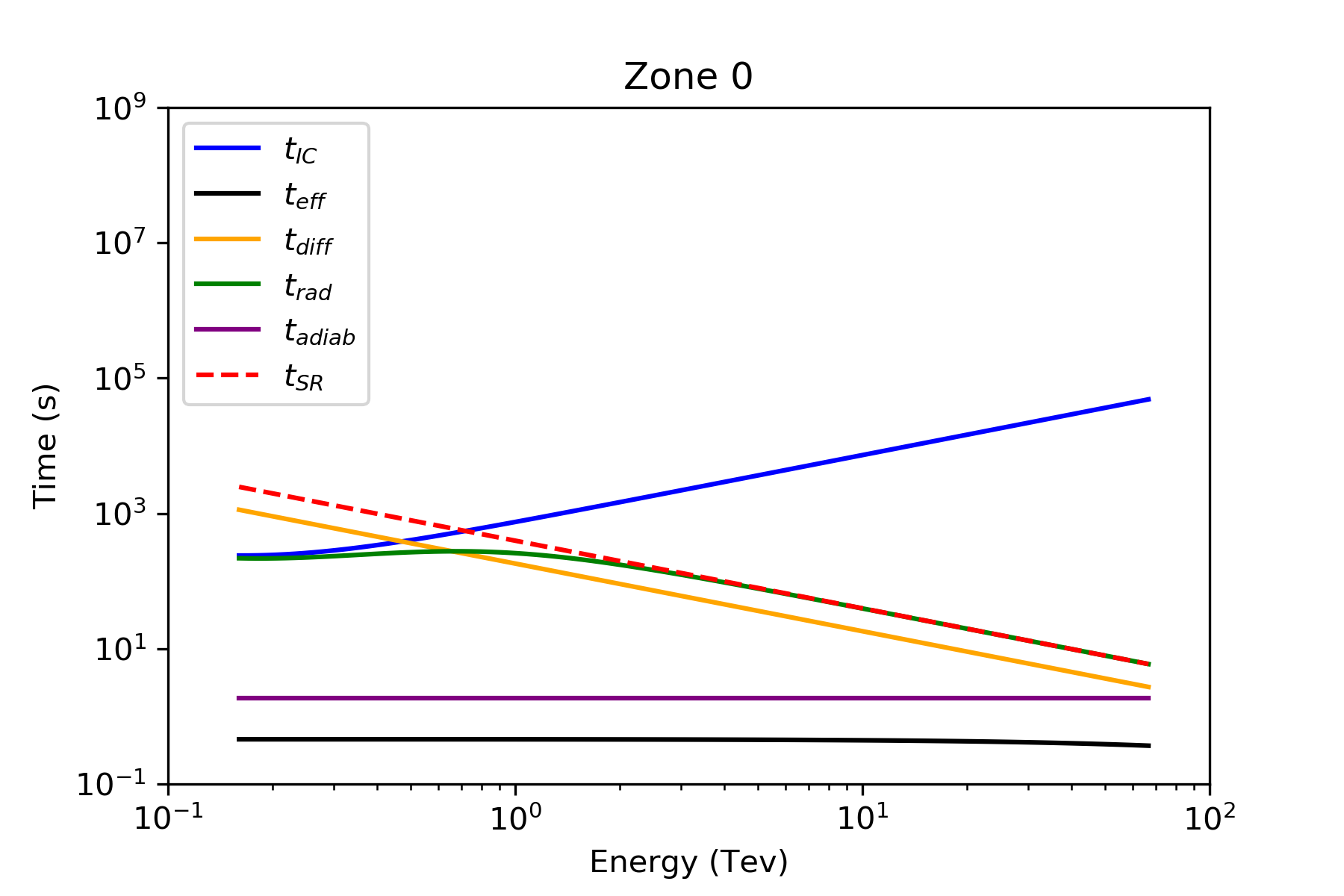}{0.5\textwidth}{(a)}
              \rightfig{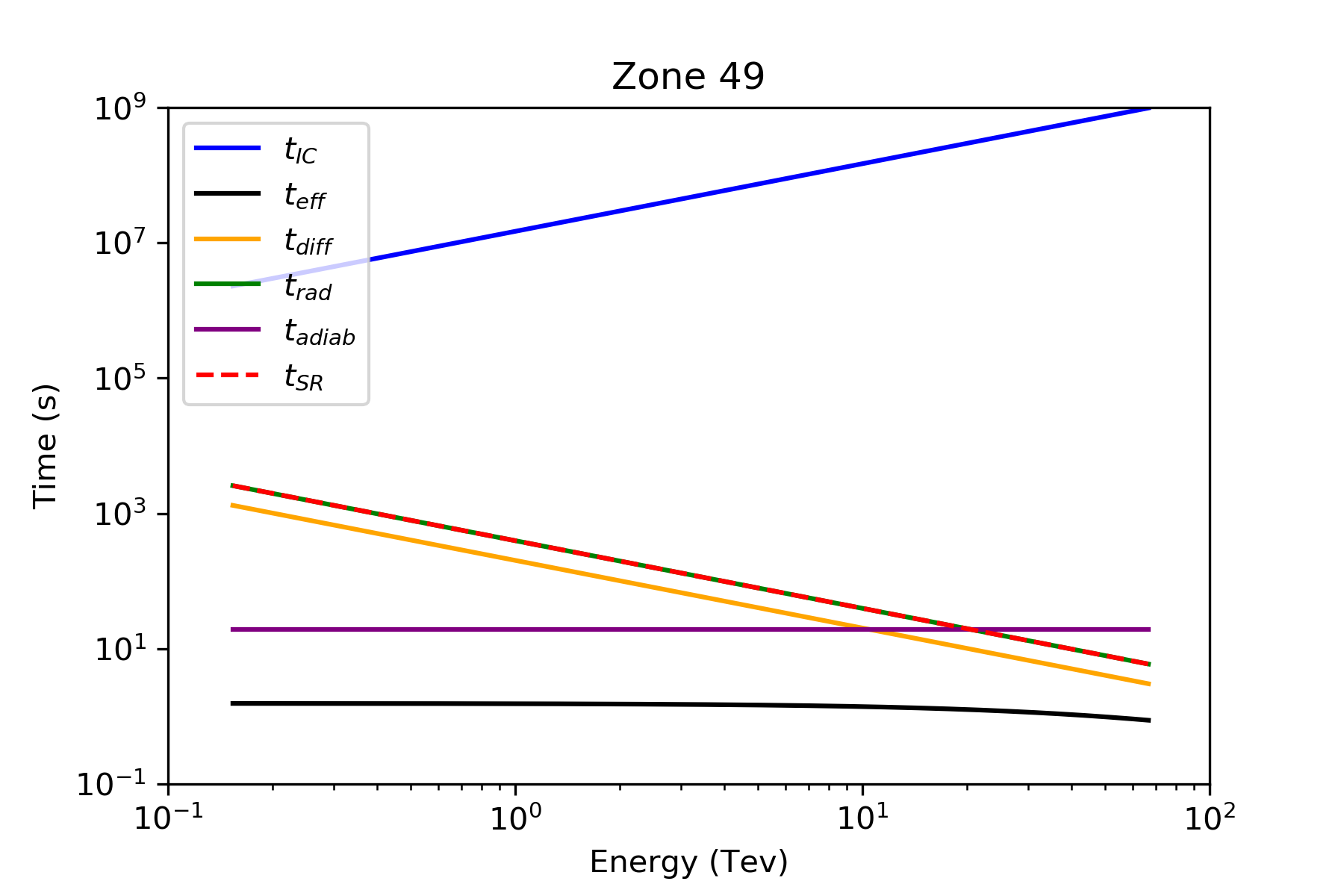}{0.5\textwidth}{(b)}
             }
    \caption{Figures depicting the timescales versus particle energy for RB PSR~J1723$-$2837 (see Table~\ref{tab1} for parameter choices). (a): The first zone near the shock nose. (b) the last zone at the shock periphery. Note the dramatic increase in the IC timescale as the soft-photon density decreases.}
\end{figure}

%%%%%%%%%%%%%%%%%%%%%%%%%%%%%%%%%%%%%%%%%%%%%%%%%%%%%%%%%%%%%%%%%%%%%%%%%%%%%%%%%%%%%%%%%%%%%%%%
\subsection{Case Study: Parameter Variation for PSR~J1723$-$2837} \label{subsec:parameter_study}
%%%%%%%%%%%%%%%%%%%%%%%%%%%%%%%%%%%%%%%%%%%%%%%%%%%%%%%%%%%%%%%%%%%%%%%%%%%%%%%%%%%%%%%%%%%%%%%%
In what follows, we indicate the effect on model predictions of varying a single parameter at a time (as indicated in the Figures below) while keeping others fixed (for a list of parameters, see Table~\ref{tab1}), and discuss the physical reasons for this behavior. In the plots in this Section, we include spectral measurements from soft X-ray observatories \citep[\textit{Chandra} and \textit{XMM}-Newton,][]{2014ApJ...781....6B}, \textit{NuSTAR} \citep{kong2017nustar} and \textit{Fermi} LAT \citep{hui2014exploring}, represented using black spectral butterflies.  We also include continuum sensitivies for \textit{AMEGO} (3-years, 20\% duty cycle) in the MeV band, and H.E.S.S.\ and CTA\footnote{Cherenkov Telescope Array (CTA) web page: {\sl https://www.cta-observatory.org/science/}} in the $>100$ GeV band. As an ensemble, these sensitivities facilitate comparison with the models so as to identify the main impacts of varying different parameters on source visibility. We note that for efficiency, we used a lower resolution for the photon energy grid for our parameter study than for the source-specific cases. This led to some numerical instability in the high-energy tail of the IC spectrum. { In the plots shown in this Section, the SEDs show the flux for PSR~J1723$-$2837 with the IBS around the pulsar at orbital phase of $\Omega_{\rm b} t = 180^{\circ}$}.

\begin{figure}[h!] 
\label{fig:B-field}
\gridline{\leftfig{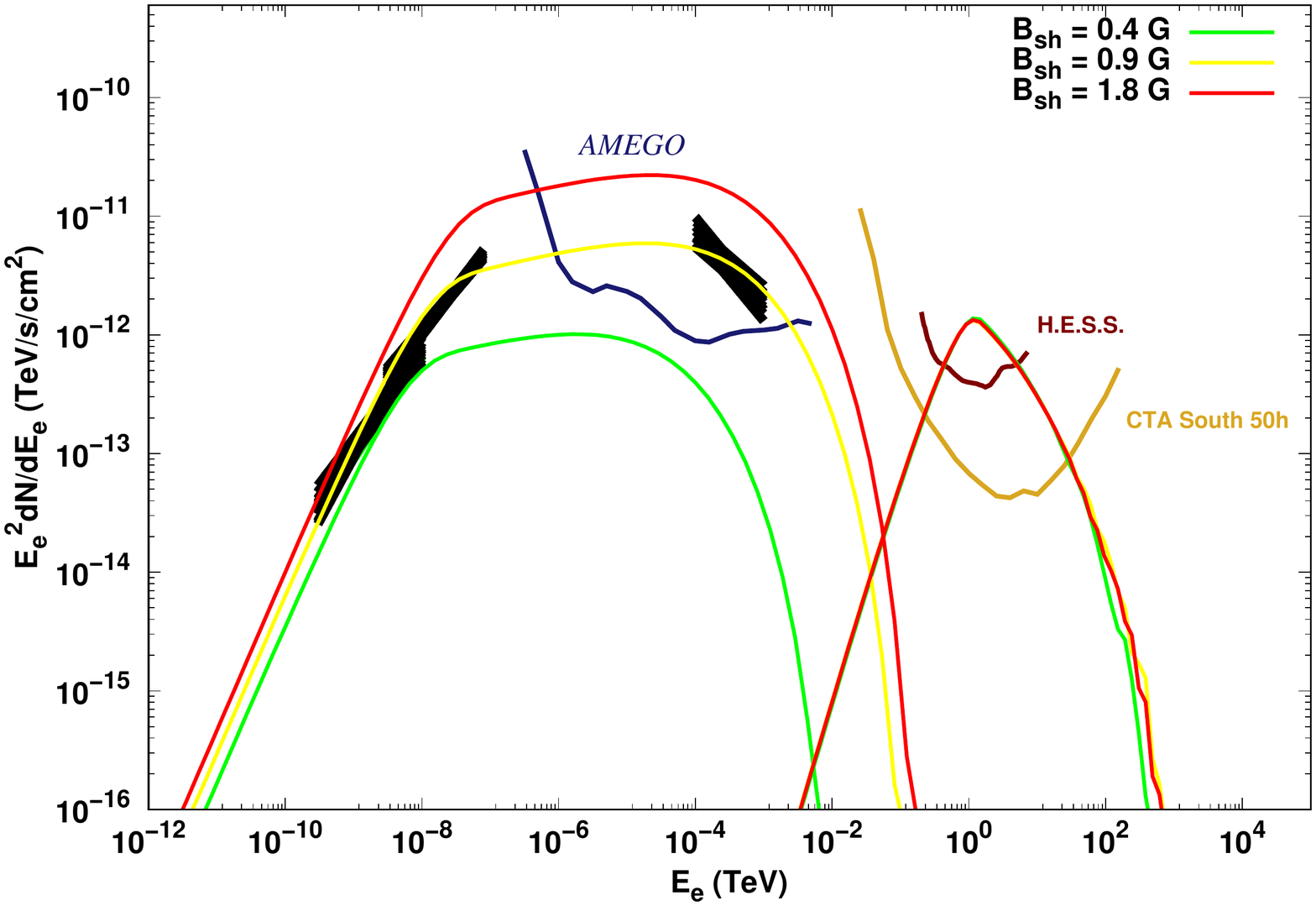}{0.5\textwidth}{(a)}
          \rightfig{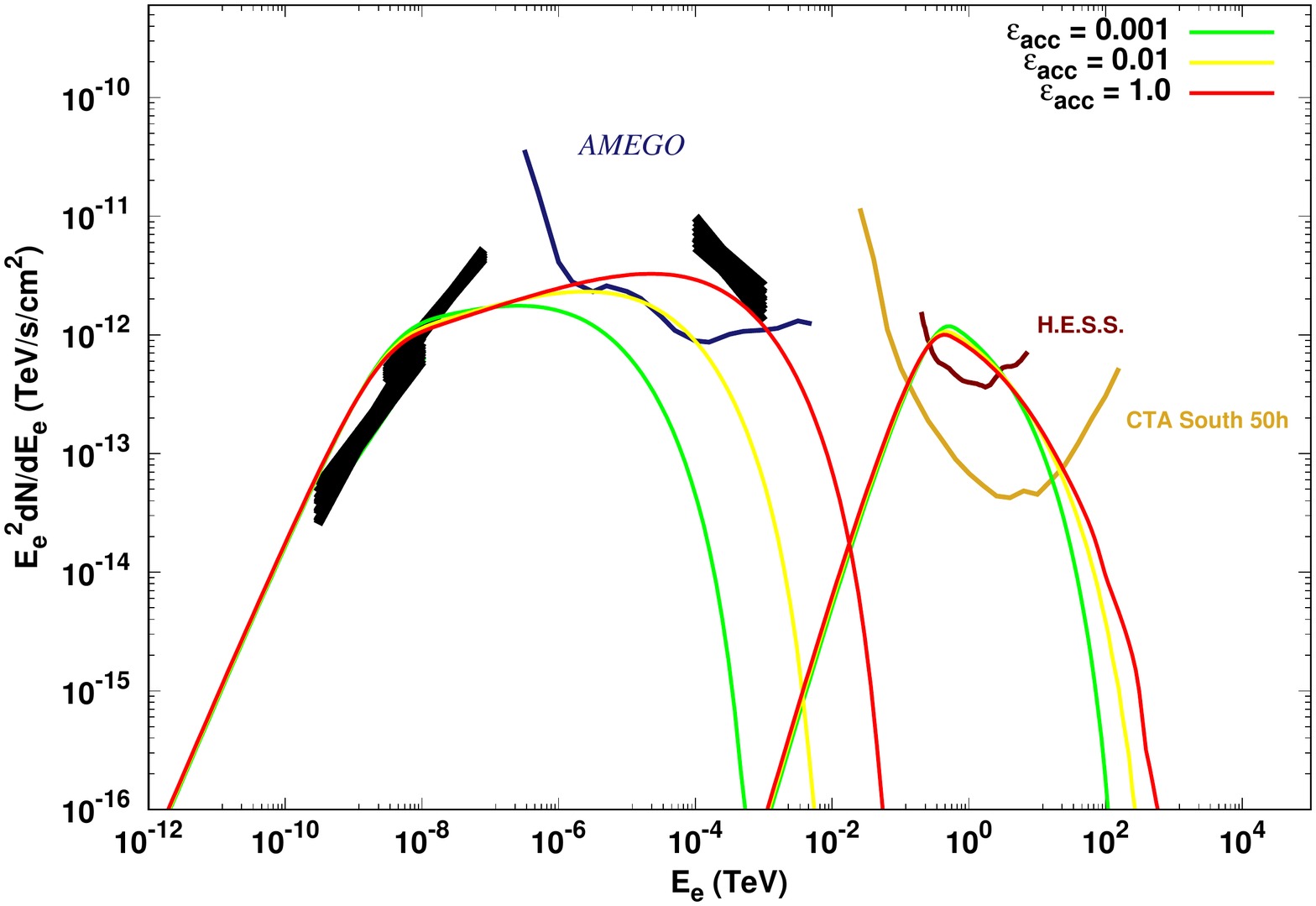}{0.5\textwidth}{(b)}
         }
\caption{Model SED plots for PSR J1723$-$2837 depicting the effect of varying (a) the $B$-field strength $B_{\rm sh}$ and (b) acceleration efficiency $\epsilon_{\rm acc}$ that limits $\gamma_{\rm max}$.}
\end{figure}

The magnetic field at the shock, $B_{\rm sh}$, is a crucial parameter for fitting the X-ray spectral data, with which we calibrate our model. As can be seen in Fig.~5, slight changes to $B_{\rm sh}$ result in large changes to the SR flux since $\dot{\gamma}_{\rm SR} \propto B_{\rm sh}^2$. The IC flux is not significantly altered, since the SR loss rate typically dominates that of IC. The acceleration efficiency $\epsilon_{\rm acc}$ is used to calculate the maximum energy of particles, $\gamma_{\rm e,\,max}^{\rm acc} \propto \epsilon_{\rm acc}^{1/2}$. It can be seen in Fig.~5 that this parameter determines the maximum energy cutoff for the SR component, and can be constrained by GeV-band LAT data. It also sets an upper limit on the IC high-energy cutoff. 

\begin{figure}[h!]
    \gridline{\leftfig{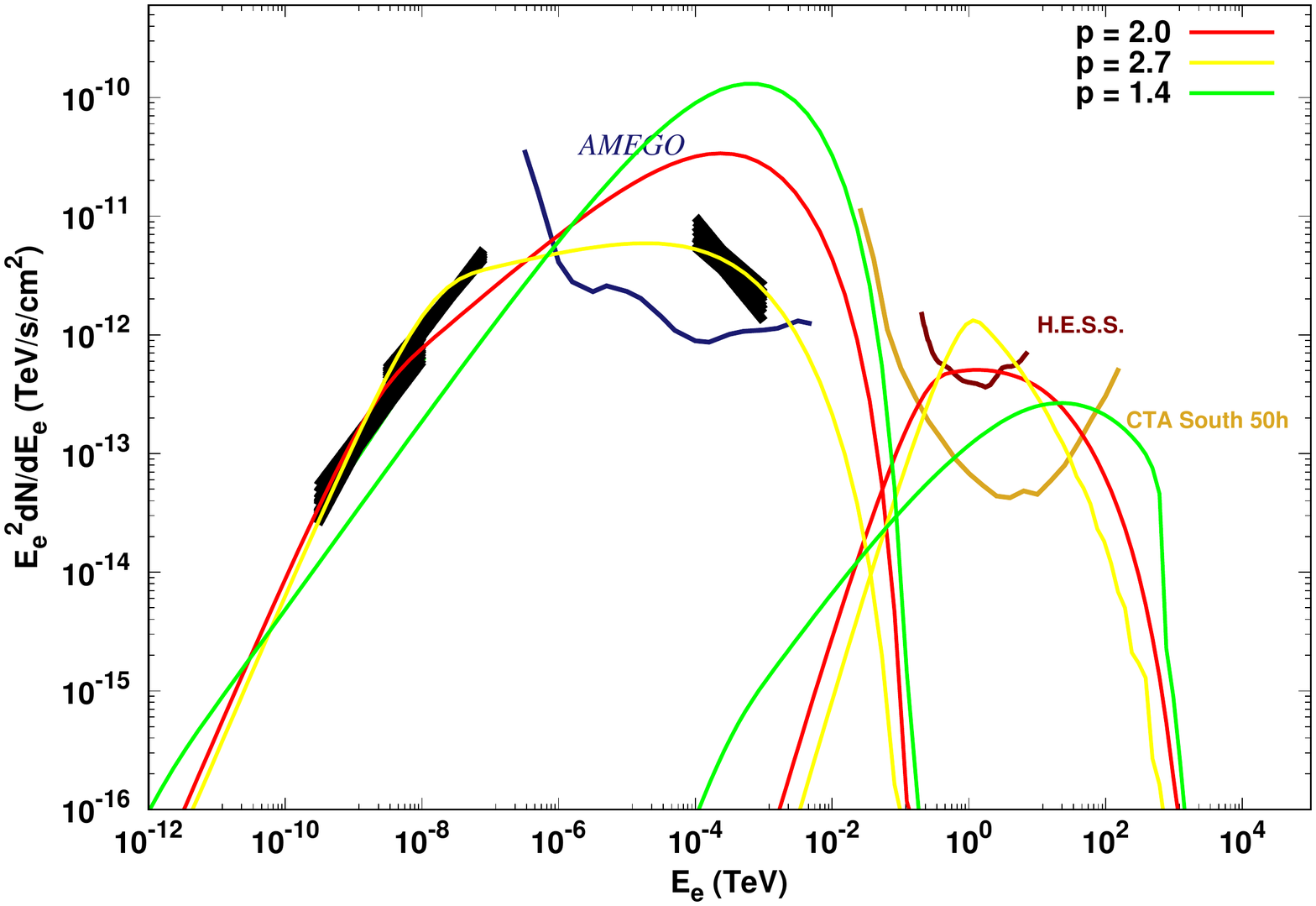}{0.5\textwidth}{(a)}
          \rightfig{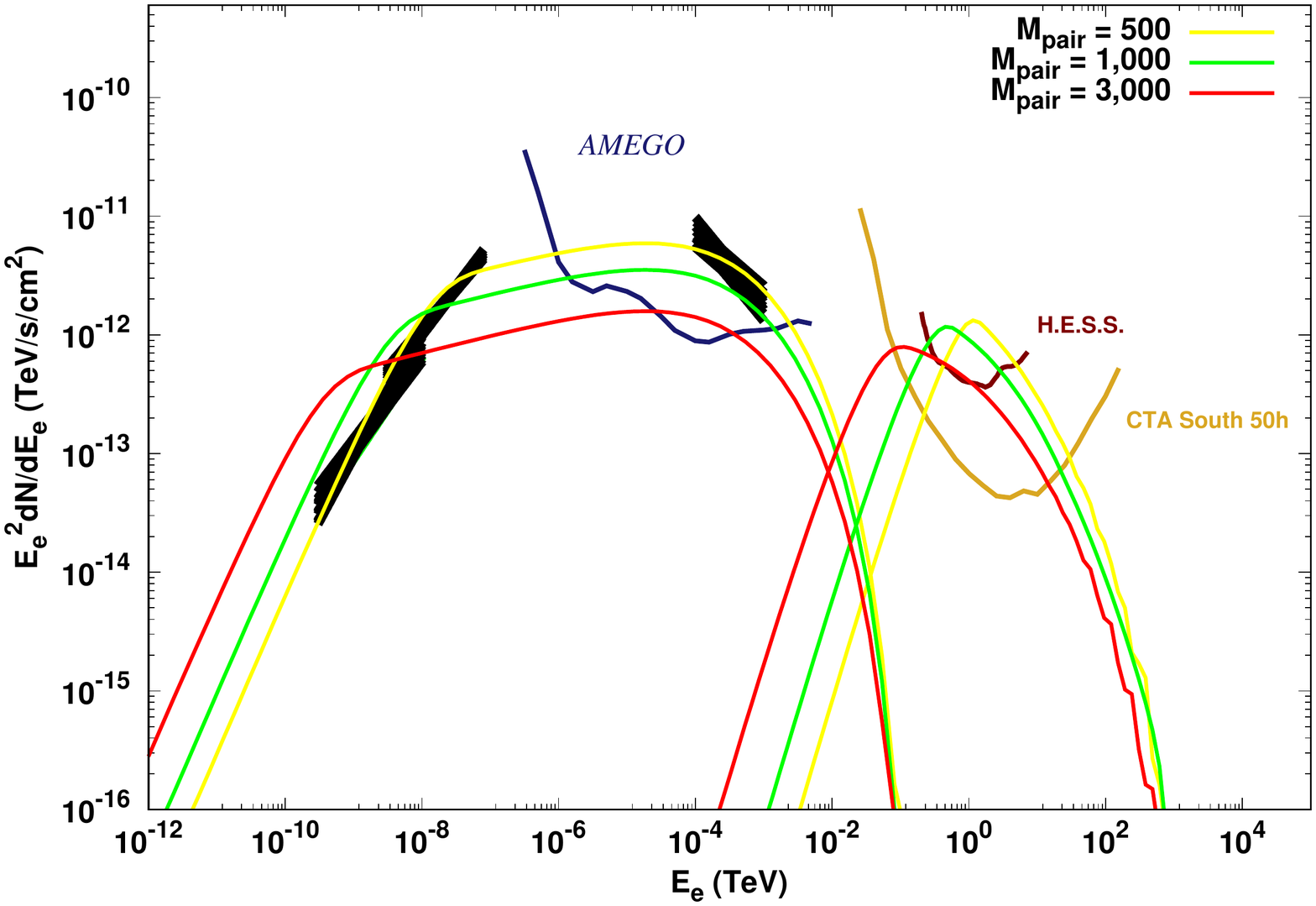}{0.5\textwidth}{(b)}
         }
\caption{Model SED plots for PSR J1723$-$2837 depicting the effect of varying (a) injection spectral index $p$ and (b) pulsar pair multiplicity $M_{\rm pair}$.}
\label{fig:spectral_Mpair}
\end{figure}
The particle injection spectral index $p$ is another crucial parameter and is constrained by the X-ray spectral data. As Fig.~\ref{fig:spectral_Mpair} indicates, decreasing $p$ (hardening the particle spectrum) also leads to altering the particle spectrum's low-energy cutoff $\gamma_{\rm e,min}$ by Eq.~(\ref{eq:Ndot_GJ_QPSR}). This parameter also plays an important role in determining the peak IC flux and VHE spectral shape. The pair multiplicity $M_{\rm pair}$ determines how many particles need to share the total available power (see Eq.~[\ref{eq:Q}]). Thus, increasing $M_{\rm pair}$ results in broader SR and IC spectra since more particles share the available power and thus $\gamma_{\rm min}$ is lower in this case \citep{Venter2015}. While we vary individual parameters in this survey stage of our presentation, it should be noted that there exist parameter correlations and degeneracies. Moreover, the parameter variations may also be coupled. For example, the conditions promoting an increase in the number of cascade generations and therefore higher $M_{\rm pair}$ should yield lower $\gamma_{\rm min}$ and $\gamma_{\rm max}$
and significant changes to the value of $p_{\rm pair}$
\citep{Daugherty1982,Harding2011,Timokhin2015}; this is discussed further in association with Fig.~\ref{fig:Emin fits}.

\begin{figure}[h!]
    \gridline{\leftfig{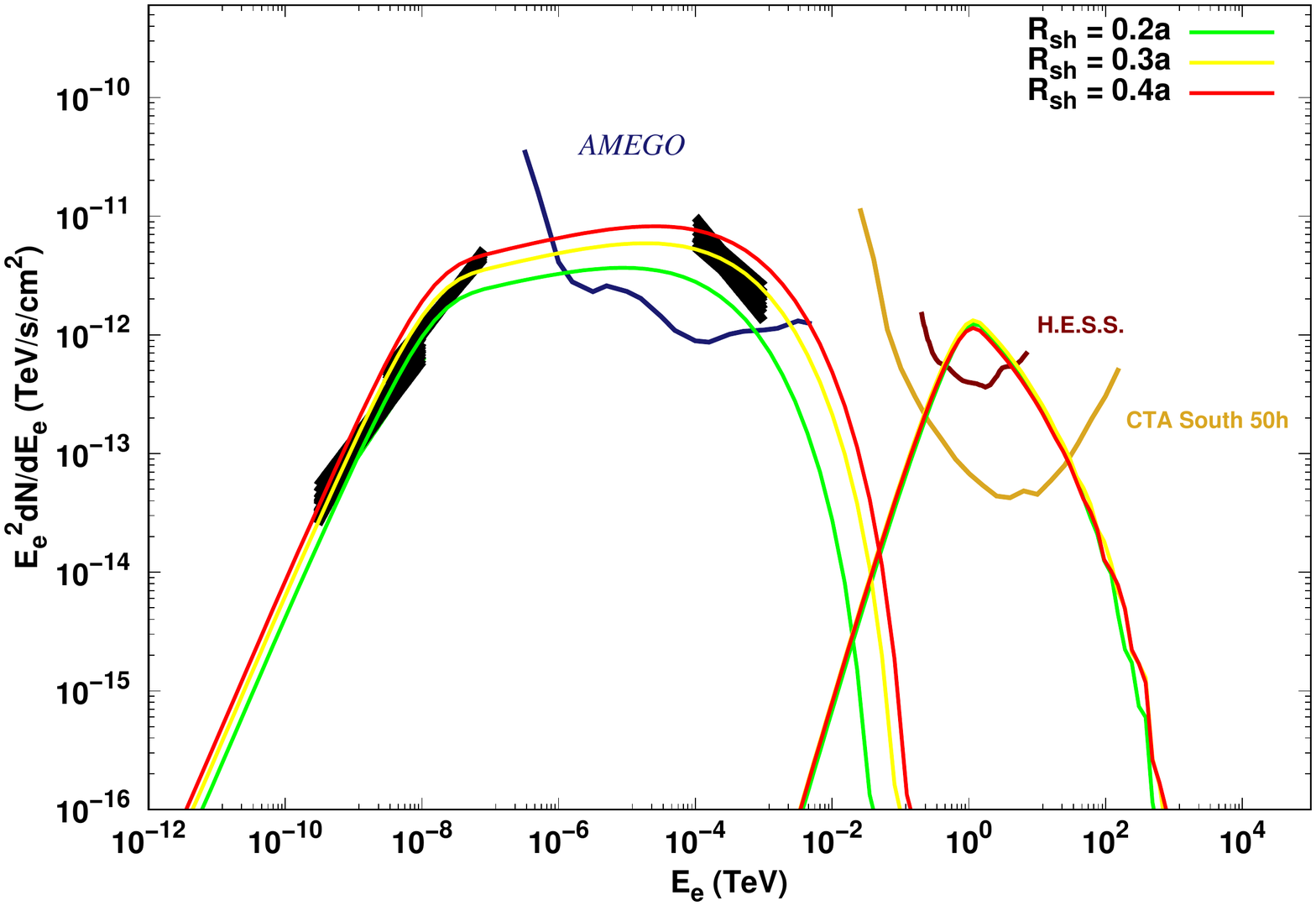}{0.51\textwidth}{(a)}
            \rightfig{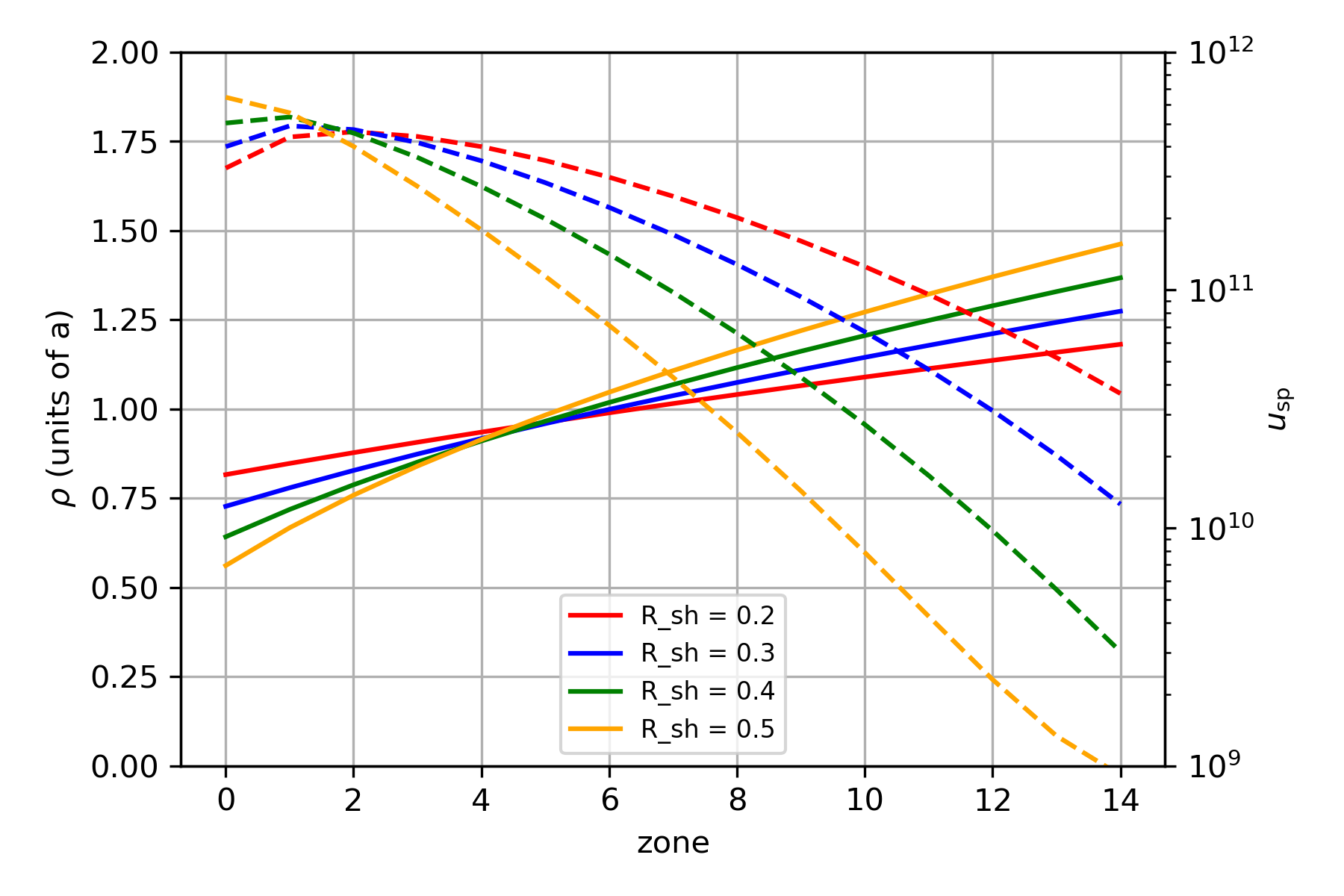}{0.49\textwidth}{(b)}
            }
\caption{Model SED plots for PSR J1723$-$2837 depicting (a) the effect of varying the shock radius $R_{\rm sh}$ and (b) the zone-to-companion distance $\rho$ (solid lines) as well as the value of $u_{\rm sp}$ (dashed lines) versus zone for the RB case. 
}
\label{fig:mass_ratio_shock_radius}
\end{figure}

The distance of the IBS from either the pulsar (RB case) or companion (BW case) is set by $R_{\rm sh}$. 
Varying this parameter only has a significant effect on the SR flux for this RB case. This is because an increase in $R_{\rm sh}$, but keeping the number of spatial zones fixed, leads to physically larger zones. Thus, while the same number of particles are injected into a particular zone (cf.\  Eq.~[\ref{eq:Q_PSR}]), since the solid angle remains constant, the residence time of particles in that zone is longer, since this scales linearly with $R_{\rm sh}$. An increased number of particles therefore leads to a higher SR flux. On the other hand, the reason that we do not see a significant change in IC flux is more complicated. One would naively expect the IC flux to increase for a larger value of $R_{\rm sh}$, because the shock and associated energetic pairs are then closer to the source of soft photons, if one assumes that $u_{\rm sp} \sim 1/R_{\rm sh}^2$. Also, the size of the spatial zones is larger, thus increasing the residence time of the particles in each zone, as indicated above. However, the scaling of $u_{\rm sp}$ is associated with a cosine rule used to calculate the distance from the companion to a specific zone along the shock, leading to a more complicated behavior for $u_{\rm sp}$. The zone-to-companion distance is given by $\rho = \sqrt{R_{\rm sh}^{2} + a^{2} - 2 a R_{\rm sh} \mu_{\rm mid}}$, where $\mu_{\rm mid}$ is the angle associated with the middle of a specific zone (cf.\ Eq.~[\ref{eq:rho}]). This causes $u_{\rm sp}$ to be larger at zones close to the shock nose (as would be expected for a simple $1/R_{\rm sh}^2$ scaling) but then it drops off much quicker than $1/R_{\rm sh}^2$ for zones farther from the shock nose. This is indicated in Fig.~7b. The net result is an approximate cancellation of these effects of first increasing and then decreasing the IC flux, with the IC flux then dropping only slightly for an increase in $R_{\rm sh}$.

\begin{figure}[h!]
    \gridline{\leftfig{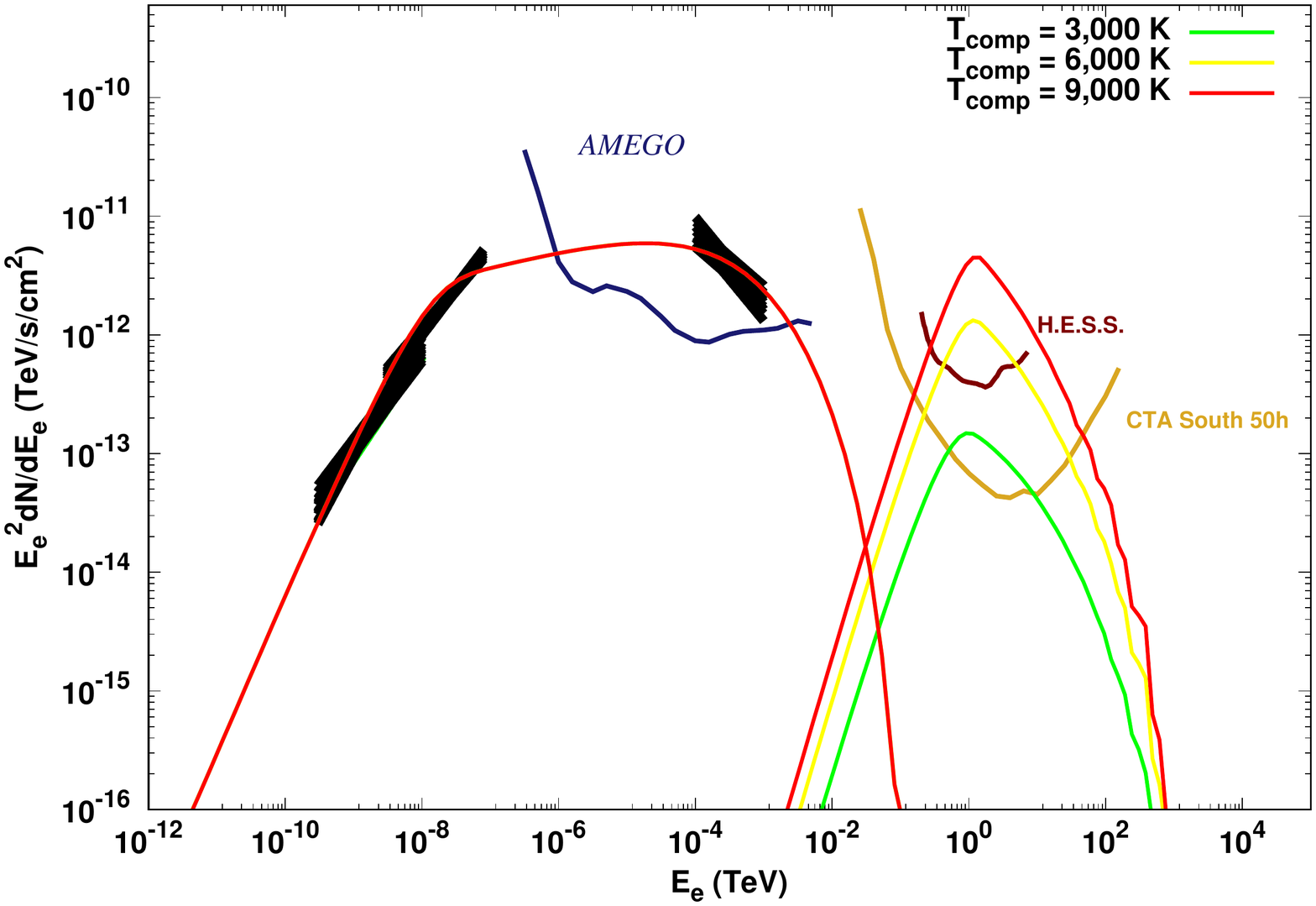}{0.49\textwidth}{(a)}
            \rightfig{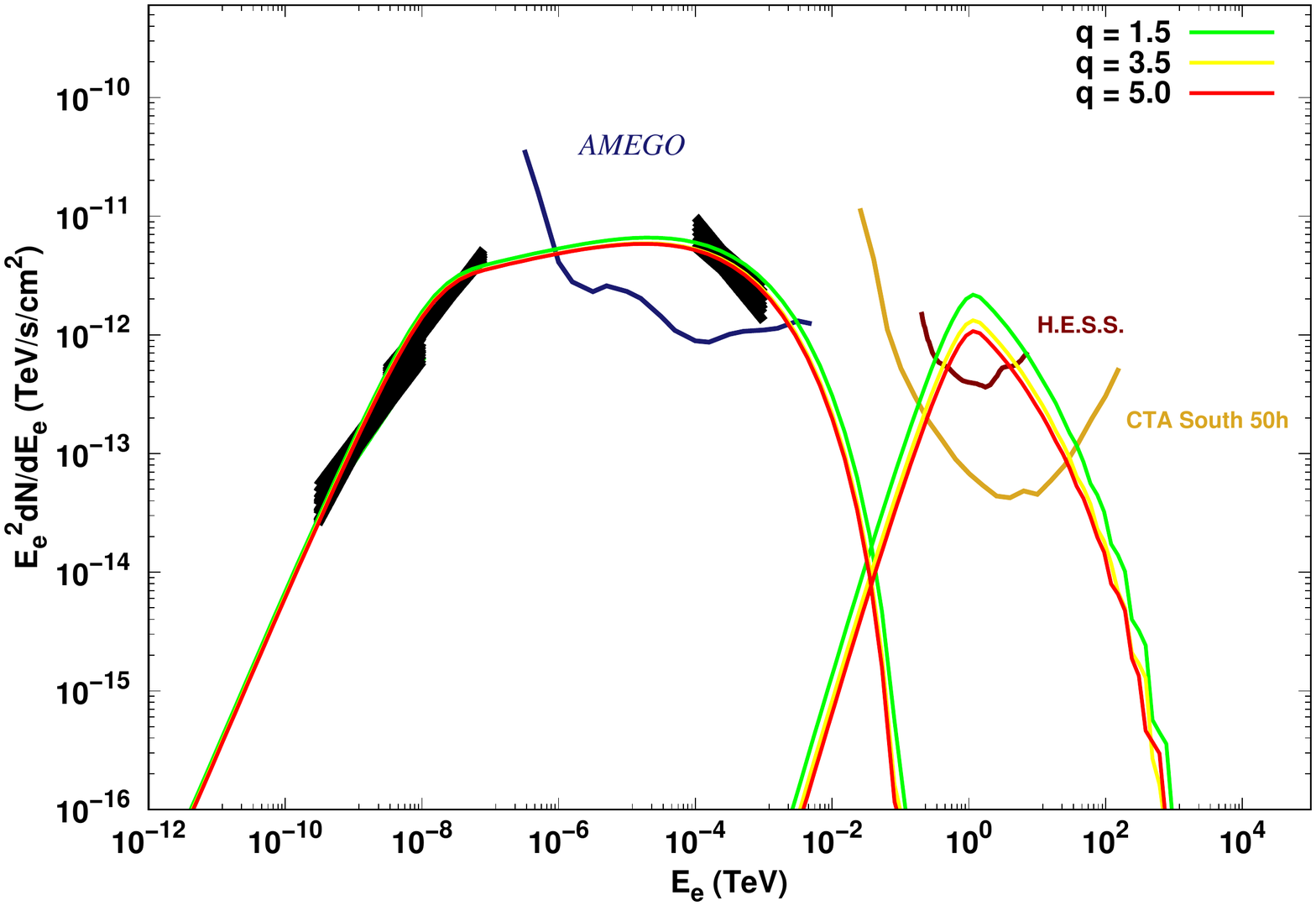}{0.49\textwidth}{(b)}
            }
\caption{SED plots for PSR J1723$-$2837 depicting the effect of varying the (a) companion temperature $T_{\rm comp}$ and (b) stellar mass ratio $q$.}
\label{fig:Tcomp&Bulk}
\end{figure}

The companion's day-side temperature, $T_{\rm comp}$, only has a major influence on the IC flux (and dominates the contribution from the CMB). This dependence is also seen in Eq.~(\ref{eq:Ruppel}) and~(\ref{IC_T}), where the IC flux scales\footnote{The energy-integrated \textit{total} photon number density scales as $n_{*, \rm total}\propto T_{\rm comp}^3$. However, the \textit{spectrum} $n_{*}(\epsilon_{*})$ is normalized using an input value for the \textit{total} energy density $u\propto T_{\rm comp}^4$. Therefore $n_{*}(\epsilon_{*})$ scales as $[\exp{(\epsilon_{*}/k_{\rm B} T_{\rm comp}}) - 1]^{-1}$; see Appendix~\ref{app:nBB}.} as $[\exp{(\epsilon_{*}/k_{\rm B} T_{\rm comp}}) - 1]^{-1}$. This relation has two limits - for low energies the flux is linearly proportional to the temperature. At high energies, the flux becomes proportional to $\exp{(-\epsilon_{*}/k_{\rm B} T_{\rm comp})}$. These limits explain the monotonic but non-linear dependence of the IC flux on the temperature. 

The mass ratio $q$ is used to calculate the inter-binary separation between the MSP and its companion, with $a \sim [(1+q)/q]^{1/3}$. This parameter is also used to calculate the radius of the companion using $R_{*}/a \approx 0.46 (1 + q)^{-1/3}$ \citep{1971ARA&A...9..183P}. { This characteristic size is a good approximation for many BWs and RB companions, which are generally bloated and far brighter than main sequence stars of similar mass. For example, J1723--2847 and others have Roche ``filling-factor" fractions near unity \citep[see Table~6 in][]{2019ApJ...872...42S}.}  Thus, smaller values of $q$ imply a slightly larger orbital separation, but also a larger $R_{*}$. This implies that the soft-photon energy density should stay roughly the same, but a slightly longer residence time is expected for particles in the larger spatial zones (since these scale linearly with $a$ and thus with $R_{\rm sh}$), which leads to a slight increase in the IC flux as $q$ is lowered, as seen in Fig.~8b.

\begin{figure}[h!]
    \gridline{\leftfig{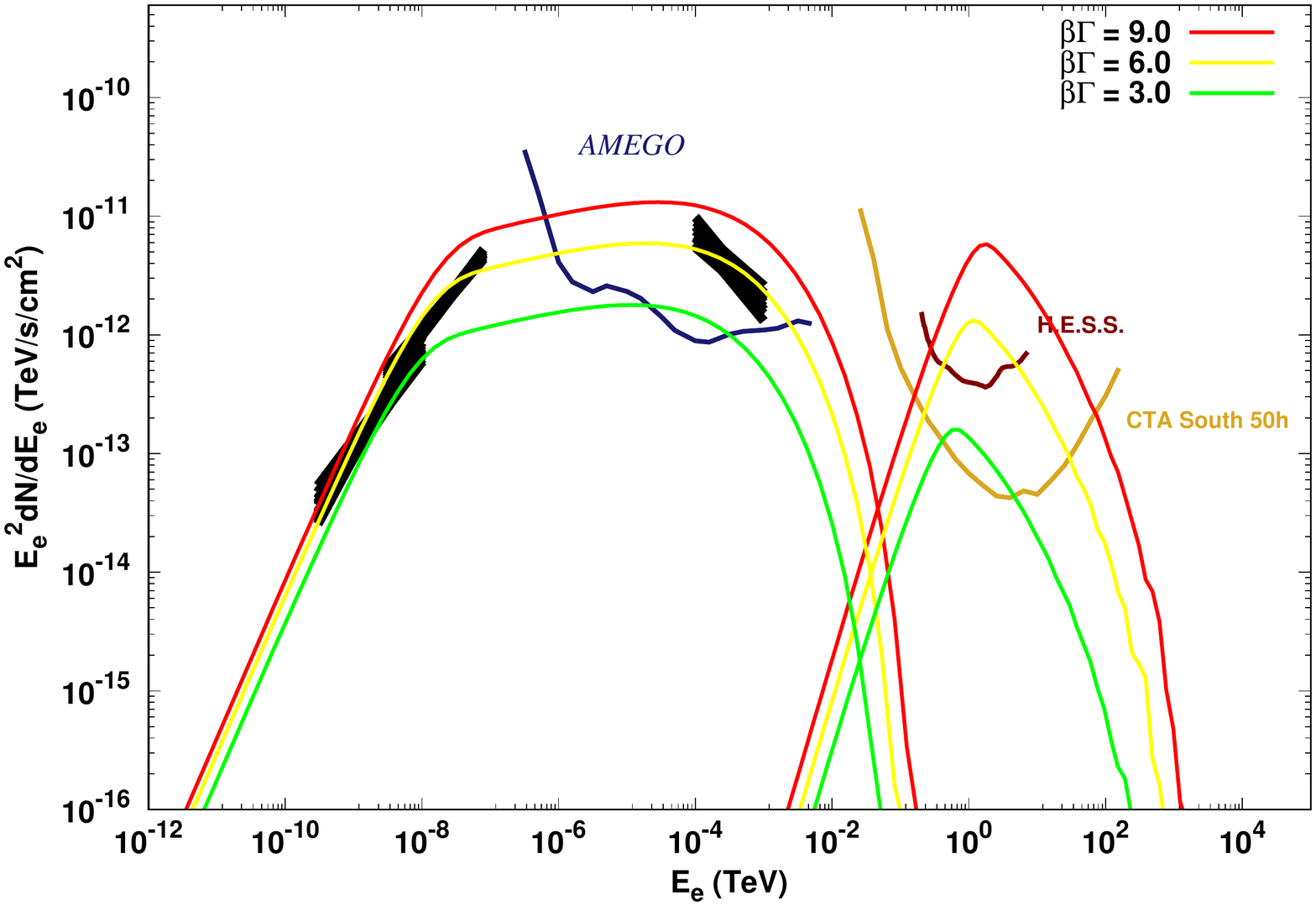}{0.5\textwidth}{(a)}
            \rightfig{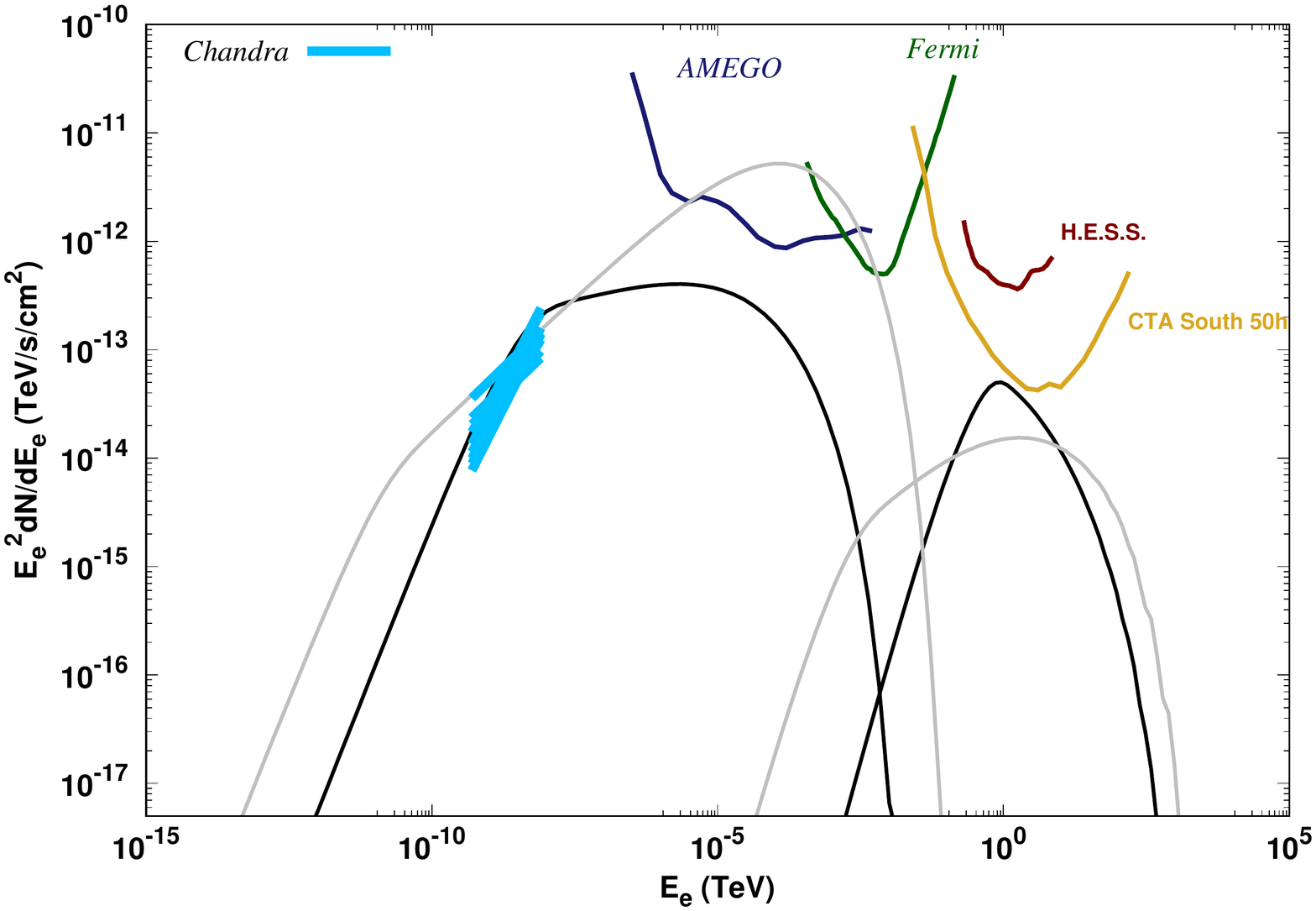}{0.5\textwidth}{(b)}
            }
\caption{Model SED plots (a) for PSR J1723$-$2837 depicting the effect of varying $\beta \Gamma$ and (b) for PSR~J2339$-$0533 indicating two different fitting scenarios: (1) large $\gamma_{\rm min}$ so that we fit X-ray data using the intrinsic single-particle SR spectral slope of 4/3 (black line) and (2) using a lower $\gamma_{\rm min}$ so that the X-ray spectrum is matched by varying the particle spectral index $p$ (gray line).}
\label{fig:Emin fits}
\end{figure}

A change in the relativistic bulk flow momentum of particles along the shock tangent, $\beta\Gamma$, impacts the velocity at which the particles are flowing along the IBS and thus will have an effect on the factor $\delta^3$ with which emission is boosted into the observer frame. Thus, a higher bulk momentum leads to higher fluxes, which would demand an accompanying adjustment, for example lowering the magnetic field, in order to describe the data well.

Next we note that there are two distinct physical scenarios associated with different values of the minimum particle energy not yet distinguishable by the data. Thus, ambiguity exists in the injection parameters used to describe the X-ray data we use to anchor our model. Fig.~9 shows the two different scenarios we considered, namely, (1) low pair multiplicity ($M_{\rm pair}=500$) and soft particle injection spectrum ($p=2.7$), and (2) high pair multiplicity ($M_{\rm pair}=3000$) and harder particle injection spectrum ($p=2.0$), represented by the black and gray spectra respectively. 

These parameter combinations lead to a relatively high and low value for $\gamma_{\rm min}$, respectively. For the black spectra we may match the X-ray data with the intrinsic single-particle SR spectral slope of 4/3, while for the gray spectra the X-ray data demands a hard $p$. The latter scenario yields an upper limit on $\gamma_{\rm min}$.

%%%%%%%%%%%%%%%%%%%%%%%%%%%%%%%%%%%%%%%%%%%%%%%%%%%%%%%%%%%%%%%%%%%%%%%%%%%%%%%%%%%%%%%%%%%%%%%%
\subsection{Application of the Model to Individual Sources}  \label{subsec:source_specific}
%%%%%%%%%%%%%%%%%%%%%%%%%%%%%%%%%%%%%%%%%%%%%%%%%%%%%%%%%%%%%%%%%%%%%%%%%%%%%%%%%%%%%%%%%%%%%%%%

In this work, we constrain our model parameters by using a manual fitting procedure to simultaneously reproduce both SED and light curve data. An automated minimization or multi-dimensional parameter sampler is deferred to a future work. For the SED, we first anchor the models to the empirical X-ray spectral fits of the nonthermal power-law component (and \textit{Fermi} LAT data if available), and then we predict the IC flux of the system. As a first approximation for the light curve, we match the available light curve data making sure the peak multiplicity (i.e., the number of peaks in the light curve) and position in phase roughly coincide with that found in the literature for these sources. We note that we do not include a (free) constant background term in our models, leading to artificially high modulation fractions in this case. Despite these simplifications, matching the light curves yields additional constraints on $\theta_{\rm max}$ and $\beta \Gamma$. {The value of $\theta_{\rm max}$ is allowed to vary (see Table~\ref{tab1} for source model values), truncating the hemispherical shock into a cap.  This latitude in the choice of $\theta_{\rm max}$ is desirable since it affords an improved ability to fit the light curves: $\theta_{\rm max}$ affects the peak multiplicity} (typically a smaller value effectively leads to a shallower shock geometry, leading to double peaks) and  enables us to produce either a single peak or double peaks. {The value of $\beta \Gamma$ (in combination with $\theta_{\rm max}$) has an effect on the width of the light curves via the Doppler boosting}, with higher $\beta \Gamma$ values decreasing the width and thus lead to sharper peaks. In the parameter source-specific cases illustrated below, we indicate whether the double peaks of each system is located at the pulsar inferior conjunction, i.e., pulsar between companion and observer (ICDP) or the pulsar superior conjunction, i.e., companion between pulsar and observer (SCDP). In the ICDP (typically the RB) case, we assume that the shock is wrapped around the pulsar, otherwise in the SCDP case, we assume that the shock is wrapped around the companion (typically the BW case). All model parameters of our illustrative scenarios are detailed in Table~\ref{tab1}. { In the cases shown below, the SEDs show the flux for sources with the IBS around the pulsar at orbital phase of $\Omega_{\rm b} t = 180^{\circ}$ and the flux for sources with the IBS around the companion at $\Omega_{\rm b} t = 0^{\circ}$}.

\begin{table}[h!]
\caption{Model Parameters for Illustrative Cases.}
\footnotesize
\label{tab1}
\scalebox{0.90}{%
\hskip-2.5cm\begin{tabular}{lcccccccc}
\hline
                               &                         &B1957+20 &B1957+20  & J1723-2837  & J1723-2837 & J2339$-$0533 & J1311$-$3430 \\
                               &                         & (A) & (B) & (Black) & (Gray) &  & Quiescent (Flaring) \\
                               \hline
\multicolumn{1}{c}{Parameters} & Symbols                 & \multicolumn{6}{c}{Values}                                                                                                                                          \\ \hline
Orbital separation\footnote{ This is a derived value using $a = (G M_{\rm NS} M_{\rm comp} P_{\rm b}^2(1+q)/4 \pi^2 q)^{1/3}.$}                 & $a$ ($\times 10^{11}$ cm)            & 1.95 & 1.95     & 2.90       & 2.90 & 1.25        & 0.597                                \\
Orbital period                 & $P_{\rm b}$ (hr)            & 9.17  & 9.17     & 14.8  & 14.8   & 4.6        & 1.56   \\
Mass ratio                     & $q$                     & 70  & 70       & 3.5   & 3.5   & 18.2       & 180    \\
Pulsar mass                    & $M_{\rm psr}$ ($M_{\odot}$) & 1.7 & 1.7      & 1.7  & 1.7    & 1.7        & 1.7       \\
Pulsar radius                  & $R_{\rm psr}$ ($\times 10^{6}$ cm)          & 1.0 & 1.0    & 1.0 & 1.0 & 1.0     & 1.0   \\
Pulsar period                  & $P$ (ms)                & 1.60 & 1.60    & 1.86   & 1.86   & 2.89       & 2.56   \\
Pulsar period derivative       & $\dot{P}$ ($\times 10^{-20}$ s/s)         & 1.7 & 1.7  & 0.75  & 0.75  & 1.4    & 2.1 \\
Moment of inertia of pulsar    & $I$ ($\times 10^{45}$)                    & 1.0 & 1.0  & 1.0  & 1.0  & 1.0     & 1.0 \\
Inclination angle              & $i$ ($^\circ$)           & 65 & 85      & 40   & 40    & 54         & 60 \\
Distance                       & $d$ (kpc)               & 1.40  & 1.40    & 0.93  & 0.93   & 1.10       & 1.4     \\
Companion temperature          & $T_{\rm comp}$ (K)          & 8,500 & 8,500    & 6000  & 6000   & 6000       & 12000 (45000)\\ \hline
Shock radius                   & {$R_{\rm sh}/a$}  & 0.4 & 0.2      & 0.4  & 0.4   & 0.3        & 0.4  \\
Magnetic field at shock        & $B_{\rm sh}$ (G)            & 1.2 & 7.0     & 0.8   & 0.8   & 0.8        & 1.0 (1.2)   \\
Pair multiplicity              & $M_{\rm pair}$              & 9000 & 9000    & 400  & 2000    & 500       & 600 (3000)\\
Maximum particle conversion efficiency  & $\eta_{p}$  & 0.6 & 0.8 & 0.9 & 0.8 & 0.8 & 0.8 (0.9) \\
Acceleration efficiency        & {$\epsilon_{\rm acc} = r_g/\lambda $}        & 0.001 & 0.001   & 0.08  & 0.001   & 0.01      & 0.1 (0.01)\\
Index of injected spectrum     & $p$                & 2.5 & 2.4      & 2.5  & 1.4    & 1.8        & 1.4 (1.8) \\
Bulk flow momentum             & $(\beta\Gamma)_{\rm max}$   & 4.0  & 5.0      & 6.0  & 7.0     & 7.0        & 5.0 (8.0)\\
Maximum shock angle            & $\theta_{\rm max}$          & 61 & 68       & 60  &50   & 55         & 65 (60)\\ \hline \hline
\end{tabular}%
}
\end{table}

%%%%%%%%%%%%%%%%%%%%%%%%%%%%%%%%%%%%%%%%%%%%%%%%%%%%%%%%%%%%%%%%%%%%%%%%%%%%%%%%%%%%%%%%%%%%%%%%
\subsubsection{PSR~J1723$-$2837 (RB) -- ICDP}
%%%%%%%%%%%%%%%%%%%%%%%%%%%%%%%%%%%%%%%%%%%%%%%%%%%%%%%%%%%%%%%%%%%%%%%%%%%%%%%%%%%%%%%%%%%%%%%%

\begin{figure}[h!]
\gridline{\leftfig{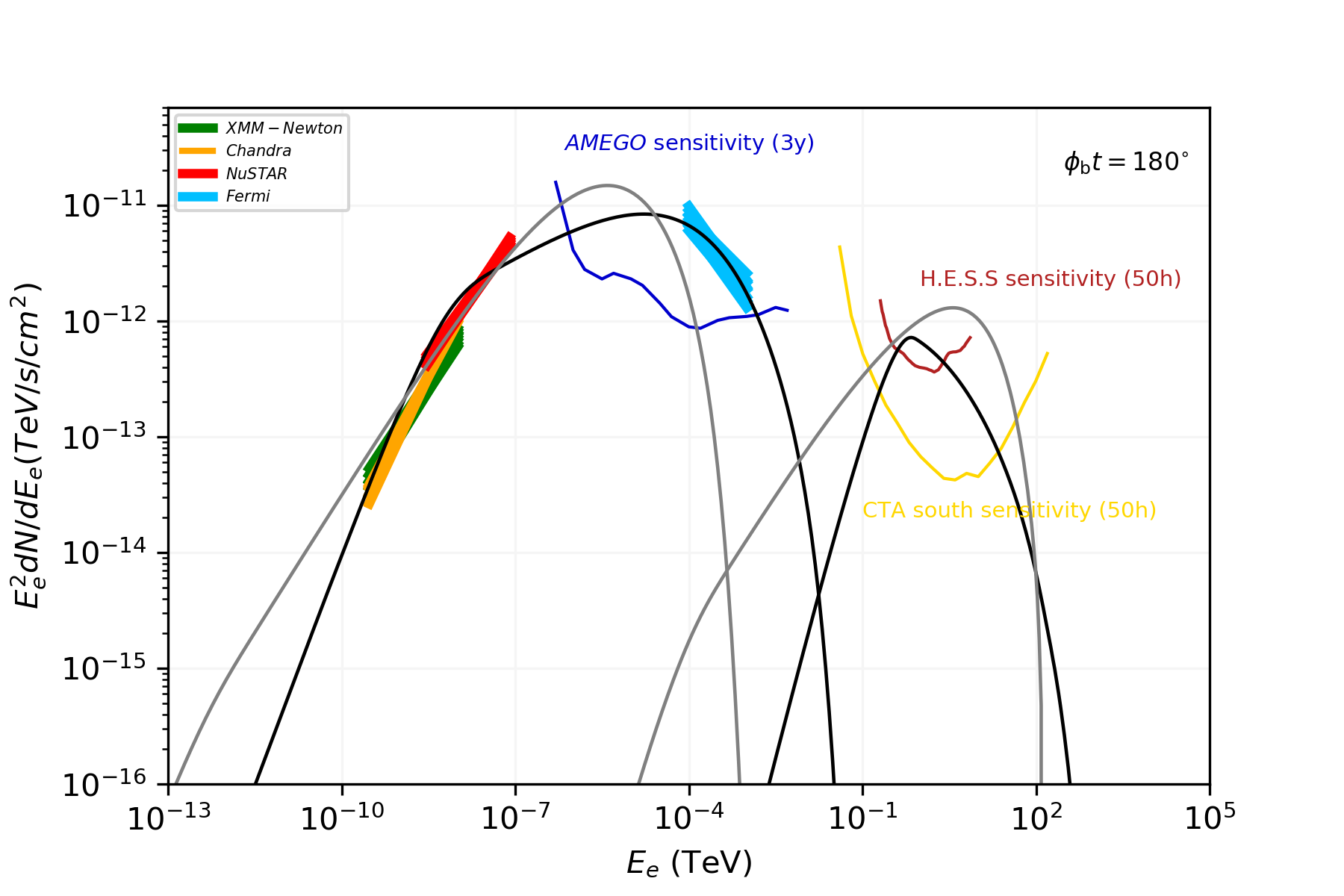}{0.50\textwidth}{(a)}
          \rightfig{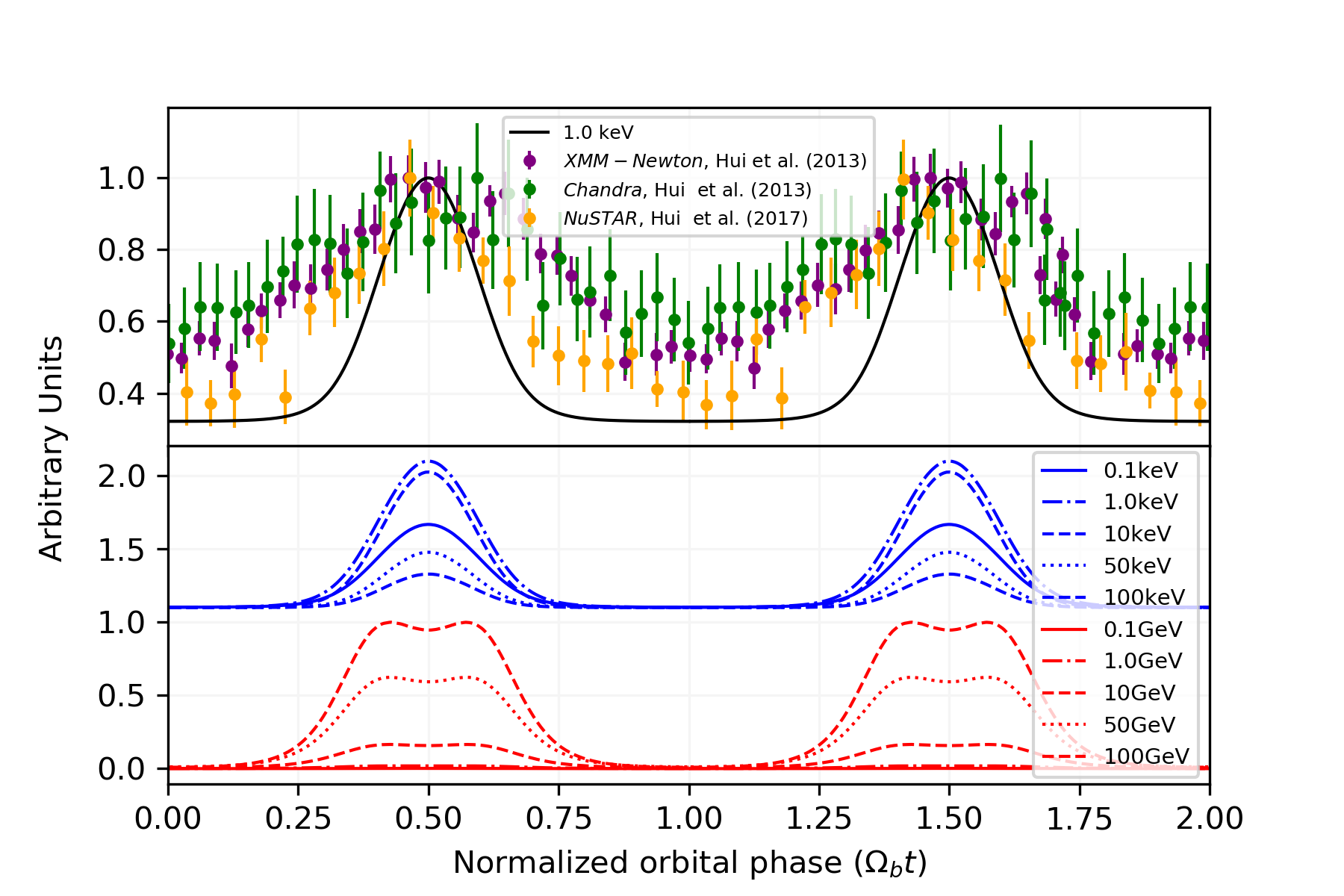}{0.50\textwidth}{(b)}
         }
\caption{Plot for PSR~J1723$-$2837 indicating the (a) SED where we show two cases, one matching {\it Fermi} LAT spectral data (black line) and the other not (gray line), and (b) energy-dependent light curves. The light curves on the right correspond to the black SED model on the left. The top panel (black line) is a model light curve for an energy close to that of the observed  X-ray one. The blue (SR) and red (IC) lines in the bottom panel illustrate energy-dependent pulse shapes, for energies as indicated in the legend.}
 \label{fig:J1723}
\end{figure}

PSR~J1723$-$2837 is an ICDP RB system. We adopt a pulsar mass $M_{\rm NS} = 1.7 M_{\odot}$, period $P = 1.86$ ms and $\dot{P} = 7.61\times 10^{-21}$. This system is located at a distance of $d = 0.93$~kpc and has an orbital period of $P_{\rm b} = 14.8$~hr. In \citet{crawford2013psr} the inclination angle was derived from radial velocity data and assuming a pulsar mass range of $1.4 - 2.0 M_{\odot}$, which gives a inferred companion mass range of $0.4-0.7M_{\odot}$. These mass ranges then yield a orbital inclination angle of $30^{\circ} - 41^{\circ}$. For this work we choose $i = 40^{\circ}$ and $q=3.5$, which implies that the companion has a mass of $M_{\rm comp} \sim 0.57 M_{\odot}$. Optical spectroscopy indicates the companion to have a temperature of $T_{\rm comp} \sim 6000$~K \citep{crawford2013psr}. We have used \textit{XMM}-Newton, \textit{Chandra, NuSTAR} and \textit{Fermi} LAT spectral data \citep{kong2017nustar,hui2014exploring} to anchor our models for this system. We are able to match the X-ray data for PSR~J1723$-$2837 in two distinct ways. Both fits require a minimum energy break $\gamma_{\rm e,min}$ and cannot be described with the intrinsic single-particle SR slope of 4/3. The black line is a model that has a very soft spectral index ($p=2.5$) and low pair multiplicity ($M_{\rm pair} = 400$). These parameter choices enable us to match the \textit{Fermi} LAT data that may be associated with this source. Additionally, a second model is shown without matching the \textit{Fermi} LAT data (gray line) as no orbital modulation has been demonstrated in this component. This model has a very hard spectral index ($p=1.4$) and a much larger pair multiplicity ($M_{\rm pair} = 2000$). Both illutrative models indicate that this source may be detectable by \textit{AMEGO}, H.E.S.S., and CTA. In particular, CTA measurements of the IC component above 100 GeV would enable constraints mainly on the bulk Lorentz factor at the intrabinary shock and the temperature of the companion. Further in the future, a mission like \textit{AMEGO} that is sensitive in the MeV band would probe the maximum Lorentz factor and magnetic field strength in the shock environs.

%%%%%%%%%%%%%%%%%%%%%%%%%%%%%%%%%%%%%%%%%%%%%%%%%%%%%%%%%%%%%%%%%%%%%%%%%%%%%%%%%%%%%%%%%%%%%%%%
\subsubsection{J1311$-$3430 (BW) -- ICDP}
%%%%%%%%%%%%%%%%%%%%%%%%%%%%%%%%%%%%%%%%%%%%%%%%%%%%%%%%%%%%%%%%%%%%%%%%%%%%%%%%%%%%%%%%%%%%%%%%

\begin{figure}[h!]
\gridline{\leftfig{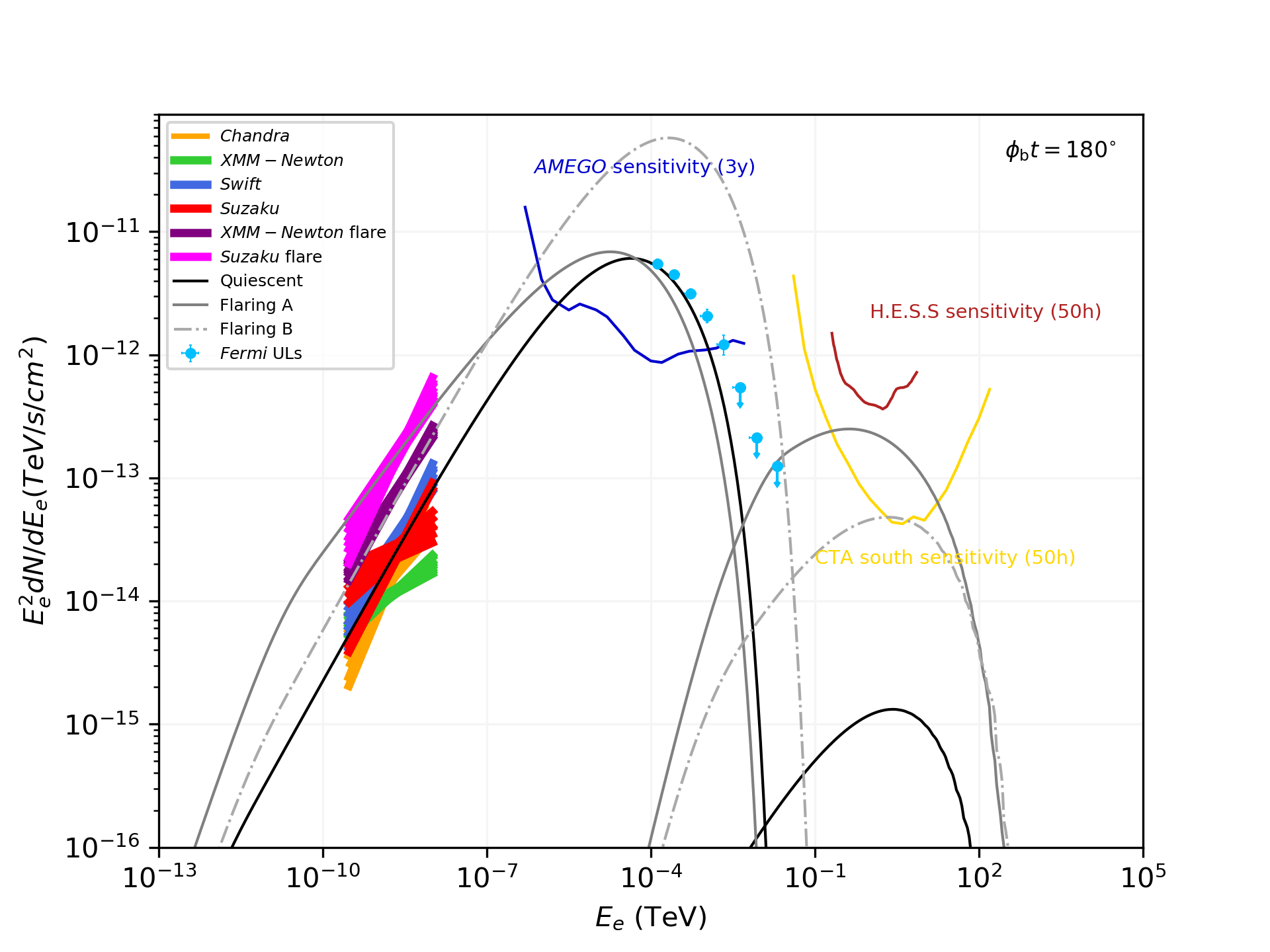}{0.5\textwidth}{(a)}
          \rightfig{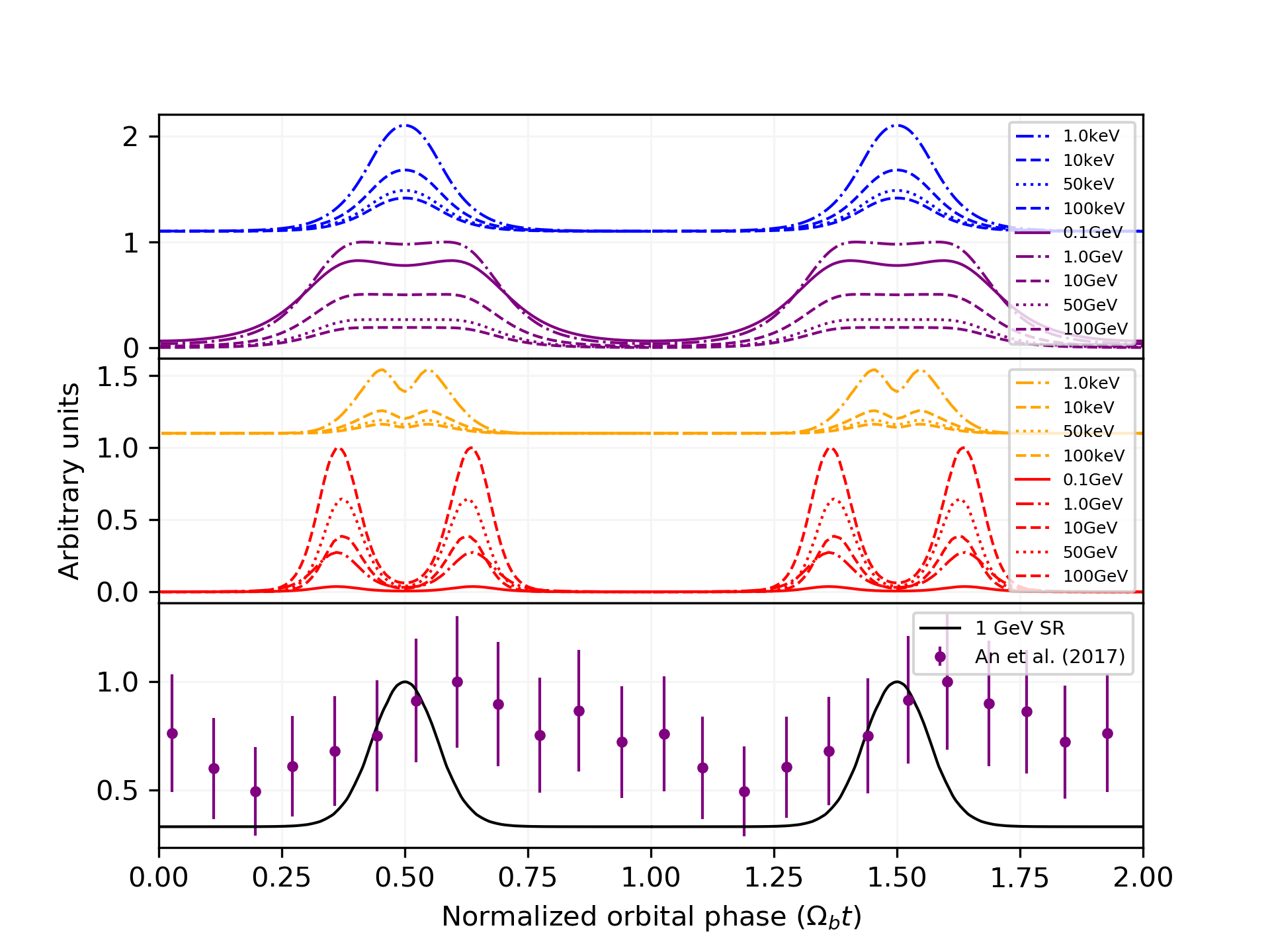}{0.5\textwidth}{(b)}
         }
\caption{Plot for PSR~J1311$-$3430 depicting the (a) SED and (b) energy-dependent light curves for the quiescent (blue and purple) and flaring (orange and red) states. The dashed gray line represents an SED model for the flaring state in which  all parameters are identical to the quiescent state (black line), except the companion temperature and magnetic field at the shock. The solid gray line represents an alternative model to the flaring-state spectral data. We do not exhibit light curve data in the right-hand panel due to their low quality. The light curves should only be taken as a qualitative indication of the predicted double-peak shape.}
\end{figure}

PSR~J1311$-$3430 is a BW system, but we interpret the orbital modulation in the X-ray band as shown in \citet{an2017high}, as implying that the shock is wrapping around the pulsar. This implies high magnetic fields ${\cal{O}}(10^4$ G$)$ on the companion as a source of the pressure balance and shock orientation. For our modeling, we adopt a pulsar mass $M_{\rm NS} = 1.7 M_{\odot}$, period $P = 2.56$ ms and $\dot{P} = 2.1\times 10^{-20}$. This system is located at a distance of $d = 1.40$~kpc, exhibiting an orbital period of $P_{\rm b} = 1.56$~hr and an inclination of $i \approx 60^{\circ}$, as inferred from light curve fitting described in \citet{romani2012psr}. The companion has inferred surface temperatures of $T_{\rm comp} = 12000$~K and $T_{\rm comp} = 40000$~K, respectively, in the quiescent and flaring states as reported by \citet{romani2015spectroscopic}. We adopt these two characteristic temperatures as case studies on quiescent or flaring states. Since \texttt{UMBRELA} is a time-independent approach, the illustrative models on the two states should not be overinterpreted. A companion mass of $M_{\rm comp} \sim 0.01 M_{\odot}$ \citep{an2017high} yields a particularly high mass ratio of $q \sim 180$. To model the quiescent state (black line) we employ \textit{XMM}-Newton, \textit{Chandra}, \textit{Suzaku}, \textit{Swift}-XRT and \textit{Fermi} LAT spectral data \citep{an2017high} to constrain as many of our model parameters as possible. For the flaring state we use the associated \textit{XMM}-Newton and \textit{Suzaku} spectral data obtained by \citet{an2017high}. We consider two possible scenarios for the flaring state as shown in Fig.~11. The first is for a higher companion temperature, pair multiplicity and bulk flow along the shock tangent (solid gray line). For the second, we keep all parameters identical to the quiescent state, and only increase the magnetic field at the IBS to $B_{\rm sh} = 3$~G and the companion temperature again to $T_{\rm comp} = 45000$~K (dashed gray line). It can be seen that this significant rise in temperature has a large effect on the predicted IC flux from this system because of its dependence of temperature as described in Section~\ref{subsec:parameter_study} in the context of Fig.~\ref{fig:Tcomp&Bulk}. Once again the energy-dependent SR and IC light curve shapes are indicated, both for the quiescent and flaring state. One can see that the predicted peaks become more widely separated as energy is increased. The SR and IC pulse shapes are similar, but there are minor differences. {Our model is able to describe the GeV modulations reported by \citet{an2017high} for the $1-100$~GeV bin and for the off-pulse pulsar phases reasonably well, although our SR-band peaks are not completely aligned with the data.} Future MeV and TeV data may constrain different scenarios for the character, and perhaps physical origin (i.e., magnetic reconnection in the companion magnetosphere) of the optical/X-ray flares.

%%%%%%%%%%%%%%%%%%%%%%%%%%%%%%%%%%%%%%%%%%%%%%%%%%%%%%%%%%%%%%%%%%%%%%%%%%%%%%%%%%%%%%%%%%%%%%%%
\subsubsection{J2339$-$0533  (RB) -- ICDP}
%%%%%%%%%%%%%%%%%%%%%%%%%%%%%%%%%%%%%%%%%%%%%%%%%%%%%%%%%%%%%%%%%%%%%%%%%%%%%%%%%%%%%%%%%%%%%%%%

\begin{figure}[h!]
\gridline{\leftfig{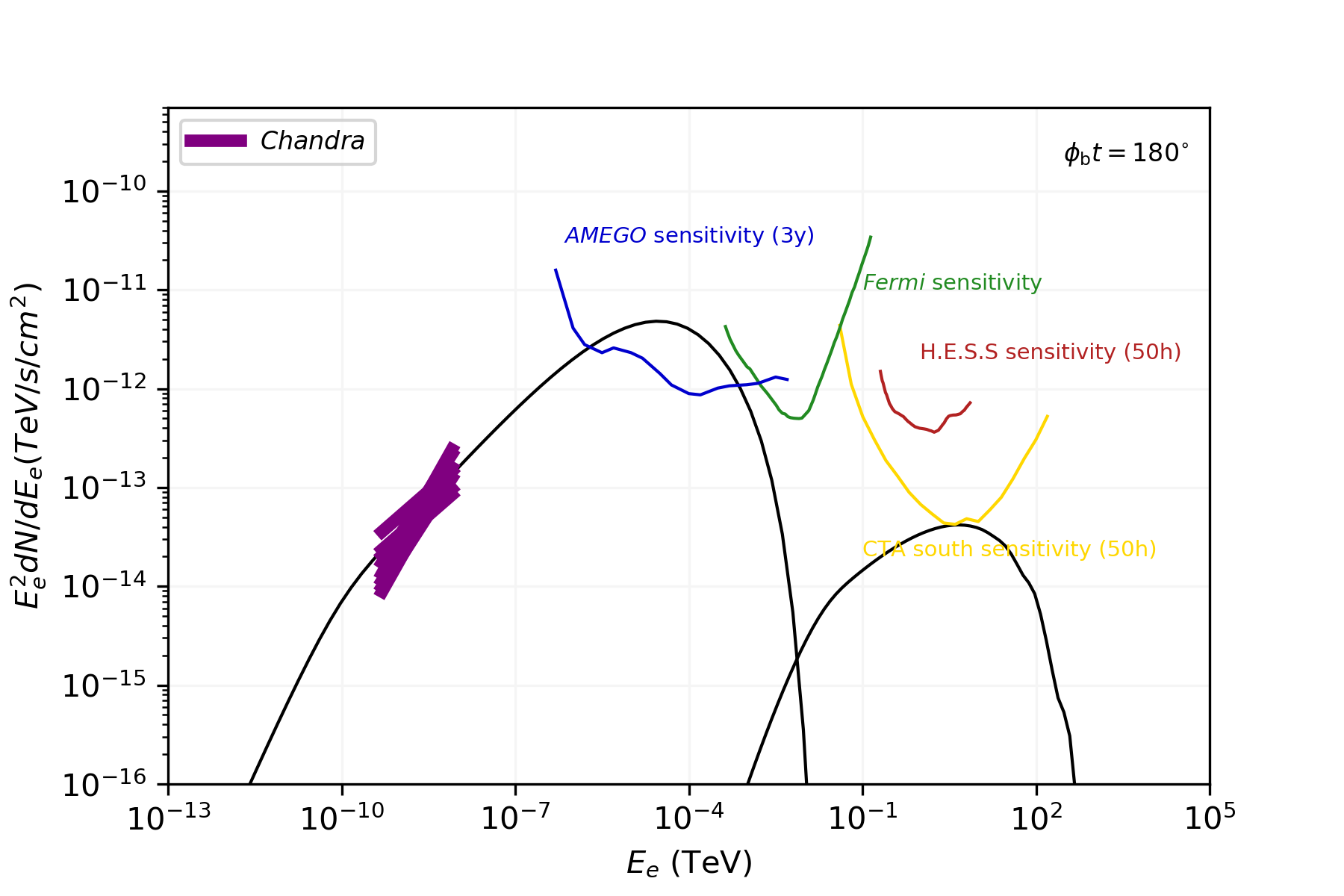}{0.51\textwidth}{(a)}
          \rightfig{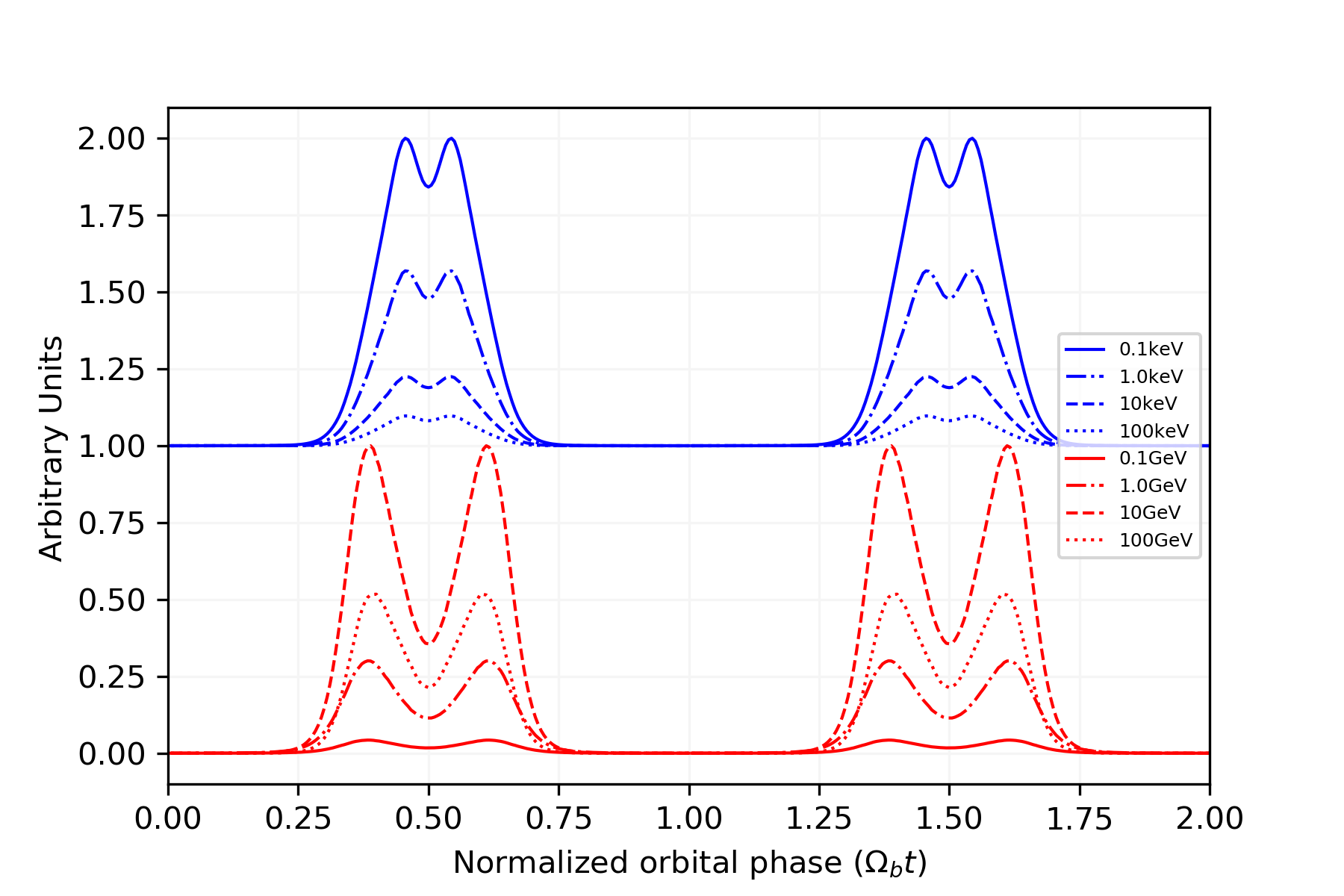}{0.51\textwidth}{(b)}
         }
\caption{The model (a) SED and (b) energy-dependent light curves of PSR~J2339$-$0533.}
\end{figure}

PSR~J2339$-$0533 is an ICDP RB system. The companion is non-degenerate with a temperature of $T_{\rm comp} \sim 6000$~K. The mass ratio is $q \sim 18.2$, which implies a companion mass of $M_{\rm comp} \sim 0.1 M_{\odot}$ \citep{2011ApJ...743L..26R}.
We use a pulsar mass of $M_{\rm NS} = 1.7 M_{\odot}$, period $P = 2.88$ ms and $\dot{P} = 1.41\times 10^{-20}$, distance $d= 1.10$~kpc, orbital period $P_{\rm b} = 4.6$~hr, and an inclination of $i = 54^{\circ}$ as reported by \citep{2019ApJ...879...73K}. 
We make use of publicly reported \textit{Chandra} spectral fits from \citet{2011ApJ...743L..26R}.
Note that we are able to match the X-ray data for PSR~J2339$-$0533 in two model scenarios, as was shown in Fig.~9(b). The model in Fig.~12(a) is similar to the gray models in Fig. 9(b) but with slightly different parameters (cf.\ Table~$\ref{tab1}$) leading to a slightly higher VHE flux and better match to the X-ray spectrum (we chose to incorporate a high minimum energy break in the particle spectrum). This model has a particle spectral index of $p=1.8$ and low pair multiplicity ($M_{\rm pair} = 500$). The model indicates that this source may be detectable by \textit{AMEGO}. 

We noted in  Section~\ref{sec:Intro} that \citet{2020arXiv200507888A} recently reported an orbitally-modulated gamma-ray light curve that is offset by $180^\circ$ in binary phase from the X-ray double-peaked light curve. This implies that this GeV component measured by \textit{Fermi} LAT must arise from a different energetic leptonic population than the IBS emission that we consider in this work. In this work, our gamma-ray and X-ray light curves are predicted to be phase-aligned, as they originate from the same underlying particle population. Moreover, our predicted SR spectrum cuts off before the \textit{Fermi} LAT energy range for the parameters we used, and thus our predicted SR light curves will only be detectable by \textit{AMEGO} in this case. We leave the modelling of the extra spectral component that has now been measured by \textit{Fermi} LAT to a future work.

%%%%%%%%%%%%%%%%%%%%%%%%%%%%%%%%%%%%%%%%%%%%%%%%%%%%%%%%%%%%%%%%%%%%%%%%%%%%%%%%%%%%%%%%%%%%%%%%
\subsubsection{J1959+2048 (B1957+20) (BW) -- SCDP }
%%%%%%%%%%%%%%%%%%%%%%%%%%%%%%%%%%%%%%%%%%%%%%%%%%%%%%%%%%%%%%%%%%%%%%%%%%%%%%%%%%%%%%%%%%%%%%%%

\begin{figure}[h!]
\gridline{\leftfig{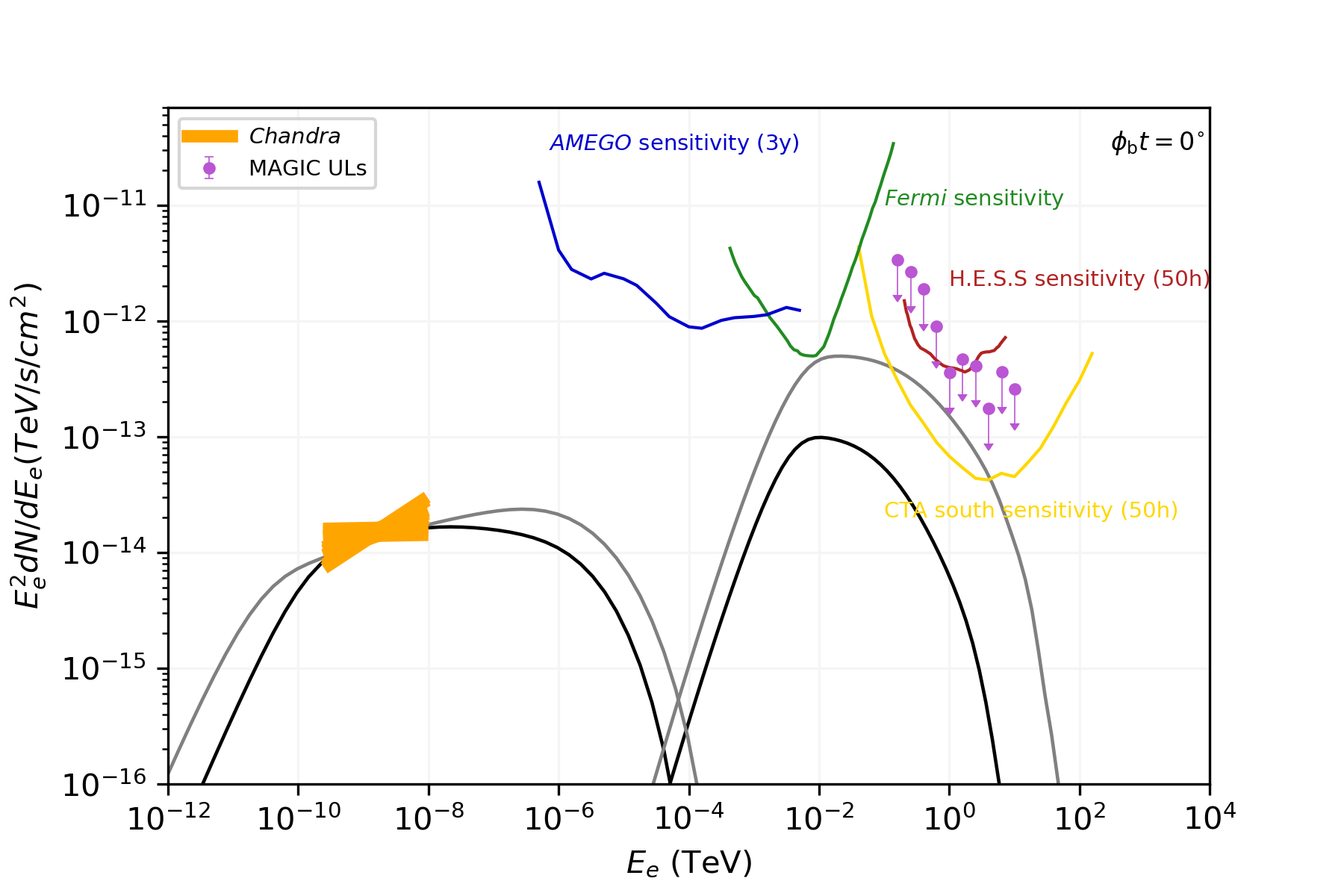}{0.51\textwidth}{(a)}
          \rightfig{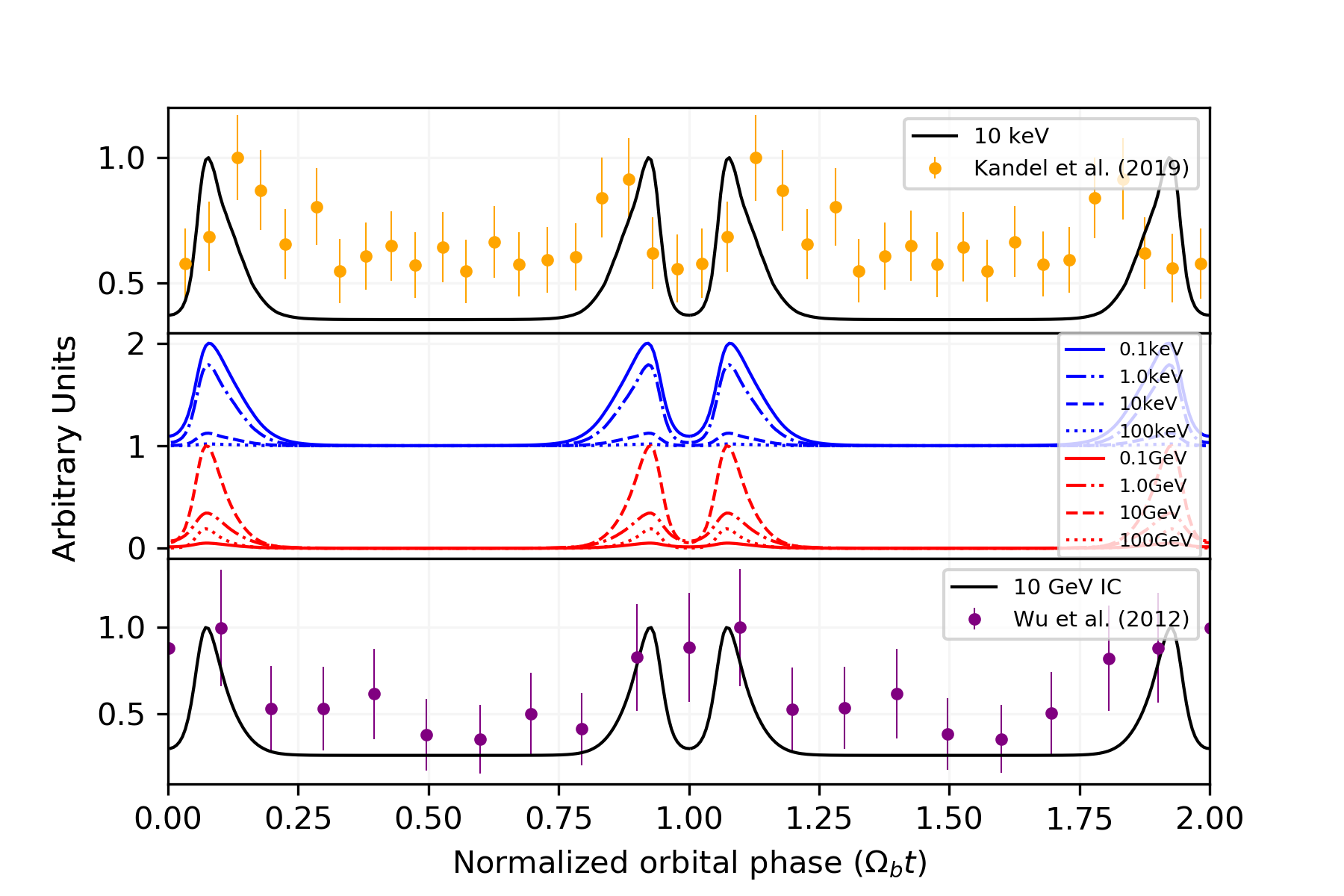}{0.5\textwidth}{(b)}
         }
\caption{Plot for PSR~B1957+20 depicting the model (a) SED for both $i = 65^{\circ}$ (gray) and $i = 85^{\circ}$ (black), and (b) energy-dependent light curves for $i = 85^{\circ}$. The bottom panel shows how the model predictions compare to modulated GeV emission reported in \citet{Wu2012}.}
 \label{fig:B1957}
\end{figure}

PSR~B1957+20 is the original and famous BW system \citep{fruchter1988millisecond}. We adopt $M_{\rm NS} = 1.7 M_{\odot}$, period $P = 1.60$ ms, and $\dot{P} = 1.69\times 10^{-20}$. The companion has a temperature of $T_{\rm comp} \sim 8500$~K \citep{reynolds2007light,2011ApJ...728...95V}. 
This system is located at a distance of $d \sim 1.40$~kpc, has an orbital period of $P_{\rm b} = 9.17$~hr, and an inclination of either $i = 65^{\circ}$ or $i = 85^{\circ}$ \citep{reynolds2007light,2011ApJ...728...95V,johnson2014constraints,Wadiasingh2017}. 
Considering the lower limit on the mass of the companion, $M_{\rm comp} \gtrsim 0.03 M_{\odot}$ \citep{reynolds2007light}, we use $M_{\rm comp} \sim 0.03 M_{\odot}$, yielding a mass ratio $q \sim 70$. We model \textit{Chandra} data from \citet{2012ApJ...760...92H}. 
We are able to match the X-ray spectral data for PSR~B1957+20 for both inclination angles. Both models require a low-energy break in the particle spectrum and the X-ray data cannot be described with the intrinsic single-particle SR slope of 4/3. The gray line ($i = 65^{\circ}$) is a model that has a very soft particle spectral index ($p=2.5$) and very high pair multiplicity ($M_{\rm pair} = 9000$). The black model has similar $p$ and $M_{\rm pair}$ values, but for a much higher magnetic field at the shock $B_{\rm sh} = 7$~G and smaller shock radius $R_{\rm sh}=0.2a$ compared to $B_{\rm sh} = 1.5$~G and $R_{\rm sh} = 0.4a$ for the $i = 65^{\circ}$ case. The IC component satisfies the VHE upper limits obtained by MAGIC \citep{ahnen2017observation}. Our predicted light curves are double-peaked and provide a reasonable match to the X-ray data from \citet{2019ApJ...879...73K}, which are not fine-tuned to our fixed energy of 10 keV. One way to improve the fit would be a lower the bulk flow magnitude; alternatively, one could diminish the latitudinal variation of the bulk Lorentz factor constituted by Eq.~(\ref{eq:bulk_flow}). This would modify the Doppler beaming in an array of directions, reducing the ``pulse fraction'' i.e., the amplitudes between maximum and minimum light. The considerable scatter in the light curve data in Fig.~\ref{fig:B1957} (and also in Fig.~\ref{fig:J1723} for PSR~J1723$-$2837) limit the insights to be gleaned from taking such a step currently. {\citet{Wu2012} reported a low-significance modulation above 2.7~GeV at the orbital period using \textit{Fermi} LAT data. Although we can fit these GeV light curve data invoking the shock IC spectral component, our model predicts double peaks and not single peaks. This perhaps suggests a non-shock emission component, from a different electron population than that of the IBS, for example the IC scattering of optical photons by leptons in the upstream pulsar wind. }

%%%%%%%%%%%%%%%%%%%%%%%%%%%%%%%%%%%%%%%%%%%%%%%%%%%%%%%%%%%%%%%%%%%%%%%%%%%%%%%%%%%%%%%%%%%%%%%%
\section{SUMMARY AND OUTLOOK}\label{sec:Conclusion}
%%%%%%%%%%%%%%%%%%%%%%%%%%%%%%%%%%%%%%%%%%%%%%%%%%%%%%%%%%%%%%%%%%%%%%%%%%%%%%%%%%%%%%%%%%%%%%%%

The single-peaked or double-peaked orbitally-modulated X-ray emission, along with orbital-phase and frequency-dependent radio eclipses of MSPs observed in spider binary systems, imply the existence of an IBS. This shock is believed to be a site of particle acceleration, with the predicted X-ray SR being Doppler-boosted due to a bulk flow of plasma along the shock tangent. These bulk motions include a dependence on the distance from the pulsar-companion axis, with an increase in speed away from the shock nose to its periphery. In this study, we used our newly developed \texttt{UMBRELA} code to predict energy-dependent light curves and phase-resolved spectra from these binary systems. This multi-zone code solves a simplified transport equation that includes diffusion, convection and radiative energy losses in an axially-symmetric, steady-state approach. The emissivities for SR and IC, including Doppler beaming, are then calculated for each spatial zone as the particles move along the shock surface. Modeling the expected SR and IC emission from the intrabinary shock and constraining our model parameters using observational data on these sources enabled us to investigate the underlying physics of these systems. 

In this work, we present models of the spectra and light curves in the X-ray through VHE band from two BW (PSR~J1311$-$3430, PSR~B1957+20) and two RB (PSR~J1723$-$2837, PSR~J2339$-$0533)  sources.  We employed reported X-ray spectral fits for all four of these systems to approximately anchor our SR spectra, enabling us to make a more robust prediction for the expected HE to VHE flux. We ascertained that the general character of our model light curves is fairly similar to the observed data for PSR~B1957+20 and PSR~J1723$-$2837, the systems most archetypal with among the best determinations of their orbital flux modulations.

We find that the predicted light curves and spectral models are currently quite unconstrained solely by X-ray and optical information. More broadband detections are needed, especially by VHE instruments, in order to better constrain the model parameter space. This prompted us to explore the parameter space, finding that the most important parameters for reproducing the SR spectra are the injection spectral index ($p$), pair multiplicity ($M_{\rm pair}$), magnetic field at the shock ($B_{\rm sh}$), and acceleration efficiency ($\epsilon_{\rm acc}$). Additionally, detectability of these sources by instruments such as H.E.S.S.\ and the future CTA telescope is highly dependent on the companion temperature ($T_{\rm comp}$), the bulk flow momentum of particles along the shock tangent ($\beta \Gamma$) and also the spectral index $p$. {We thus find that BWs and RBs may be a promising class of VHE targets for detecting modulated IC flux, especially for nearby, flaring sources.} Constraining the IC emission by future data from ground-based Cherenkov telescopes may probe the particle acceleration at the shock as well as the pulsar wind content. Likewise, measurement of the SR component and its curvature in the MeV band will constrain the efficiency of particle acceleration in the termination shock.

In future, we will improve our fitting procedure by incorporating an automated minimization or multi-dimensional parameter sampler. The use of Markov chain Monte Carlo techniques is an obvious future path. { Likewise, measurements of MeV--GeV orbitally-modulated emission components \citep[see Table~2 in][]{2020arXiv200403128T} can provide important constraints for our models.} Following the findings of \cite{2020arXiv200507888A} for PSR~J2339$-$0533, we will also incorporate non-shock emission components in our modeling, e.g., anisotropic IC emission from a cold, upstream wind. We will furthermore implement an improved shock geometry, shock sweepback, spatially-dependent acceleration and injection, refine our transport calculations, and use our code to make predictions for the expected HE to VHE flux for several more spider binaries. One could also calculate the escaping positron flux from these systems that may contribute to the local positron excess seen above 10~GeV \citep{adriani2015time, Venter2015}. MHD calculations may further elucidate the wind-wind or wind-magnetosphere interactions, which will give more insight into the underlying physics found in these unique systems, and the plasma's spatial variations along the intrabinary shock surface.
Continued observation of these binaries in both the time and energy domains will provide improved model constraints and aid in the scrutiny of pulsar wind physics in BW and RB contexts.

\acknowledgments
We acknowledge helpful discussions with Valenti Bosch-Ramon, Michael Backes, and Carlo van Rensburg. C.J.T. vdM.\ is supported by the National Astrophysics and Space Science Programme (NASSP). Z.W.\ is supported by the NASA postdoctoral program. This work is based on the research supported wholly / in part by the National Research Foundation of South Africa (NRF; Grant Numbers 87613, 90822, 92860, 93278, and 99072). The Grantholder acknowledges that opinions, findings and conclusions or recommendations expressed in any publication generated by the NRF supported research is that of the author(s), and that the NRF accepts no liability whatsoever in this regard. A.K.H.\ acknowledges the support from the NASA Astrophysics Theory Program. C.V., A.K.H., and M.G.B.\ acknowledge support from the \textit{Fermi} Guest Investigator Program. This work has made use of the NASA Astrophysics Data System.

\appendix
%%%%%%%%%%%%%%%%%%%%%%%%%%%%%%%%%%%%%%%%%%%%%%%%%%%%%%%%%%%%%%%%%%%%%%%%%%%%%%%%%%%%%%%%%%%%%%%%
\section{Particle Transport}\label{app_Transport}
%%%%%%%%%%%%%%%%%%%%%%%%%%%%%%%%%%%%%%%%%%%%%%%%%%%%%%%%%%%%%%%%%%%%%%%%%%%%%%%%%%%%%%%%%%%%%%%%

In this Appendix, we simplify Eq.~(\ref{eq:transport}) which is repeated here for convenience:
\begin{eqnarray}
\frac{\partial N_{\rm e}}{\partial t}  =  -\vec{V}\cdot\left(\vec{\nabla}N_{\rm e}\right) + \kappa(\gamma_{\rm e})\nabla^2N_{\rm e}\nonumber
 + \frac{\partial}{\partial \gamma_{\rm e}}\left(\dot{\gamma}_{\rm e,tot}N_{\rm e}\right)-\left(\vec{\nabla}\cdot\vec{V}\right)N_{\rm e}+Q.
\end{eqnarray}
We make a number of assumptions. First, we assume that $N_{\rm e}$ has only co-latitudinal and energy dependence, with $N_{\rm e}\propto\mu=\cos\theta$ (i.e., to model the fact that we expect to have a high density of particles at the shock nose, but that this density decreases as particles move along the shock tangent and eventually escapes from the last zone). The conventional definition of co-latitude is such that it increases counter-clockwise.  Our definition is such that $\theta = -\theta^\prime$, with $\theta^\prime$ the standard definition of co-latitude. This implies (neglecting $\partial^2N_{\rm e}/\partial\mu^2$, since $N_{\rm e}$ is assumed to be linear in $\mu$)
\begin{eqnarray}
 \frac{\partial N_{\rm e}}{\partial\theta^\prime} & = & -\sin\theta\frac{\partial N_{\rm e}}{\partial\mu}\frac{\partial\theta}{\partial\theta^\prime} = \sin\theta \frac{N_{\rm e}}{\mu}=\tan\theta N_{\rm e},\\
 \frac{\partial^2 N_{\rm e}}{\partial(\theta^\prime)^2} & = & \frac{\partial}{\partial\mu}\left(\frac{\partial N_{\rm e}}{\partial\mu}\frac{\partial \mu}{\partial\theta}\right)\frac{\partial \mu}{\partial\theta}\left(\frac{\partial\theta}{\partial\theta^\prime}\right)^2 = \frac{\partial N_{\rm e}}{\partial\mu}\frac{\partial^2 \mu}{\partial\theta^2}\nonumber
 = -N_{\rm e}.
\end{eqnarray}
We assume that the bulk flow only has a co-latitudinal dependence $\vec{V} = c\beta({\theta})\hat{\boldsymbol{\theta}}$. The first term on the right-hand side of Eq.~(\ref{eq:transport}) then becomes (with $\theta$ increasing clockwise)
\begin{eqnarray}
-\vec{V}\cdot\left(\vec{\nabla}N_{\rm e}\right)  =  -V(\theta)\hat{\boldsymbol{\theta}}\cdot\left(\frac{1}{R_0}\frac{\partial N_{\rm e}(\theta)}{\partial\theta}\right)\hat{\boldsymbol{\theta}} \nonumber
 =  -\frac{c\beta(\theta)}{R_0} \tan\theta N_{\rm e} \equiv  -\frac{N_{\rm e}}{\tau_1},
\end{eqnarray}
with $\tau_1 = R_0/(c\beta(\theta)\tan\theta)$.

The second term on the right-hand side of equation~(\ref{eq:transport}) becomes (again neglecting $\partial^2N_{\rm e}/\partial\mu^2$)
\begin{eqnarray}
  \kappa(\gamma_{\rm e})\nabla^2N_{\rm e} & = & \frac{\kappa}{R_0^2\sin\theta^\prime}\left(\cos\theta^\prime\frac{\partial N_{\rm e}}{\partial\theta^\prime} + \sin\theta^\prime\frac{\partial^2 N_{\rm e}}{\partial(\theta^\prime)^2}\right)\\  
  & = & -\frac{\kappa}{R_0^2\sin\theta}\left(\cos\theta\left(\tan\theta N_{\rm e}\right) - \sin\theta(- N_{\rm e})\right)\nonumber\\
  & = & -\frac{2\kappa}{R_0^2}N_{\rm e}\equiv-\frac{N_{\rm e}}{\tau_{\rm diff}},
\end{eqnarray}
with $\tau_{\rm diff}=R_0^2/2\kappa$.

Let us assume that the particles lose energy due to adiabatic and radiation losses: $\dot{\gamma}_{\rm e,tot} = \dot{\gamma}_{\rm e,ad} +\dot{\gamma}_{\rm e,rad}$. The adiabatic loss (cooling) rate is given by $\dot{\gamma}_{\rm e,ad} = -\gamma_{\rm e}(\vec{\nabla}\cdot\vec{V})/3.$ Splitting the third term on the right-hand side of equation~(\ref{eq:transport}) we find 
\begin{equation}
\frac{\partial}{\partial \gamma_{\rm e}}\left(\dot{\gamma}_{\rm e,ad}N_{\rm e}\right) = -\frac{1}{3}\left(\vec{\nabla}\cdot\vec{V}\right)N_{\rm e}\equiv-\frac{N_{\rm e}}{\tau_{\rm ad}}.
\end{equation}
Using
\begin{eqnarray}
\vec{\nabla}\cdot\vec{V}  =  \frac{1}{R_0\sin\theta^\prime}\frac{\partial}{\partial\theta^\prime}\left(c\beta(\theta)\sin\theta^\prime\right)\nonumber
 =  -\frac{1}{R_0\sin\theta}\frac{\partial}{\partial\theta}(c\beta(\theta)\sin\theta)
 =  -\frac{c}{R_0\sin\theta}\left[\frac{\partial\beta}{\partial\theta}\sin\theta +\beta\cos\theta\right],
\end{eqnarray}
we obtain 
\begin{equation}
	\tau_{\rm ad} = \frac{3R_0}{c}\left[\frac{\partial\beta}{\partial\theta} +\beta\cot\theta\right]^{-1}.
\label{eq:t_ad}
\end{equation}
The penultimate term on the right-hand-side of equation~(\ref{eq:transport}) is just three times larger in absolute magnitude than the adiabatic loss term. We can thus replace this with $-N_{\rm e}/\tau_2$ and define $\tau_2=\tau_{\rm ad}/3.$ To be explicit, we kept $\tau_{\rm ad}$ and $\tau_2$ separate, but these could be combined into one term. Furthermore, the adiabatic losses and convection terms all depend on there being a non-zero bulk flow of particles, and thus share the same basic physical origin.

Let us next assume that $N_{\rm e}\dot{\gamma}_{\rm e,rad}\propto \gamma_{\rm e}^{-p}$, such that 
\begin{equation}
\frac{\partial}{\partial \gamma_{\rm e}}\left(\dot{\gamma}_{\rm e,rad}N_{\rm e}\right) = \frac{-pN_{\rm e}\dot{\gamma}_{\rm e,rad}}{\gamma_{\rm e}}\equiv-\frac{N_{\rm e}}{\tau_{\rm rad}}.
\end{equation}
If $p\approx1$, we have $\tau_{\rm rad}=\gamma_{\rm e}/\dot{\gamma}_{\rm e,rad}$.

%%%%%%%%%%%%%%%%%%%%%%%%%%%%%%%%%%%%%%%%%%%%%%%%%%%%%%%%%%%%%%%%%%%%%%%%%%%%%%%%%%%%%%%%%%%%%%%%
\section{Derivation of the Expression for the Photon Number Density in the Comoving Frame} \label{app:nBB}
%%%%%%%%%%%%%%%%%%%%%%%%%%%%%%%%%%%%%%%%%%%%%%%%%%%%%%%%%%%%%%%%%%%%%%%%%%%%%%%%%%%%%%%%%%%%%%%%

\begin{figure}[ht!]
\centering
\includegraphics[scale=0.7]{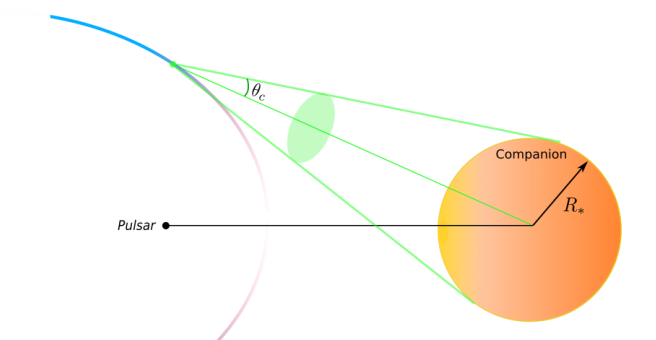} 
\caption{Schematic diagram depicting the geometry for the calculation of $n_*$.}
\end{figure}

The total soft-photon energy density in the comoving frame is \citep{bottcher2012relativistic}, 
\begin{equation}
	u^{\rm em} = \int^{\infty}_{0} d\nu^{\rm rec} \int\displaylimits_{4\pi} d\Omega^{\rm em} \frac{u^{\rm rec}_{\nu^{\rm rec}}(\Omega^{\rm rec})}{\Gamma^{4}(1 + \beta \mu^{\rm em})^{4}},
\end{equation}
\\
where $\nu$ is the photon frequency, $d\Omega$ the solid angle, $\Gamma$ the bulk flow Lorentz factor, $\beta_\Gamma=v/c$ the normalized bulk flow speed, $\mu=\vec{n}\cdot\vec{u}$ the cosine of the angle between the flow and observer directions (see Eq.~[\ref{eq:delta}]), $u_\nu(\Omega)$ the energy density per solid angle and frequency interval, $u_\nu$ the energy density per frequency interval, and $u=
\int^{\infty}_{0} d\nu\, u_{\nu}$  the total energy density; the superscripts `em' and `rec' respectively refer to the comoving (emission) and observer (lab) frame, previously indicated by primed and unprimed quantities in Section~\ref{sec:Beaming}, to conform to the usage of \citet{bottcher2012relativistic}.

We assume that the radiation field in the observer (rec) frame is isotropic,
\begin{equation}
	u^{\rm rec}_{\nu^{\rm rec}}(\Omega^{\rm rec}) = \frac{u^{\rm rec}_{\nu^{\rm rec}}}{4\pi}.
\end{equation}
\\
Noting that $\delta = [\Gamma(1 - \beta\mu^{\rm rec})]^{-1} = \Gamma(1 + \beta\mu^{\rm em})$, where $\mu^{\rm rec} = \cos (\alpha -\pi/2)$ is the angle between the velocity of a particle (tangent to the shock) and the direction of a soft photon originating from the companion (taken to be along the direction pointing from the companion center to a particular spatial zone or particle position),
with $\alpha=\pi - \lambda - \theta$ (see Fig.~\ref{fig:schematic_BW}),  we have
\begin{eqnarray}
u^{\rm em} &=&  \frac{u^{\rm rec}}{4\pi} \int d\Omega^{\rm rec} \left( \frac{d\Omega^{\rm em}}{d\Omega^{\rm rec}} \right) \delta^{-4}.
\end{eqnarray}
We assume that the isotropic BB emission is beamed into a cone of width $\mu_{\rm c} = \cos \theta_{c}$ in the observer frame:
\begin{eqnarray}
u^{\rm em}  &=&  \frac{1}{2}u^{\rm rec} \int^{1}_{\mu_{\rm c}}  d\mu^{\rm rec} \left( \delta^{2} \right) \delta^{-4}\\
  &=& \frac{1}{2} \Gamma^{2} u^{\rm rec} \int^{1}_{\mu_{\rm c}} d\mu^{\rm rec}(1 -  \beta\mu^{\rm rec})^{2}
\end{eqnarray}\\
Assuming $\mu\approx\mu_{\rm c}$ due to the beamed nature of the BB emission (as `seen' by the particle) and normalizing this expression using the BB energy density in the observer frame, we can evaluate the above integral to finally obtain an expression for the total soft-photon density in the comoving frame (divided by a factor 2 since we are only assuming radiation from the facing side of the companion)
\begin{equation}
u_{\rm sp} \approx  \frac{2 \sigma_{\rm SB} T^4 }{c} \Gamma^{2} [1 - \mu_{\rm c}](1 - \beta \mu^{\rm rec})^{2}.
\end{equation}
The expression for the spectral photon number density in the comoving frame can then be written as
\begin{equation}
n_{*}(\epsilon_*) = \frac{30 \sigma_{\rm SB} h}{(\pi k_{\rm B})^4} (1-\mu_{\rm c}) \epsilon_*^{2}\delta^{-2} \left[\exp\left(\frac{\epsilon_*}{k_{\rm B} T \delta}\right) - 1 \right]^{-1}\!\!\!\!\!\!. 
\end{equation}
A Taylor expansion of the $1- \mu_{\rm c}$ term yields,
\begin{equation}
n_{*}(\epsilon_*) = \frac{30 \sigma_{\rm SB} h}{(\pi k_{\rm B})^4} \left(\frac{R_{*}}{R_{\rm sh}}\right)^2  \epsilon^{2}_{*}\delta^{-2}\left[\exp\left(\frac{\epsilon_{*}}{k_{\rm B} T \delta}\right) - 1 \right]^{-1}\!\!\!\!\!\!,
\end{equation}
where $h$ is Planck's constant and $\epsilon_{*}$ is the photon energy. This final expression is similar to the expression for the photon number density used in \citet{dubus2015modelling}.
\\
\section{Estimate of Synchrotron-self-Compton Flux Level}\label{sec:AppSSC}
In this Section, we show that the SSC flux is expected to be orders of magnitude lower than the IC flux, and this emission process can therefore safely be neglected.
\\
\\
Let us write the normalized photon energy as 
\begin{equation}
    \chi \equiv \frac{h \nu}{m_{\rm e} c^2}.
\end{equation}
To make a rough estimate of the expected SSC flux, we focus on SR photons with energies of $E_{\rm SR} =100~{\rm MeV}= 1.6\times10^{-4}$~erg, which means that $\chi \sim 200 \gg 1$ and implies that we are in the Klein-Nishina regime. The expected SSC flux is approximated by:
\begin{equation}
    \left(\nu F_{\nu}\right)_{\rm SSC} \approx c n_{\rm ph} \frac{\sigma_{\rm KN}}{4 \pi d^2} \frac{dN}{dE_{\rm e}} E_{\rm IC},
\end{equation}
where \citep{1970RvMP...42..237B}
\begin{equation}
    \sigma_{\rm KN} \approx \frac{3}{8} \sigma_{\rm T} \left[\ln \left(2 \chi \right) +\frac{1}{2} \right] \left(\frac{1}{\chi}\right) \sim 8 \times 10^{-27}~{\rm cm}^2.\label{eq:KN}
\end{equation}
If we then consider a pulsar with a spin-down luminosity of $\dot{E}_{\rm rot} = 10^{34}$~ erg\,s$^{-1}$ and very conservatively assume that $50\%$ of $\dot{E}_{\rm rot}$ is converted into SR (in reality this number is much lower in our model fits, typically a few fractions of a percent, but this depends on the assumed parameters that determine the particle transport), we can approximate the SR luminosity as being $L_{\rm SR} \lesssim 5 \times 10^{33}$~erg\,s$^{-1}$. Assuming the size of the emission region (thus assuming the emission to come primarily from the first spatial zone) to be $A\sim 2\pi R_0^2(1-\mu_0) \sim 10^{18}$~cm$^2$, one can then estimate the photon number density as
\begin{equation}
    n_{\rm ph} \lesssim \frac{L_{\rm SR}}{c A E_{\rm SR}} \sim10^9~{\rm ph\,cm}^{-3}.
\end{equation}
Consider a source at a distance of $d = 2$~kpc. Estimating the steady-state particle spectrum to be roughly $dN/dE_{\rm e} \sim  10^{33}$~erg$^{-1}$ and assuming the typical electron energy to be $\gamma_{\rm e} \approx 2\times10^6$, we estimate the SSC flux as (using a cross section $\sigma_{\rm KN} \approx 8 \times 10^{-27}$~cm$^2$ as in Eq.~[\ref{eq:KN}])
\begin{equation}
    \left(\nu F_{\nu}\right)_{\rm SSC}\lesssim  10^{-18}~ {\rm erg\,s^{-1}\,cm^{-2}}.
    \label{SSC_flux}
\end{equation}
Next, we estimate the IC flux by considering only the scattering of soft photons produced by the companion, with an energy $E_{\rm soft} \sim 2.7kT \approx 1~{\rm eV}\sim 2 \times 10^{-12}$~erg, which means that $\chi \ll 1$,  implying that we are now in the Thompson regime. The expected IC flux is approximated by
\begin{equation}
    \left(\nu F_{\nu}\right)_{\rm IC} \approx c n_{\rm ph} \frac{\sigma_{\rm T}}{4 \pi d^2} \frac{dN}{dE_{\rm e}} E_{\rm IC},
\end{equation}
where $n_{\rm ph}$ is now given by
\begin{equation}
    n_{\rm ph} \sim \frac{u_{\rm BB}}{E_{\rm soft}} \sim \frac{1}{E_{\rm soft}}\frac{4 \sigma_{\rm SB} T^4}{c} \left(\frac{R_{*}}{r}\right)^2 \sim 3 \times 10^{11}~{\rm ph\,cm}^{-3},
 \end{equation}
 using $T \sim 6\times10^{3}$~K, $R_{*}\sim 8\times10^{10}$~cm and $r = a - R_{0} \sim 2\times10^{11}$~cm. Now using $\gamma_{\rm e}=2\times10^6$ and  $E_{\rm IC} \sim\gamma^2_{\rm e}E_{\rm soft}= 10$~erg, we estimate the IC flux to be
 \begin{equation}
     \left(\nu F_{\nu}\right)_{\rm IC} \sim  10^{-13}~{\rm  erg\,s^{-1}\,cm}^{-2}.
     \label{IC_flux}
 \end{equation}
Upon comparing Eq.~(\ref{SSC_flux}) and (\ref{IC_flux}), it is clear that the SSC flux is negligible compared to the IC flux (as can also be seen from more detailed modeling in Section~\ref{sec:Results}).

%%%%%%%%%%%%%%%%%%%%%%%%%%%%%%%%%%%%%%%%%%%%%%%%%%%%%%%%%%%%%%%%%%%%%%%%%%%%%%%%%%%%%%%%%%%%%%%%
\section{$\gamma\gamma$ absorption} \label{app:Photon_abs}
%%%%%%%%%%%%%%%%%%%%%%%%%%%%%%%%%%%%%%%%%%%%%%%%%%%%%%%%%%%%%%%%%%%%%%%%%%%%%%%%%%%%%%%%%%%%%%%%

TeV photons produced by IC scattering of the companion's soft-photon field may be may be attenuated by the same (optical) photon field. The angle-dependent two-photon pair production (Breit-Wheeler) cross section peaks at $\varepsilon_1 \varepsilon_2 (1-\mu_{12})/2 \approx 2$ where $\varepsilon_{1,2}$ are the photon energies and $\mu_{12}$ is the cosine of the interaction angle in the lab frame. Correspondingly, optical photons of a few eV preferentially interact with TeV photons for nearly all interaction angles relevant in this work. The optical depth for absoption of TeV photons is maximized for propagation directions directed toward the observer of photons traveling near pulsar superior conjunction. Such opacity is not all that relevant for Doppler-boosted photons that arise from wing locales of intrabinary shocks which surround the pulsar, as is the case for most of the sources in this work. Nevertheless, it is instructive to estimate the influence of $\gamma\gamma$ absorption here.
 
The characteristic optical depth at the peak of the cross section is $\tau_{\gamma\gamma} \sim  n_\gamma \sigma_{\gamma\gamma} a$ where $\sigma_{\gamma\gamma} \sim \sigma_T$ and $n_\gamma$ is the characteristic photon number density. Here we may estimate $n_\gamma \sim \zeta(3) (\Theta^\prime/\lambar)^3 (R_*/a)^2/\pi^2$ where $\Theta^\prime = k_b T/(m_e c^2)$ and $\lambar$ is the reduced Compton wavelength. Then (for the most optimal case), 
\begin{eqnarray}
\tau_{\gamma\gamma} \sim \alpha_{\rm f}^2 \left(\frac{8 \zeta(3)}{3 \pi} \right) \left(\frac{R_*^2}{a \lambar}\right) (\Theta^\prime)^3 
\;\sim \;  
 \begin{cases}
     0.03 \left(\frac{T}{6\times 10^3 \, \rm K}\right)^3 \left( \frac{a}{2\times10^{11} \, \rm cm} \right) \left( \frac{R_*/a}{0.3} \right)^2 \qquad \text{J1723--2837-like RB systems} \\
      0.05 \left(\frac{T}{3\times 10^4 \, \rm K}\right)^3 \left( \frac{a}{5\times10^{10} \, \rm cm} \right) \left( \frac{R_*/a}{0.07} \right)^2 \qquad \text{J1311--3430-like BW systems.}
  \end{cases}
\end{eqnarray}
The above estimate is for head-on photon angles and thus mostly operative for orbital phases near pulsar superior conjunction. For other phases, $\tau_{\gamma\gamma} $ is much smaller. Thus, we see that $\gamma\gamma$ absorption ($\propto \exp[-\tau_{\gamma\gamma}]$) may have at most a few percent influence on the TeV flux in both RBs and hot systems such as PSR~J1311$-$3430, except in exceptionally flaring states, owing to the $T^3$ scaling of the photon number density in those cases. 

%%%%%%%%%%%%%%%%%%%%%%%%%%%%%%%%%%%%%%%%%%%%%%%%%%%%%%%%%%%%%%%%%%%%%%%%%%%%%%%%%%%%%%%%%%%%%%%%
\section{Transport Timescales }\label{app:Timescales}
%%%%%%%%%%%%%%%%%%%%%%%%%%%%%%%%%%%%%%%%%%%%%%%%%%%%%%%%%%%%%%%%%%%%%%%%%%%%%%%%%%%%%%%%%%%%%%%%

\begin{figure}[h!]
\gridline{\fig{1723_timescale_Z0.png}{0.33\textwidth}{(a)}
          \fig{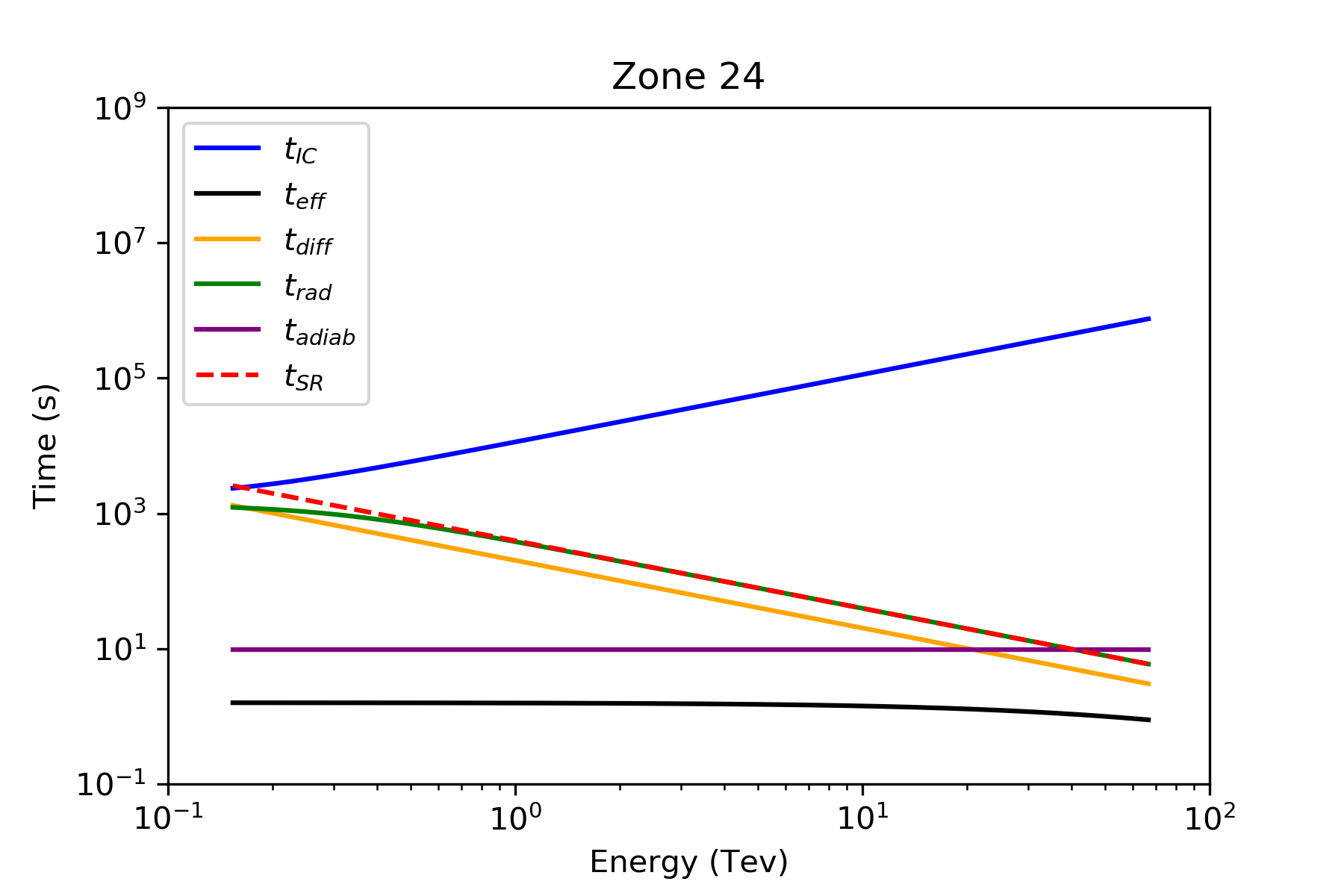}{0.33\textwidth}{(b)}
          \fig{1723_timescale_Z49.png}{0.33\textwidth}{(c)}
         }
\caption{Timescale plots for PSR~J1723$-$2837 depicting the (a) first, (b) middle and (c) last spatial zones.}
\end{figure}
In Fig.~15 we plot relevant timescales for the first, middle and final spatial zones of the shock for PSR~J1723$-$2837. These timescales were described in some detail in \S\ref{sec:Results} and the plots and discussion are repeated here for convenience and as a basis for comparison with results of other sources. The IC timescale (blue line) is much shorter for the first zone than for the last, given the reduction of soft-photon energy density with distance (increasing zone). The SR timescale (red line) is the same, given the constant shock magnetic field. The green line represents the effective radiation loss timescale ($\tau_{\rm rad}^{-1} = \tau_{\rm SR}^{-1} + \tau_{\rm IC}^{-1}$), while the yellow line indicates the diffusion timescale. This is similar for zones of similar sizes (and a constant magnetic field), as is the adiabatic timescale (purple line). Thus, the spatial (adiabatic and diffusion) timescales dominate the radiation ones in this case. As mentioned before, on the one hand convection leads to relatively short residence times of particles in particular zones, and thus reduced emission, but on the other hand, a large bulk flow speed increases the Doppler factor in each zone, compensating for this effect.
The effective timescale (black line) finds the shortest timescale between all the different processes. 

\begin{figure}[h!]
\gridline{\fig{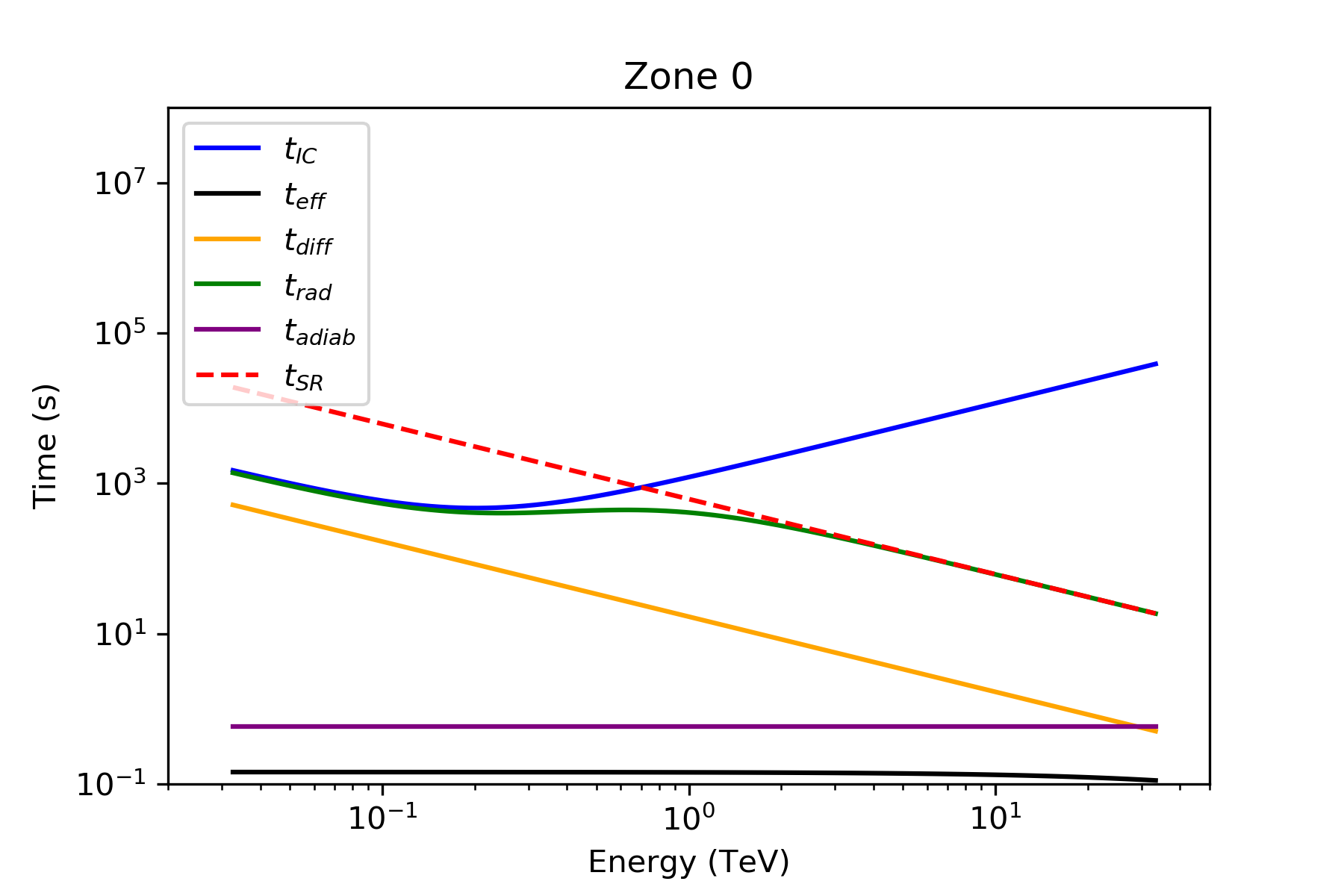}{0.33\textwidth}{(a)}
          \fig{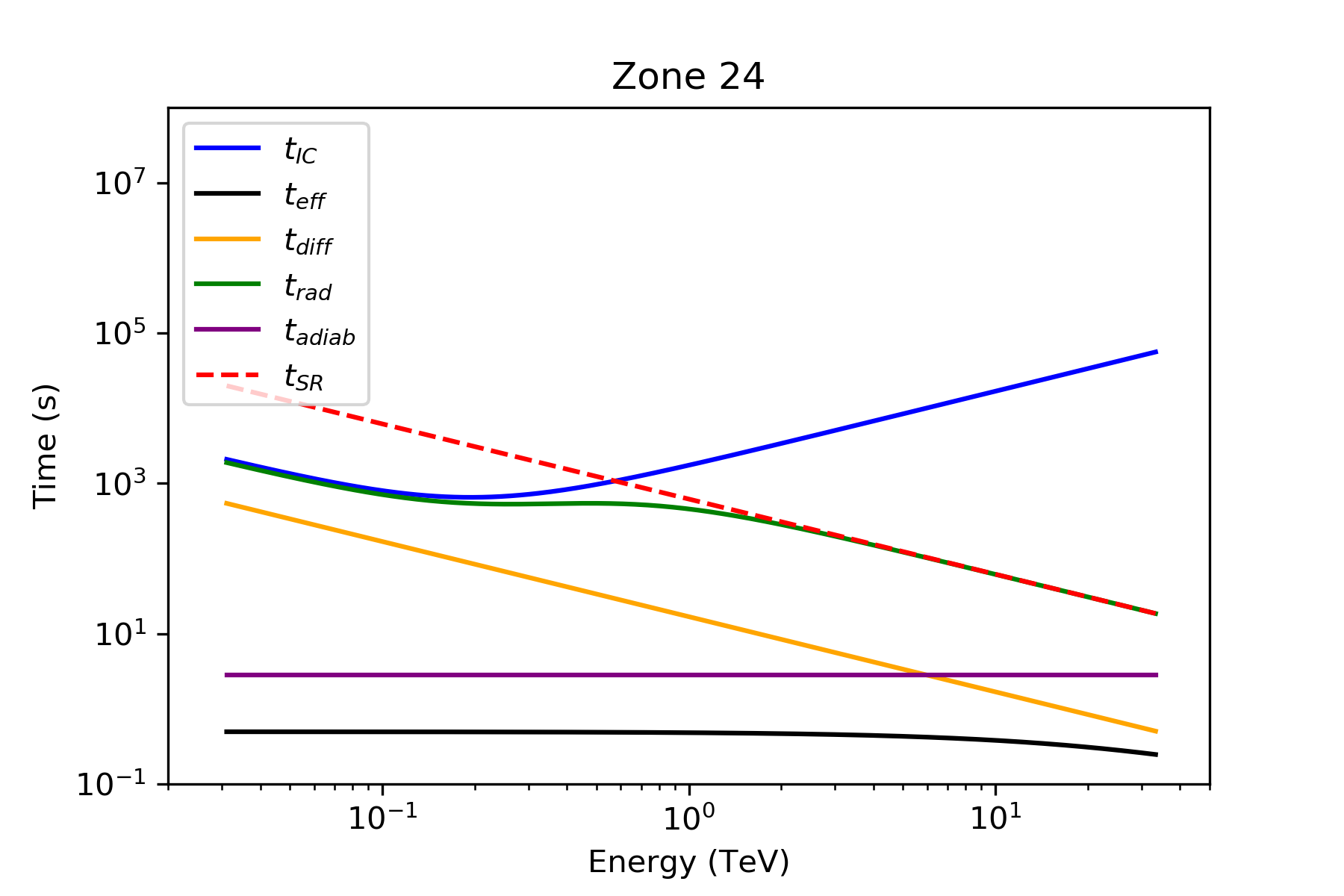}{0.33\textwidth}{(b)}
          \fig{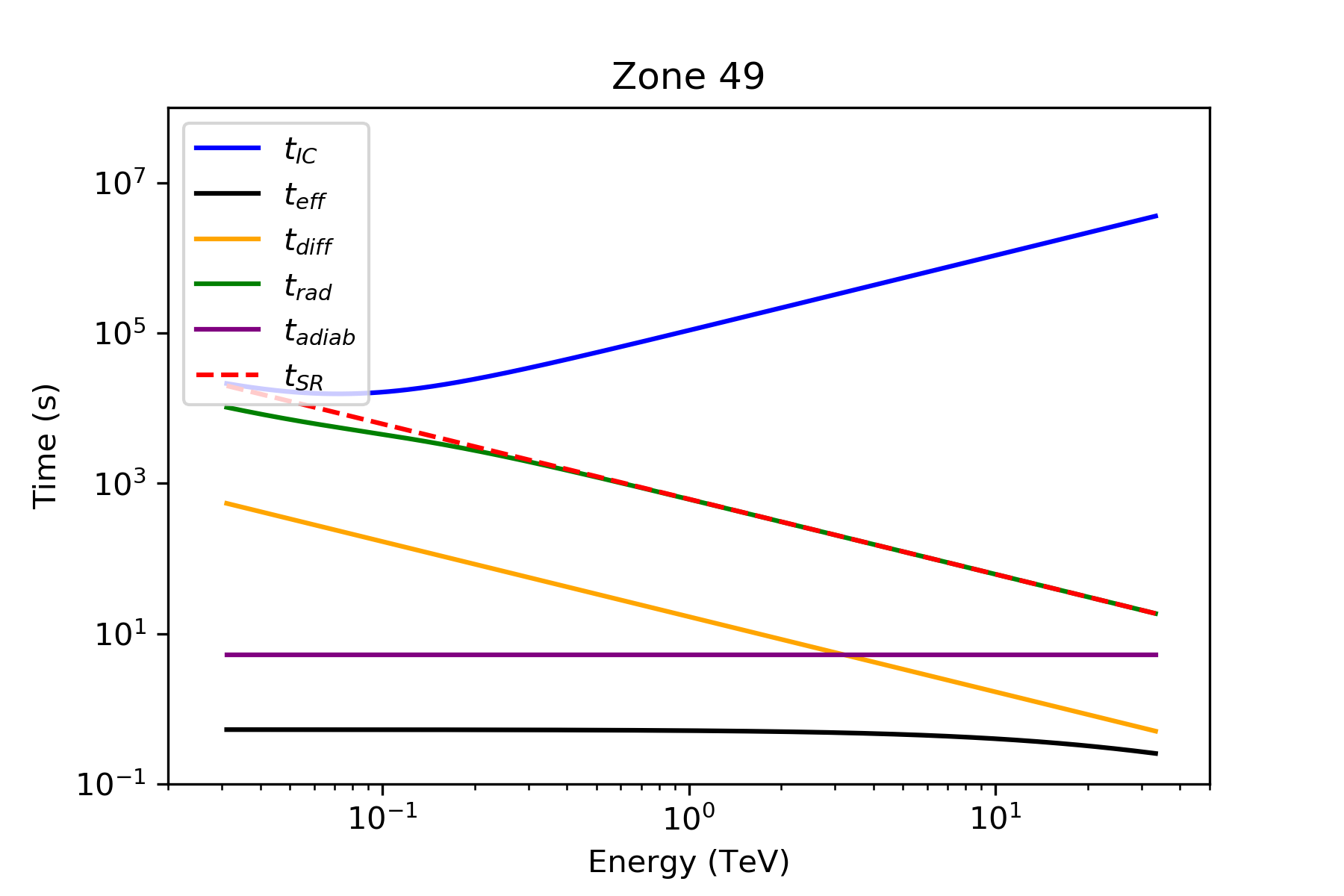}{0.33\textwidth}{(c)}
         }\label{fig:T2}
\caption{Timescale plots for PSR~J2339$-$0533 depicting the (a) first, (b) middle and (c) last spatial zones.}
\end{figure}

In Fig.~16 we plot relevant timescales for the first, middle and final spatial zones for PSR~J2339$-$0533. We note that the IC timescale (blue line) is only slightly shorter for the first zone than for the last. This is because we model this source with a smaller shock radius than for PSR~J1723$-$2837, so the particles are removed farther from the companion source of soft photons. For larger-distance zones, the particles are even farther away from the source of soft photons, giving a reduction of soft-photon energy density and thus an increase in the IC loss timescale. We see that the spatial (adiabatic and diffusion) timescales dominate the radiation ones for this source as well. The other timescales behave in a similar way to those  for PSR~J1723$-$2837.

\begin{figure}[h!]
\gridline{\fig{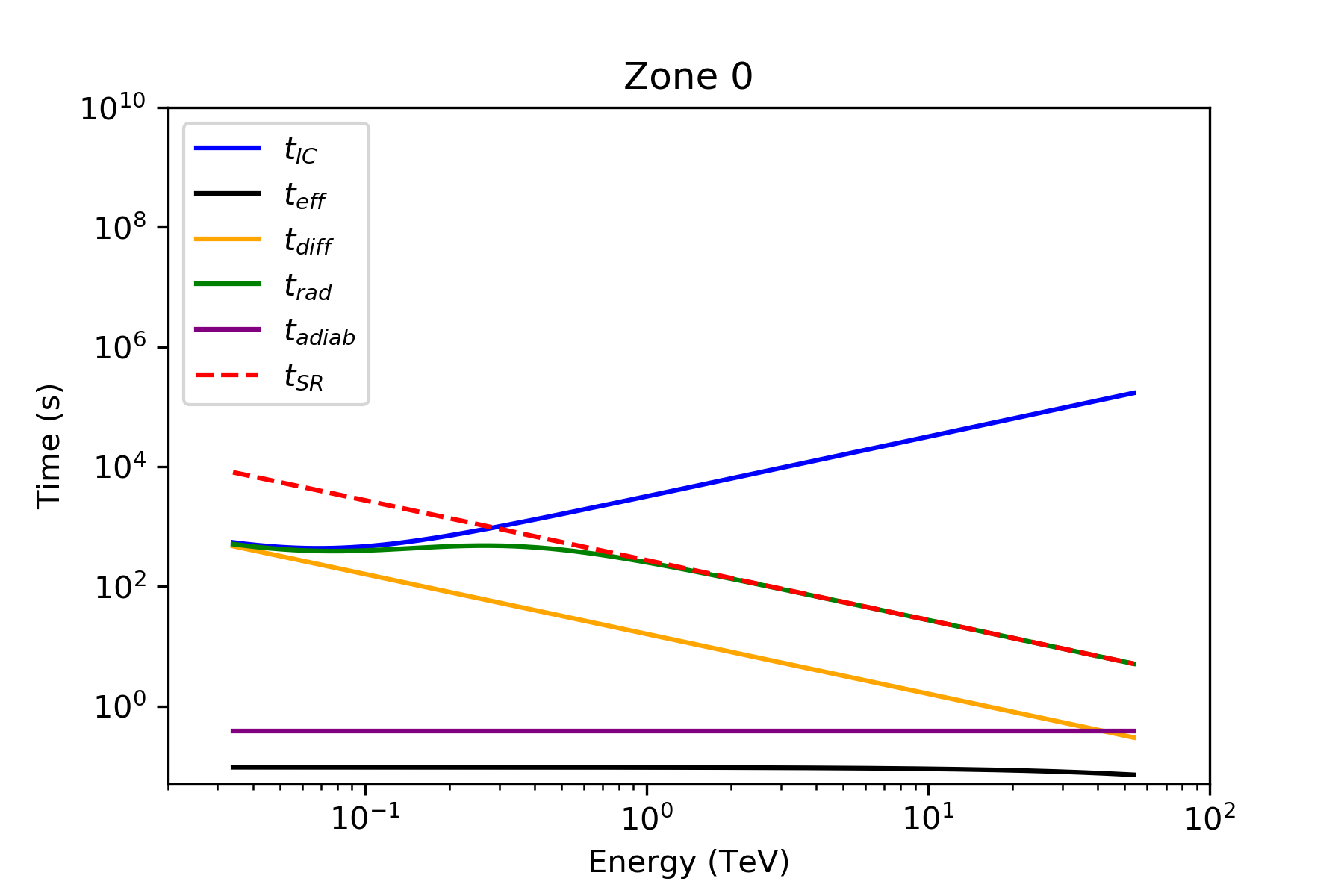}{0.33\textwidth}{(a)}
          \fig{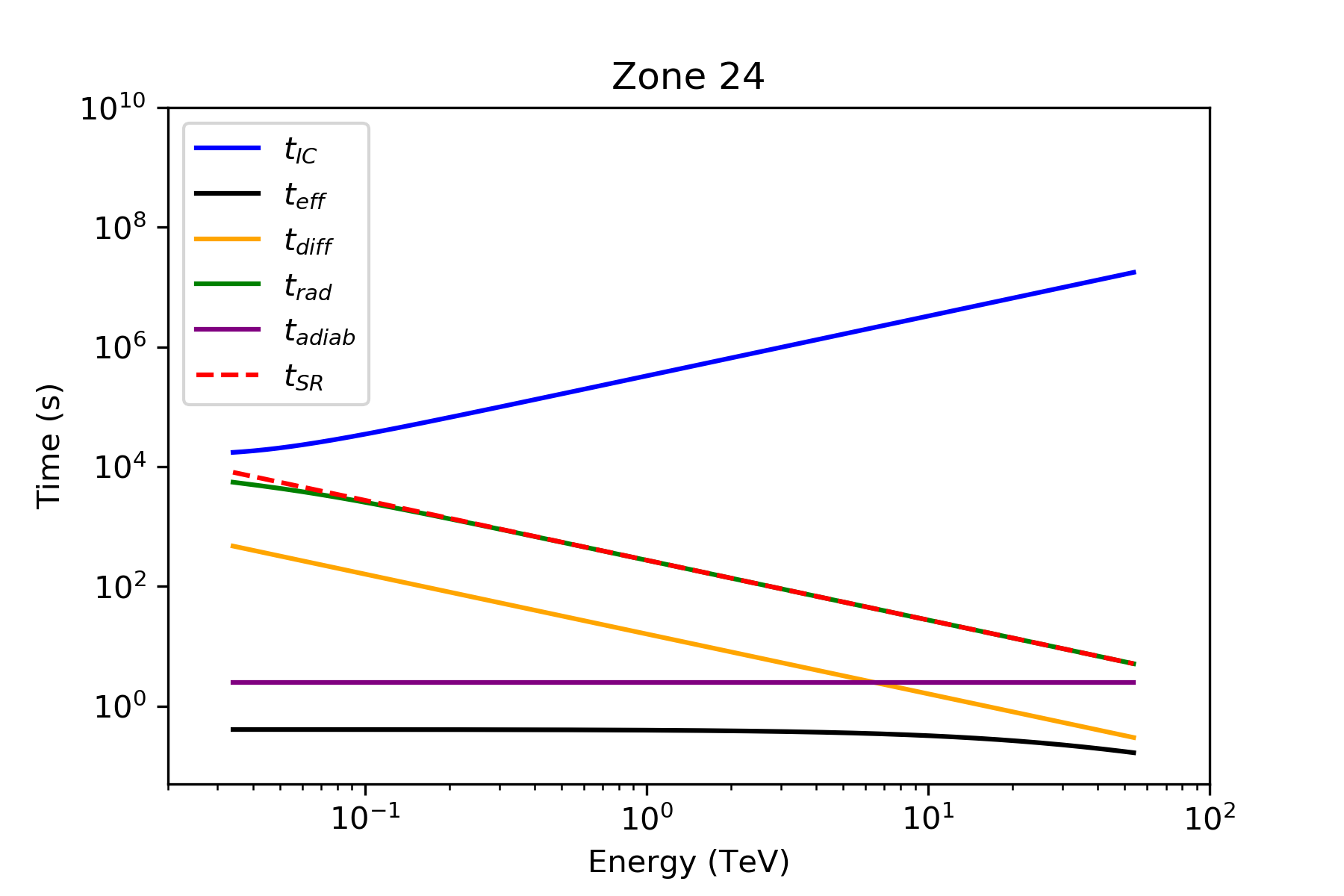}{0.33\textwidth}{(b)}
          \fig{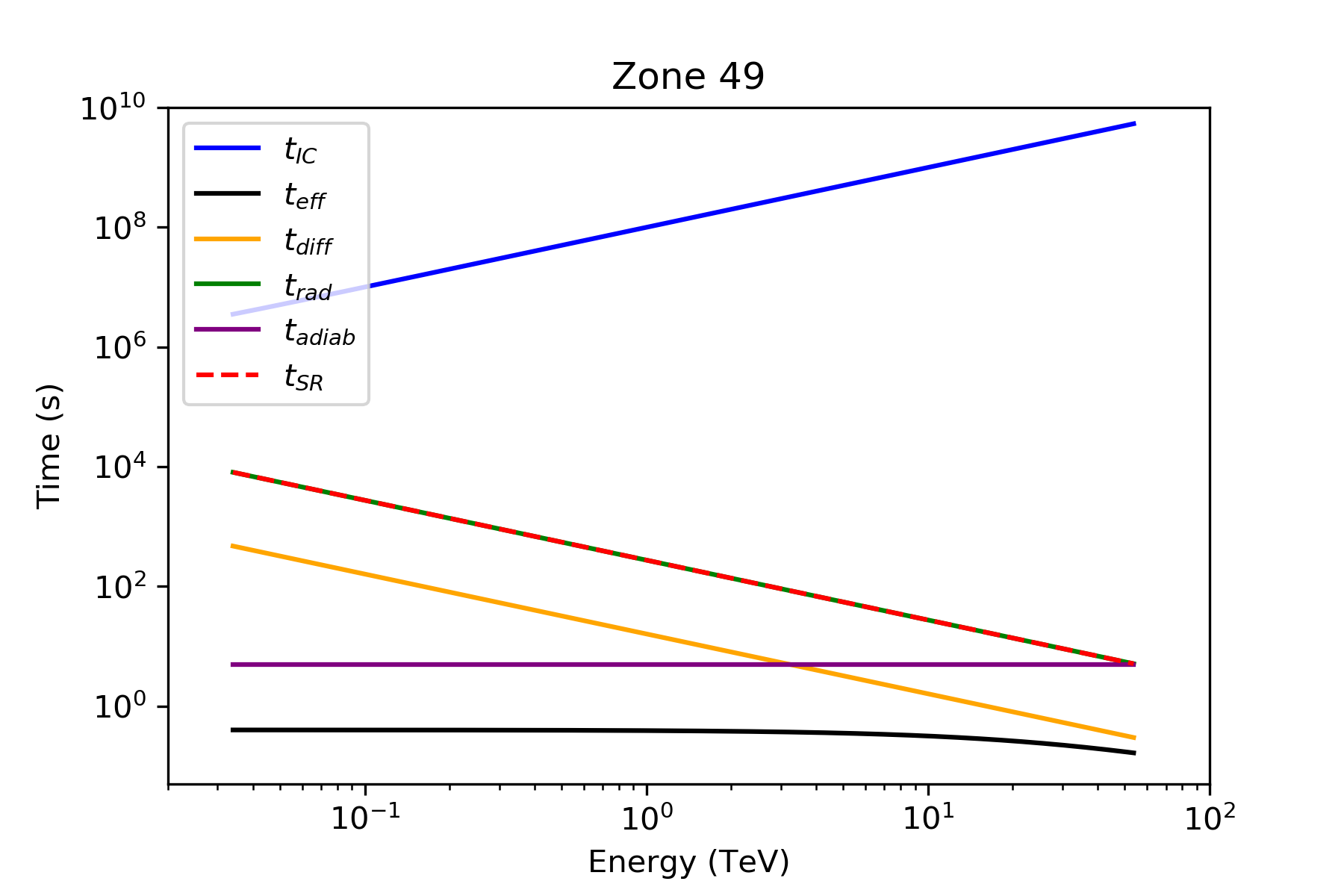}{0.33\textwidth}{(c)}
            }\label{fig:T3}
\caption{Timescale plots for the quiescent state of PSR~J1311$-$3430 depicting the (a) first, (b) middle and (c) last spatial zones.}
\end{figure}

In Fig.~17 we plot relevant timescales for the first, middle and final spatial zones for PSR~J1311$-$3430. We note that the IC timescale (blue line) is much shorter for the first zone than for the last. Note that we have modeled this source with the shock wrapping around the pulsar. The spatial timescales again dominate the radiation ones and this seems to be a characteristic for the RB systems we have modeled. The other timescales behave in a similar way to those described for PSR~J1723$-$2837.

\newpage
\begin{figure}[h!]
\gridline{\fig{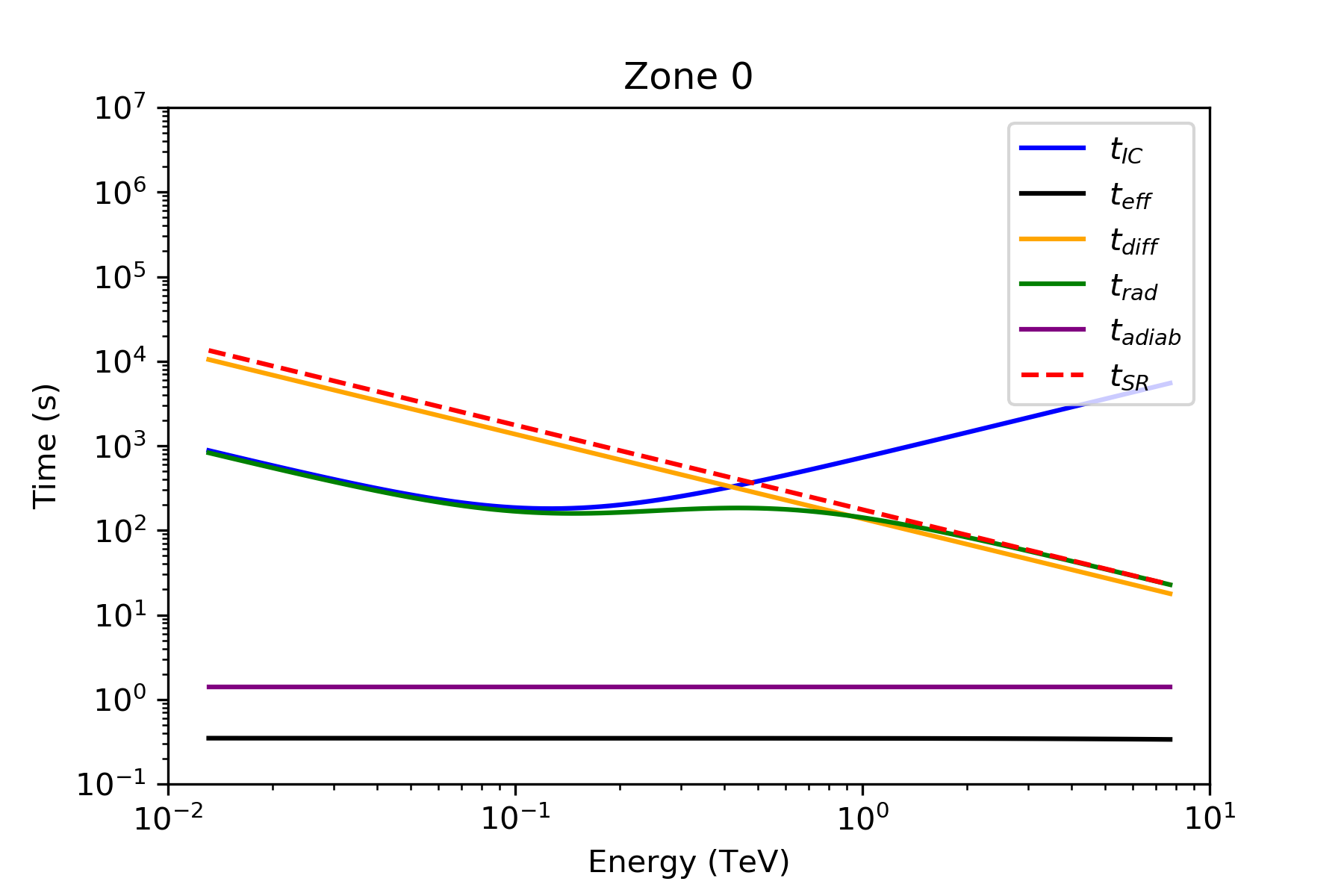}{0.33\textwidth}{(a)}
          \fig{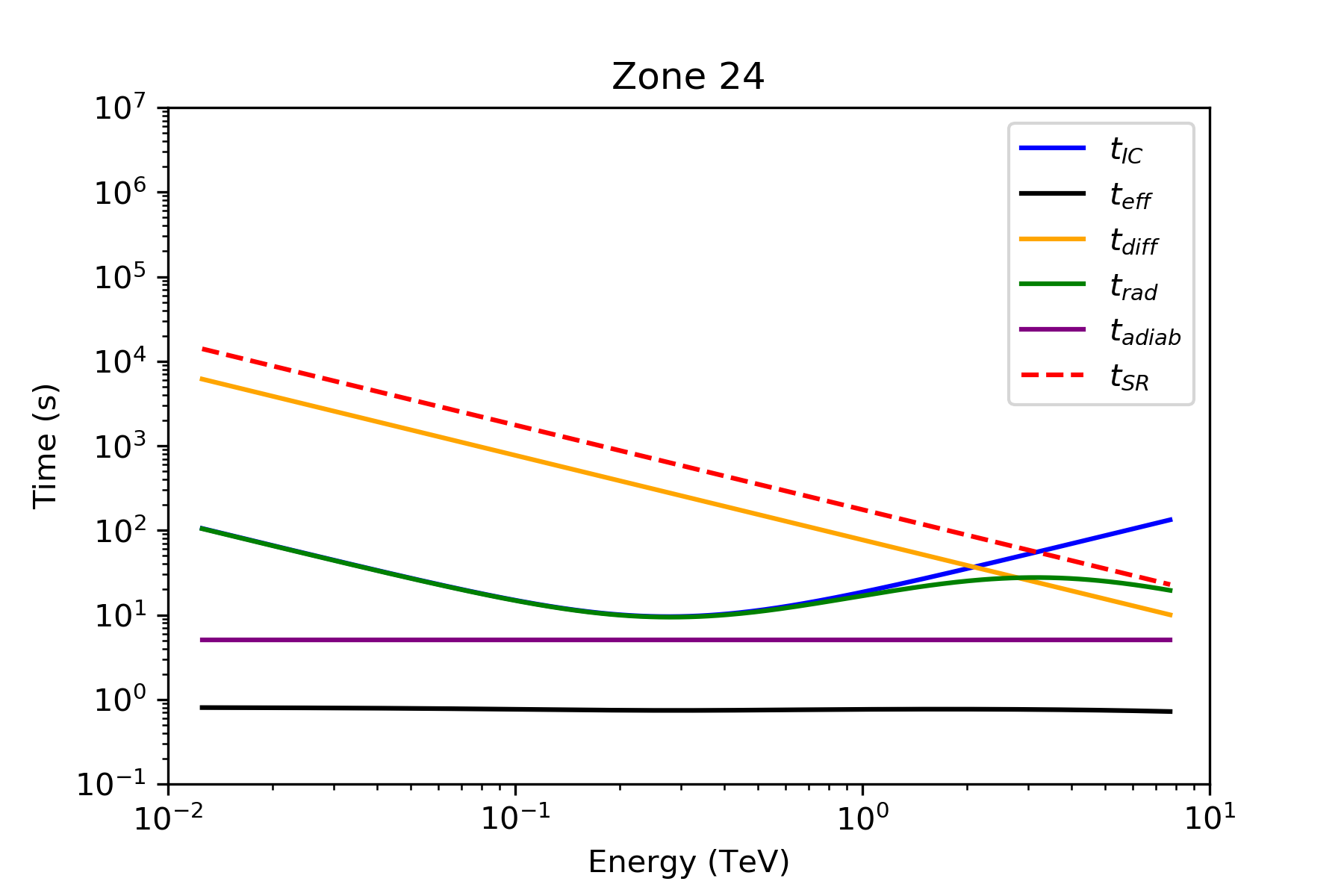}{0.33\textwidth}{(b)}
          \fig{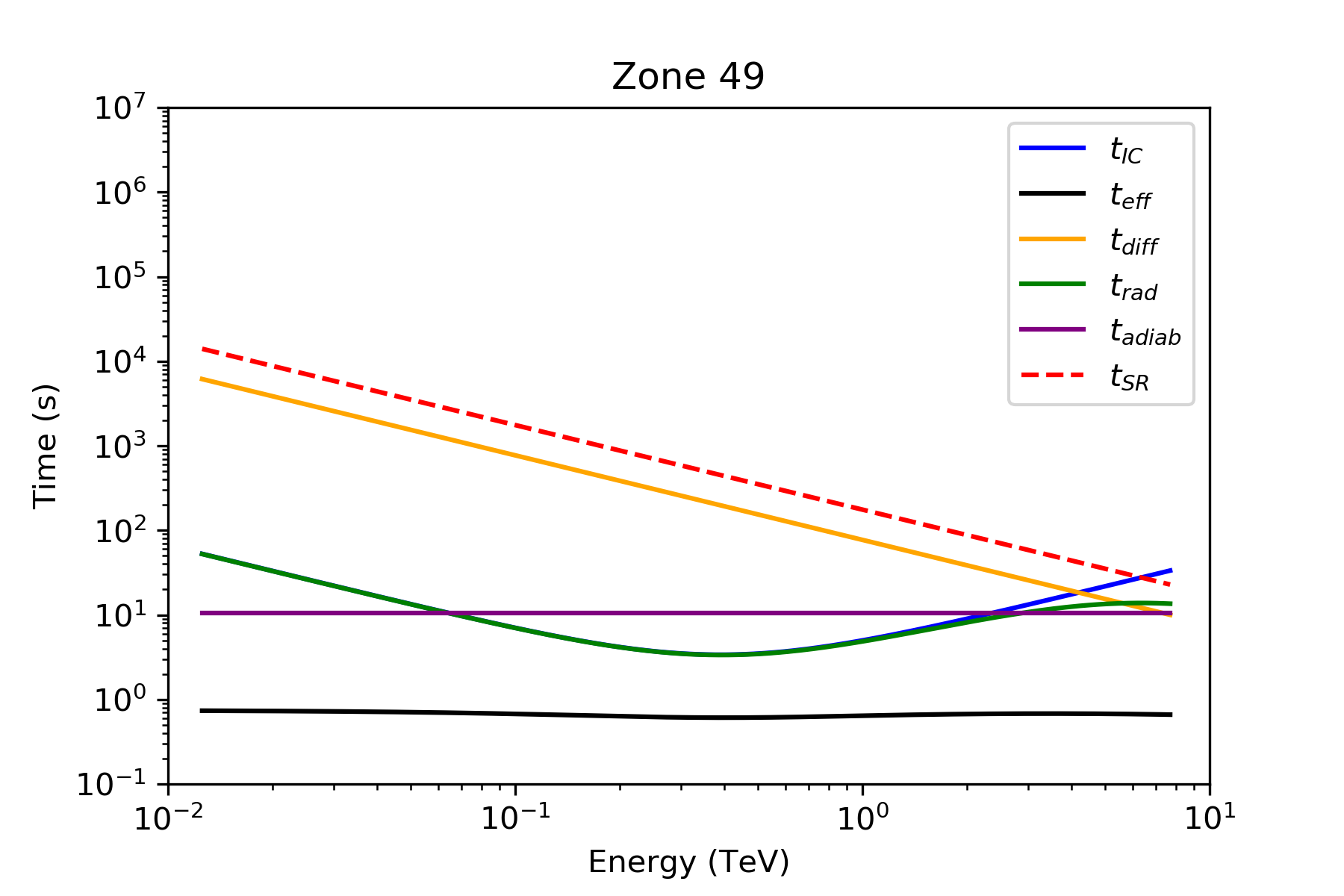}{0.33\textwidth}{(c)}
          }
\gridline{\fig{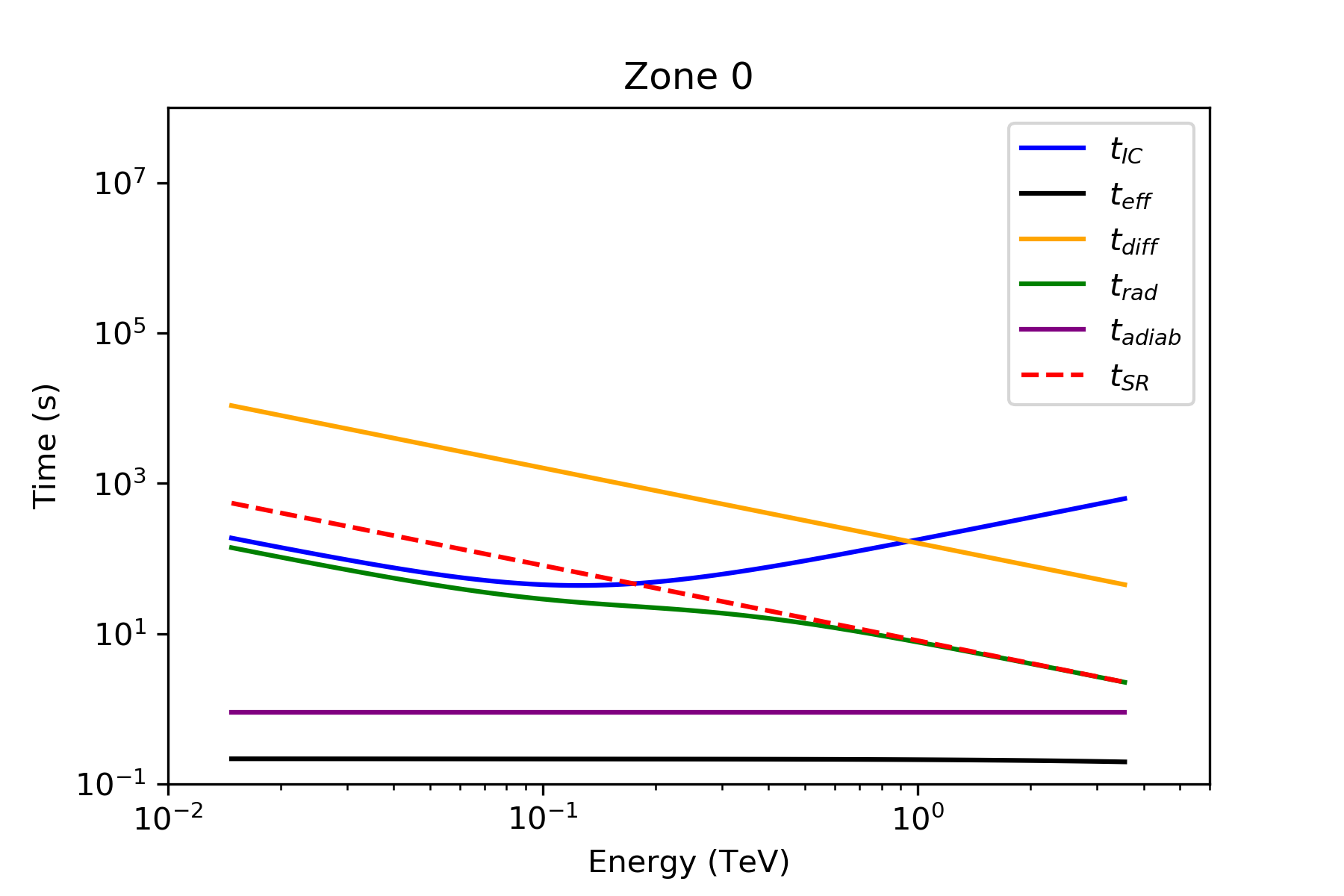}{0.33\textwidth}{(d)}
          \fig{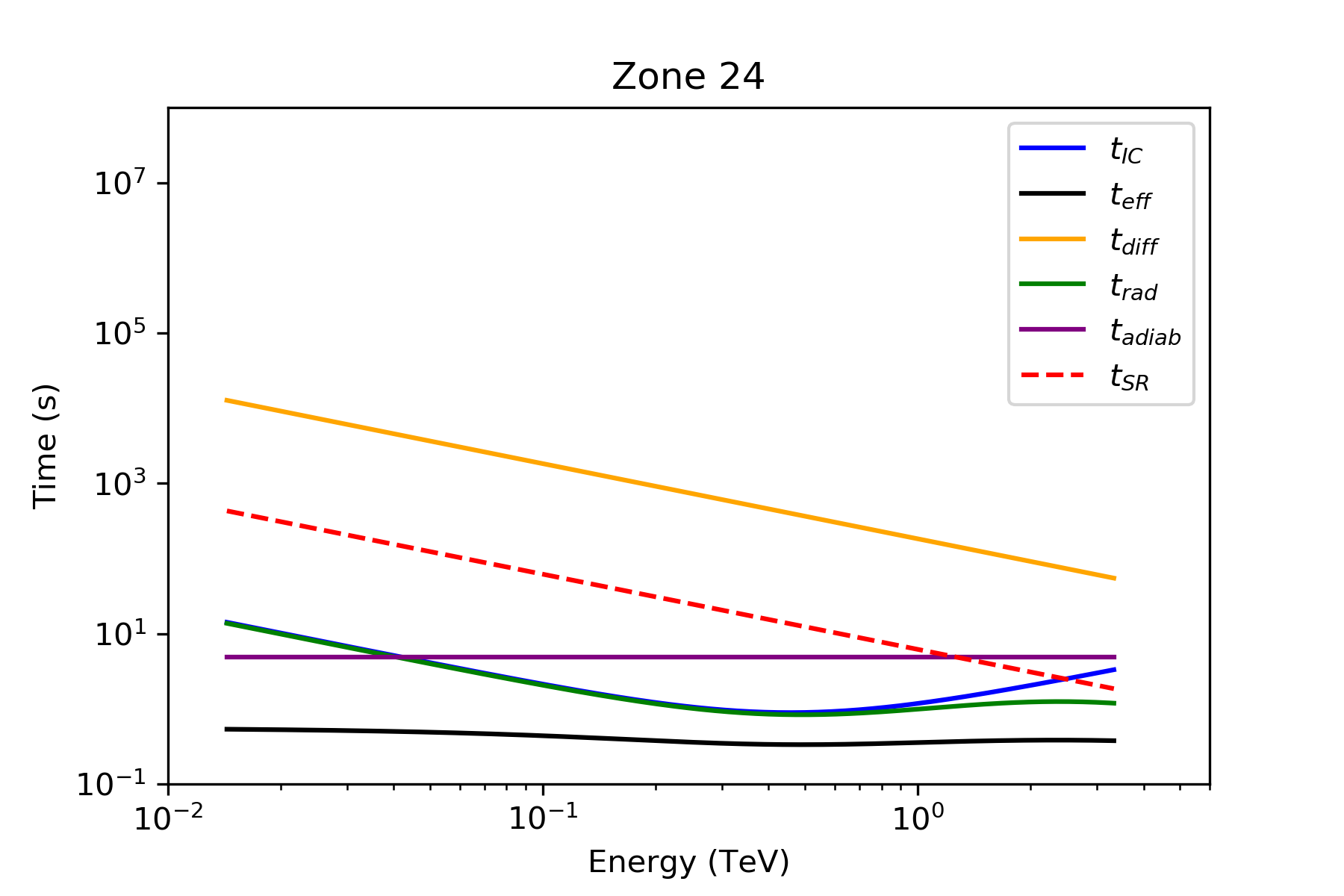}{0.33\textwidth}{(e)}
          \fig{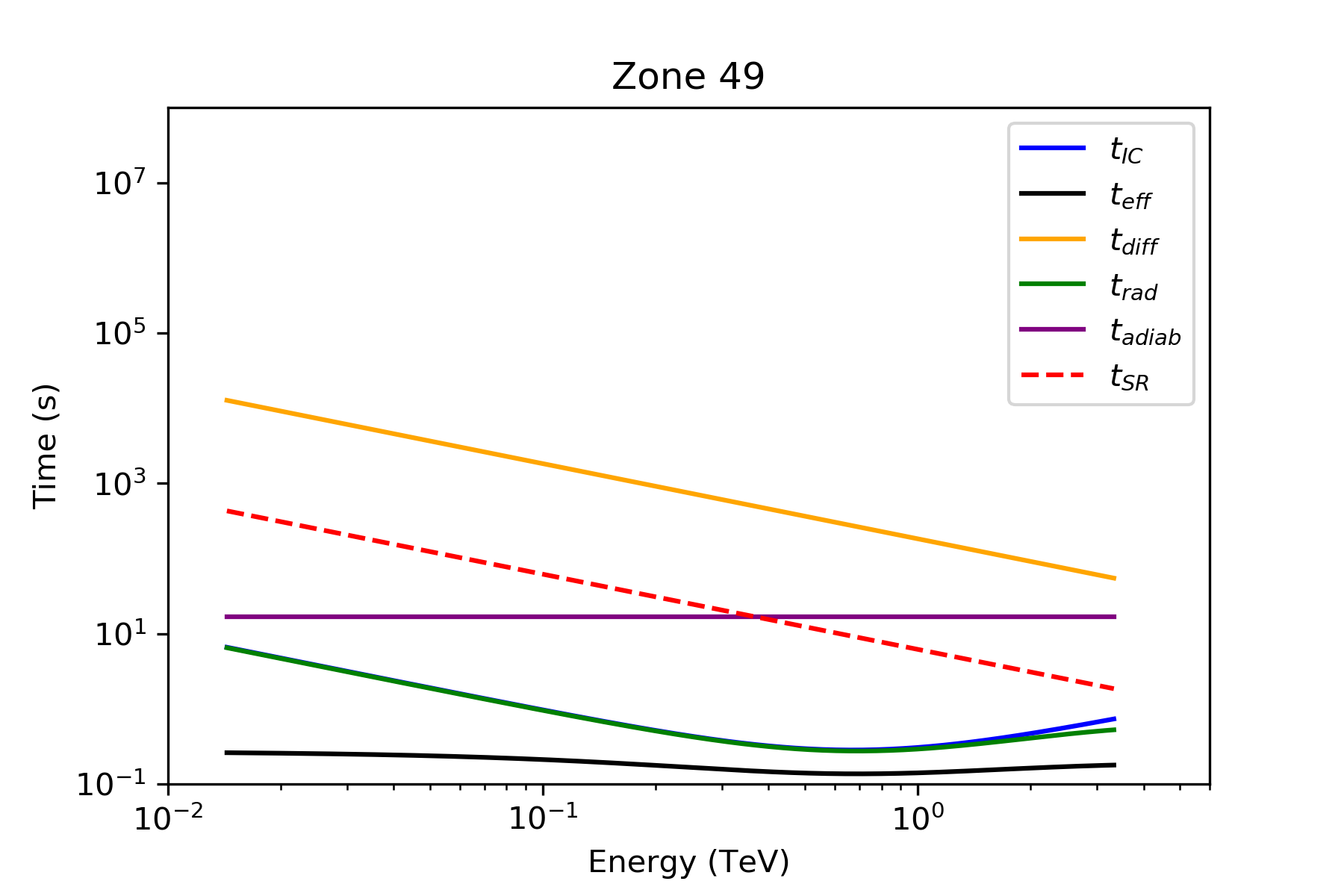}{0.33\textwidth}{(f)}
         }
         \label{fig:T4}
\caption{Timescale plots for PSR~B1957+20 depicting the (a) first, (b) middle and (c) last spatial zones for inclination angle $i = 64^{\circ}$. Similarly (d), (e) and (f) are for the alternative fit with inclination angle of $i = 84^{\circ}$.}
\end{figure}

In Fig.~18 we plot relevant timescales for the first, middle and final spatial zones for PSR~B1957+20. The IC timescale (blue line) is much longer for the first zone than for the last. This is because the shock is wrapping around the companion for this BW system and thus the shock is close to the soft-photon source for all zones. The reason for the shorter timescales for later zones is that the bulk flow speed increases for each consecutive zone, increasing the Doppler factor and thus the comoving soft-photon number density in these zones. The timescale for IC is even shorter for the higher inclination angle because in this case, $R_0$ is relatively smaller, so the shock is even closer to the companion and thus closer to the soft-photon field. The radiation timescales now dominate the spatial ones. This behaviour may thus be a distinguishing feature between BWs and RBs.

\vspace{5mm}

\bibliographystyle{aasjournal} 
\bibliography{bibfile}

\end{document}